\documentclass[12pt,preprint]{aastex}

\newcommand{\bet}{$\rm \beta~$}
\newcommand{\bets}{$\rm \beta$}

\newcommand{\ta}{$\rm \tau_{225\,GHz}~$}

\newcommand{\as}{$\rm ^{\,\prime\prime}~$}

\def\gs{\mathrel{\raise0.35ex\hbox{$\scriptstyle >$}\kern-0.6em \lower0.40ex\hbox{{$\scriptstyle \sim$}}}}
\def\ls{\mathrel{\raise0.35ex\hbox{$\scriptstyle <$}\kern-0.6em \lower0.40ex\hbox{{$\scriptstyle \sim$}}}}

\newcommand{\cm}{$\rm \,cm~$}

\newcommand{\K}{$\rm \,K~$}
\newcommand{\Ks}{$\rm \,K$}

\newcommand{\Lsun}{$\rm \,L_\odot~$}
\newcommand{\Lsuns}{$\rm \,L_\odot$}

\newcommand{\Msun}{$\rm \,M_\odot~$}

\newcommand{\DL}{$\rm D_{\tiny{L}}~$}

\newcommand{\Td}{$\rm T_d~$}
\newcommand{\Tds}{$\rm T_d$}

\newcommand{\Lfir}{$\rm L_{\mbox{\tiny{FIR}}}~$}
\newcommand{\Lfirs}{$\rm L_{\mbox{\tiny{FIR}}}$}

\newcommand{\Md}{$\rm M_d~$}

\begin{document}
\title{350$\,\mu$m Observations of Ultraluminous Infrared Galaxies at Intermediate Redshifts}
\author{M.\ Yang$^1$, T.\ R.\ Greve$^1$, C.\ D.\ Dowell$^1$ and C.\ Borys$^2$}
\affil{$^1$Caltech, Department of Physics, MC 320-47, Pasadena, CA 91125}
\affil{$^3$University of Toronto, Department of Astronomy \& Astrophysics, Toronto, ON, Canada, M5S 3H8}
\email{min@submm.caltech.edu}

\begin{abstract}
We present 350$\,\mu$m observations of 36 ultraluminous infrared galaxies (ULIRGs) 
at intermediate redshifts ($0.089 \leq z \leq 0.926$) using the Submillimeter High Angular Resolution Camera II 
(SHARC-II) on the Caltech Submillimeter Observatory (CSO). In total, 28 sources are detected at $\rm S/N \geq 3$,
providing the first flux measurements longward of 100$\,\mu$m for a statistically significant sample of 
ULIRGs in the redshift range of $0.1 \ls z \ls 1.0$. Combining 
our 350$\,\mu$m flux measurements with the existing {\it IRAS\,} 60 and 100$\,\mu$m data, 
we fit a single-temperature model to the spectral energy distribution (SED), and 
thereby estimate dust temperatures and far-IR luminosities. 
Assuming an emissivity index of \bet = 1.5, we find a median dust temperature and far-IR
luminosity of \Td = 42.8$\pm$7.1\K and log\,({\Lfirs}/{\Lsuns}) = 12.2$\pm$0.5, respectively. 
The far-IR/radio correlation observed in local star-forming galaxies is found to hold
for ULIRGs in the redshift range $0.1\ls z\ls 0.5$, suggesting that
the dust in these sources is predominantly heated by starbursts. 
We compare the far-IR luminosities and dust temperatures derived for dusty galaxy 
samples at low and high redshifts with our sample of ULIRGs at intermediate redshift. 
A general \mbox{\Lfirs-\Td} relation is observed, albeit with significant 
scatter, due to differing selection effects and variations in dust mass and grain properties. 
The relatively high dust temperatures observed for our sample compared to
that of high-$z$ submillimeter-selected starbursts with similar far-IR luminosities 
suggest that the dominant star formation in ULIRGs at moderate redshifts 
takes place on smaller spatial scales than at higher redshifts.
\end{abstract}


\section{Introduction}\label{intro}
Dust grains in the interstellar medium (ISM) of galaxies are heated by various energy sources 
in vastly different physical environments: the diffuse interstellar radiation field, active 
star formation and active galactic nuclei (AGN). The bulk of the emission from heated dust 
grains occurs at far-IR and submillimeter (submm) wavelengths, and can account for a significant, 
if not dominant, fraction of the total energy output from dust-enshrouded galaxies
where extinction and re-radiation by dust lead to high IR-to-optical flux ratios. 

In the latter category we find the population of ULIRGs ($\rm log\,(L_{\mbox{\tiny{IR}}}/L_{\odot}) \geq 12$),  
which were first discovered in large numbers in the local Universe by {\it IRAS\,} 
(Soifer et al. 1986). These galaxies harbor vast amounts of dust, responsible for 
obscuring most of the intense star formation and AGN activity that are taking place. 
Observations in the optical/near-IR as well as of rotational CO line emission revealed that ULIRGs 
tend to be associated with interacting or merging gas-rich disks, with the more luminous 
being the more disturbed and morphologically complex systems (Sanders \& Mirabel\ 1996), thus
suggesting that ULIRGs are driven by mergers in which gas is funneled towards
the central regions, fueling an intense starburst and/or a black hole.
While local ($z\ls 0.1$) ULIRGs are believed to re-enact some of the  processes relevant for
galaxy formation at high redshifts, it is essential that studies of ULIRGs over a wide redshift 
range are carried out, in order to get a complete picture of how the most massive galaxies
formed as a function of redshift.
At high redshifts ($1\ls z\ls 3$), the population of optically faint
radio sources -- which includes (sub)mm-selected galaxies (SMGs) 
uncovered by {\sc scuba} and {\sc mambo} 
(Smail et al.\ 1997; Hughes et al.\ 1998; Bertoldi et al.\ 2000; Carilli et al.\ 2001)
as well as the population of optically faint 24$\,\mu$m sources selected with {\it Spitzer} (Papovich et al.\ 2004)
-- is thought to represent the high-$z$ analogues of local ULIRGs.

In contrast to the populations of local and distant ULIRGs,
the population at intermediate redshifts ($0.1\ls z\ls 1.0$)
remains virtually unexplored at far-IR/submm wavelengths.
This intermediate redshift interval is of particular interest as it 
bridges the gap between the distant and local ULIRG populations, and
broadly marks the transition period where ULIRGs cede their position 
to the less extreme luminous infrared galaxies (LIRGs -- $\rm log\,(L_{\mbox{\tiny{IR}}}/L_{\odot}) \geq 11$)
as the dominating contributor to the co-moving IR luminosity density (Le Floc'h et al.\ 2005).
In addition, the cosmic infrared background (CIRB -- Hauser et al.\ 1998; Schlegel et al.\ 1998), 
which peaks at $\sim$150$\,\mu$m and constitutes about half of the total
energy density in the Universe, is dominated by emission from (U)LIRGs in the above
intermediate redshifts range (Lagache et al.\ 2005).

In this paper we present 350$\,\mu$m observations of a statistically significant sample of
36 ULIRGs in the intermediate redshift range $0.089 \leq z \leq 0.926$. 
Throughout this paper we adopt a flat cosmology with $\rm H_0=70\,{km\,s^{-1}\,Mpc^{-1}}$, 
$\rm \Omega_m=0.30$ and $\rm \Omega_{\Lambda}=0.70$.

\section{The Sample}\label{ffsample}
\begin{figure}
\begin{center}
\includegraphics[width=5in, angle=90]{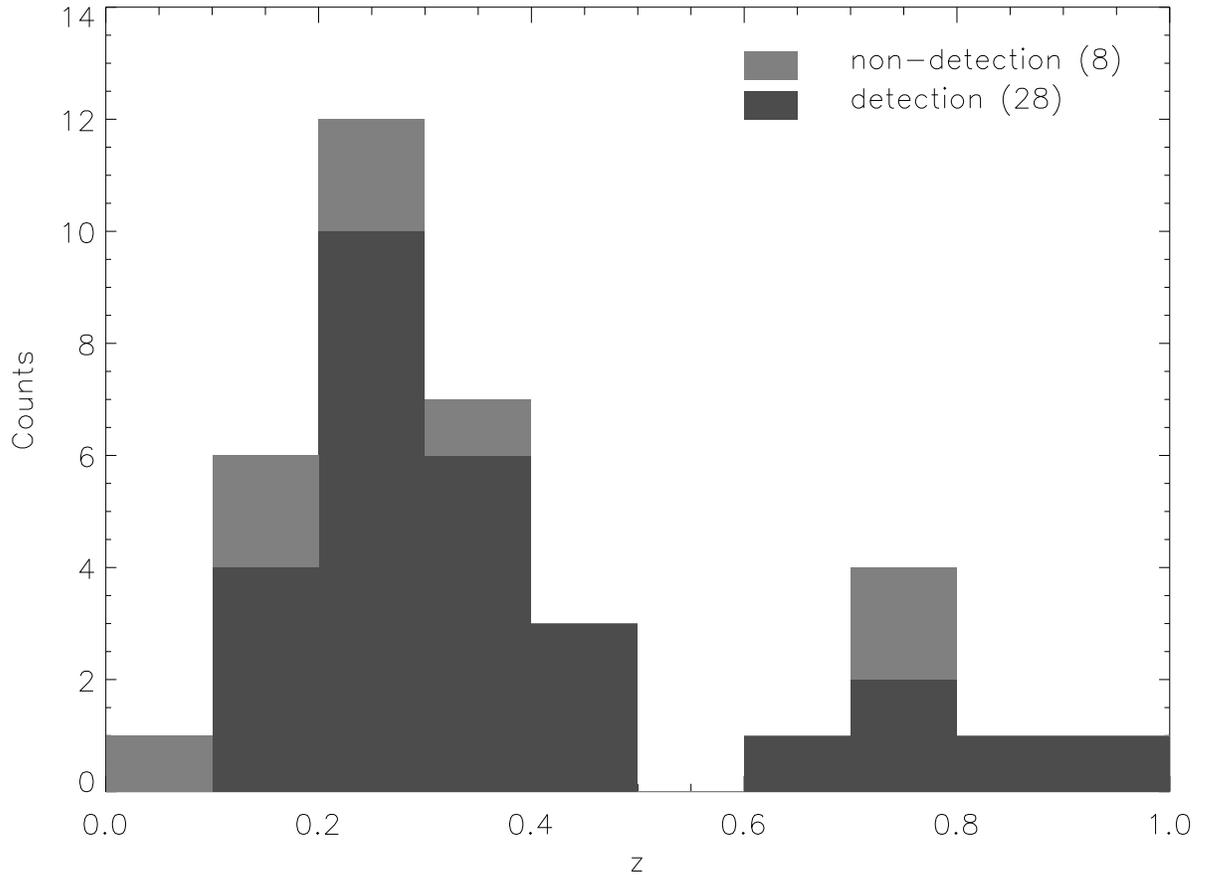}
\caption{The redshift distribution of the 36 FF sources selected for SHARC-II observations.}\label{z_histo}
\end{center}
\end{figure}

Our sample is a subset of the FSC-FIRST (FF) catalog (Stanford et al.\ 2000), 
which was compiled by cross-correlation between the IRAS Faint Source Catalog (FSC -- Moshir et al.\ 1992) and 
the Faint Images of the Radio Sky at Twenty \cm (FIRST -- Becker et al.\ 1995), 
and consists of 108 ULIRGs for which spectroscopic redshift information 
and $K$-band imaging were successfully obtained. Situated in the redshift 
range of $0.1 \ls z \ls 1$, the FF sources constitute a well 
defined and statistically significant sample of ULIRGs at intermediate redshifts. 
Only sources with secure {\it IRAS\,} detections at both 60 and 100\,$\mu$m were 
considered for 350\,$\mu$m observations, with no flux cutoff employed at either waveband. 
The fact that virtually no other far-IR/submm continuum data points exist longward of 100\,$\mu$m for the FF sources, makes
our 350$\,\mu$m flux measurements extremely valuable for attaining meaningful estimates of the SEDs.
Furthermore, the availability of the spectroscopic redshift information allows
accurate estimates of dust temperatures and far-IR luminosities.
Finally, we note that targeted observations at 350$\,\mu$m were made possible due to 
the positional accuracy provided by the existing radio observations.
This is highly desirable for far-IR/submm observations given 
the limited angular resolution currently available to telescopes 
working in this wavelength range. In total, we observed 36 sources 
in the FF catalog at 350$\,\mu$m, which are listed in 
Table \ref{target}, along with their redshifts 
and flux densities at 60\,$\mu$m, 100$\,\mu$m and 1.4\,GHz.
The redshift distribution of our sample, having a sample median of 0.366, is shown in Fig.\ \ref{z_histo}.

There are a few selection biases possibly present in the FF catalog. 
First, high-redshift galaxies are potentially underrepresented
due to the adverse $K$-correction at 60$\,\mu$m and 1.4\,GHz.
Second, since the FSC is a flux-limited sample (the 60\,$\mu$m cutoff flux density is $\sim 0.2\,$Jy -- Moshir et al.\ 1992),
the selection function is sensitive to dust temperature and biased against galaxies 
with low dust temperatures. Third, galaxies with enhanced AGN activity are probably overrepresented 
in the sample, since such sources are expected to be more luminous (Genzel et al.\ 2000).

\section{Observations}
The observations were made during a series of SHARC-II (Dowell et al.\ 2003) observing runs from 
January 2003 to September 2004 at the CSO 
under excellent weather conditions (\ta$\leq$ 0.06). 
Integration time varied from source to source, 
depending on the brightness of each source at 350$\,\mu$m, 
atmospheric opacity and sky variability. 
On average, 1.5 hours of integration time was spent on each source. 
All measurements were made by scanning the bolometer array in a Lissajous pattern 
centered on each source.
With the exception of FF\,0738$+$2856, FF\,0758$+$2851, and FF\,0804$+$3919, 
scans were taken without chopping the secondary mirror.
Pointing and calibration scans were taken on an hourly basis on strong, point-like
submillimeter sources. SHARC-II raw data - taken on both calibration and science sources -
were reduced by using the software package CRUSH (Kov\'{a}cs\ 2006), 
with the same ``crush'' option ``-deep''.
Every single calibration scan was reduced individually, 
to obtain a 350$\,\mu$m map and flux density. 
The instrument flux was compared to the true flux\footnote{Available from http://www.submm.caltech.edu/sharc/},
and a calibration factor, defined as $X_{\rm cal} = {S_{\rm true}}/{S_{\rm instrument}}$, was calculated.
We also applied the built-in CRUSH utility ``calibrate'' to the 350$\,\mu$m
calibrator map, to have a measure of the actual beam size (FWHM).
All X factors and FWHM values thereby estimated were complied for each evening, 
from which calibration scans that yielded $0.5 \leq X \leq 2$ and $\rm FWHM \leq 20^{\,\prime\prime}$
were deemed reliable, and thus selected to form a subset. Finally, the sample median of the $X_{\rm cal}$ factor subset 
is taken as the $X_{\rm cal}$ factor appropriate for use to reduce science scans taken 
on the same evening. The calibration accuracy, as measured by the standard deviation of 
the $X_{\rm cal}$ factor subset, was found to be generally within 20\%. Consistent calibrated flux measurements were also 
found between the chopped and unchopped observations.

\begin{figure*}
\begin{center}
\begin{tabular}{cccc}
\includegraphics[width=1.5in, angle=270]{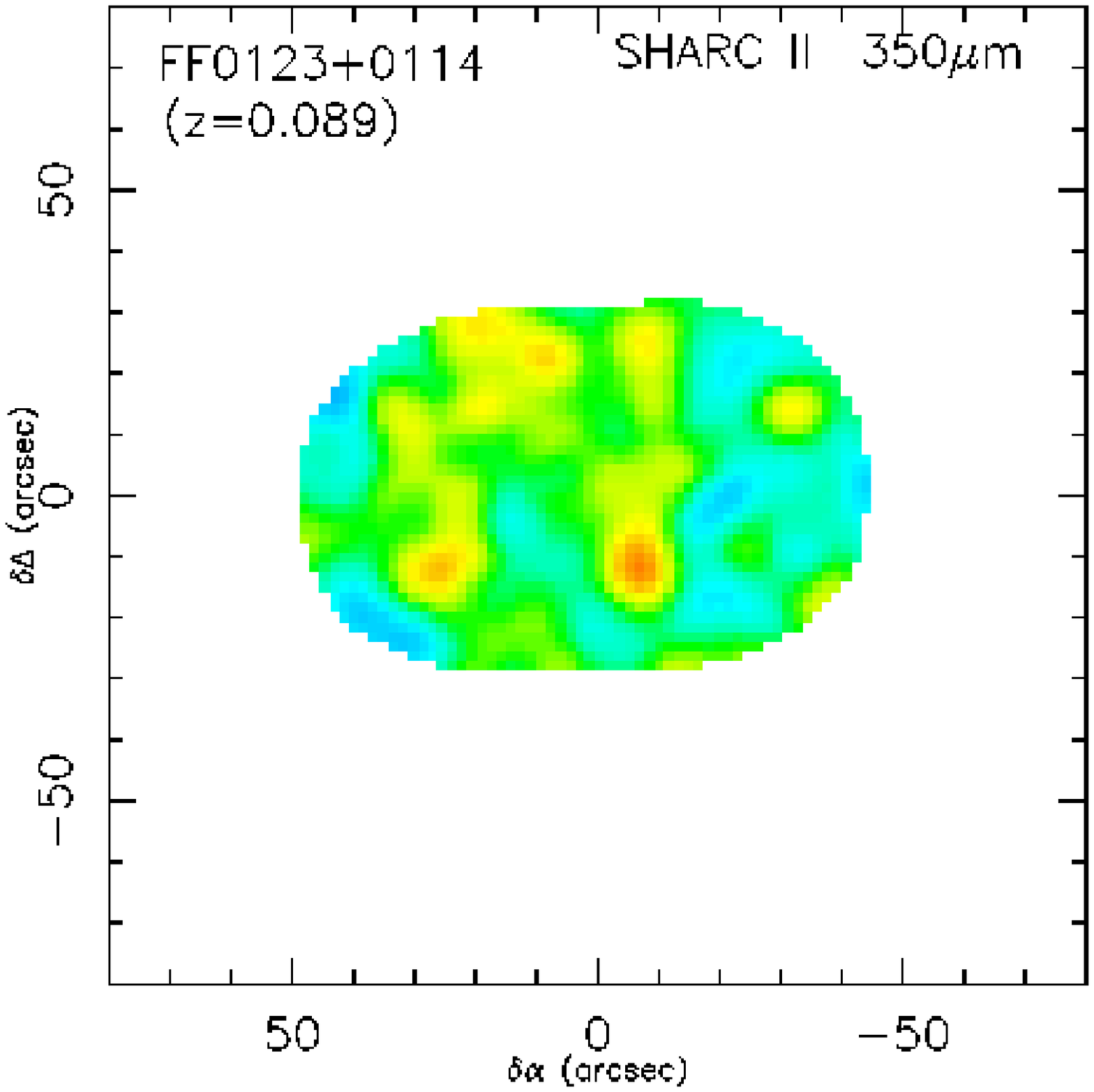}&
\includegraphics[width=1.5in, angle=270]{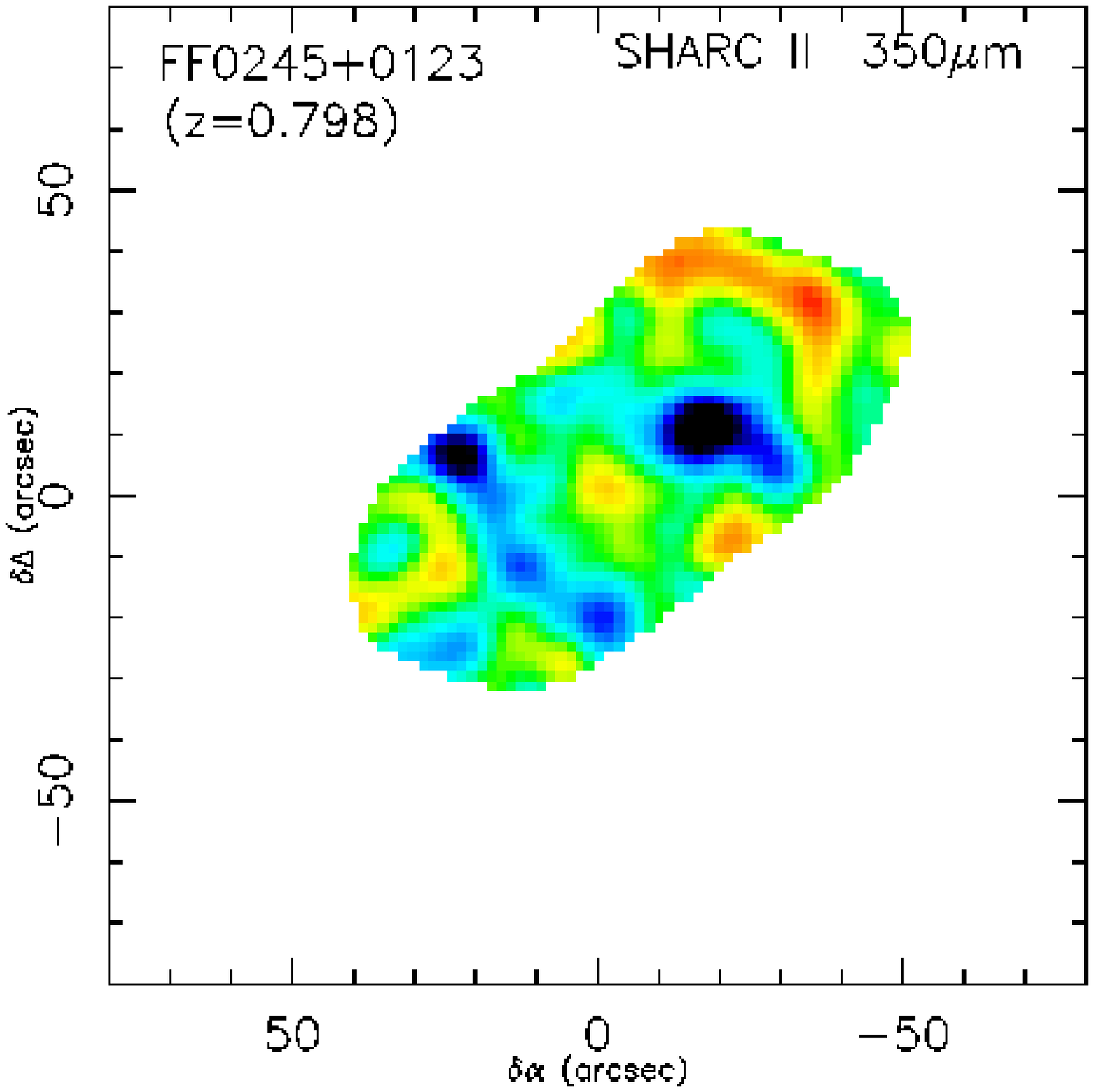}&
\includegraphics[width=1.5in, angle=270]{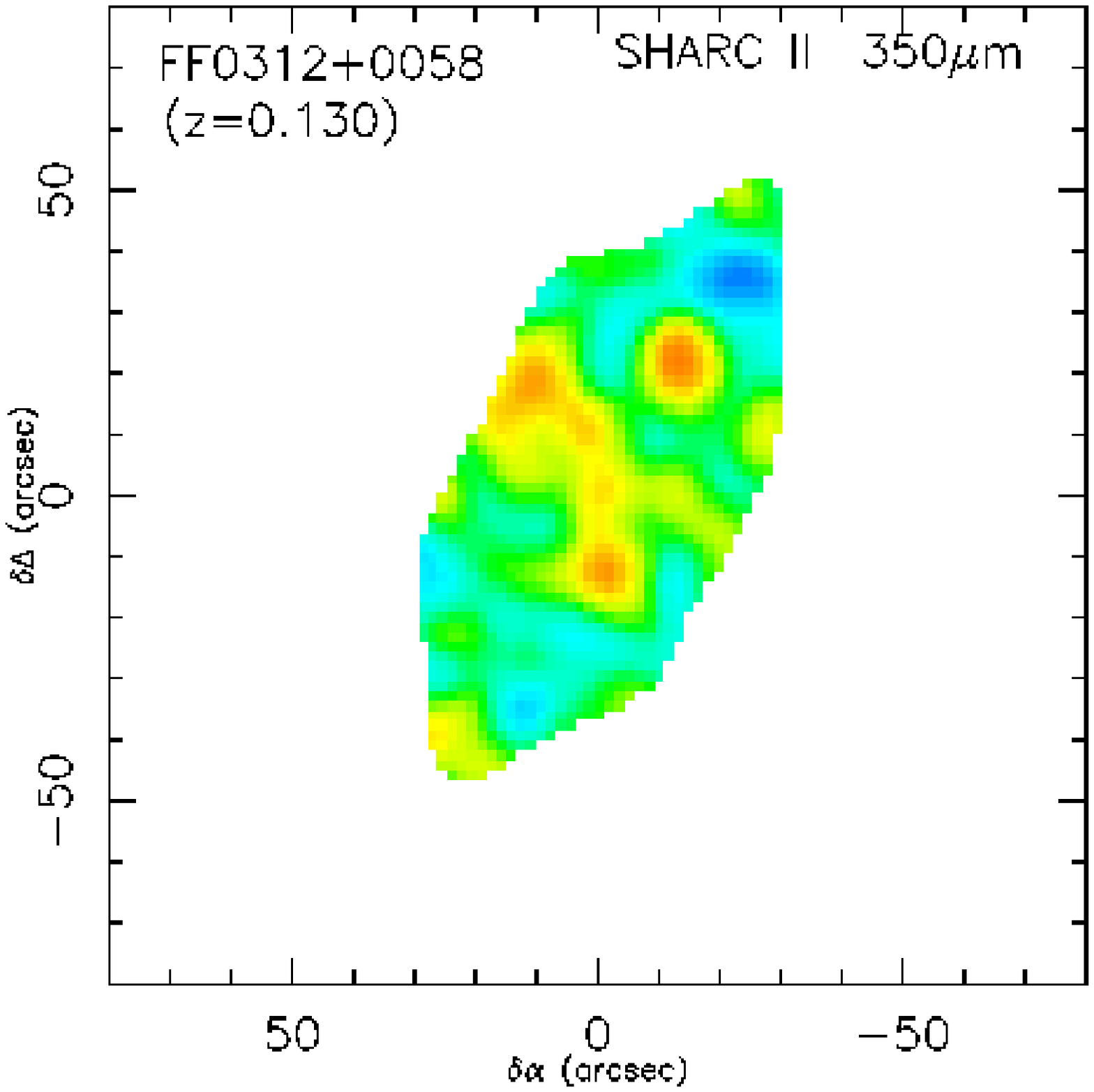}&
\includegraphics[width=1.5in, angle=270]{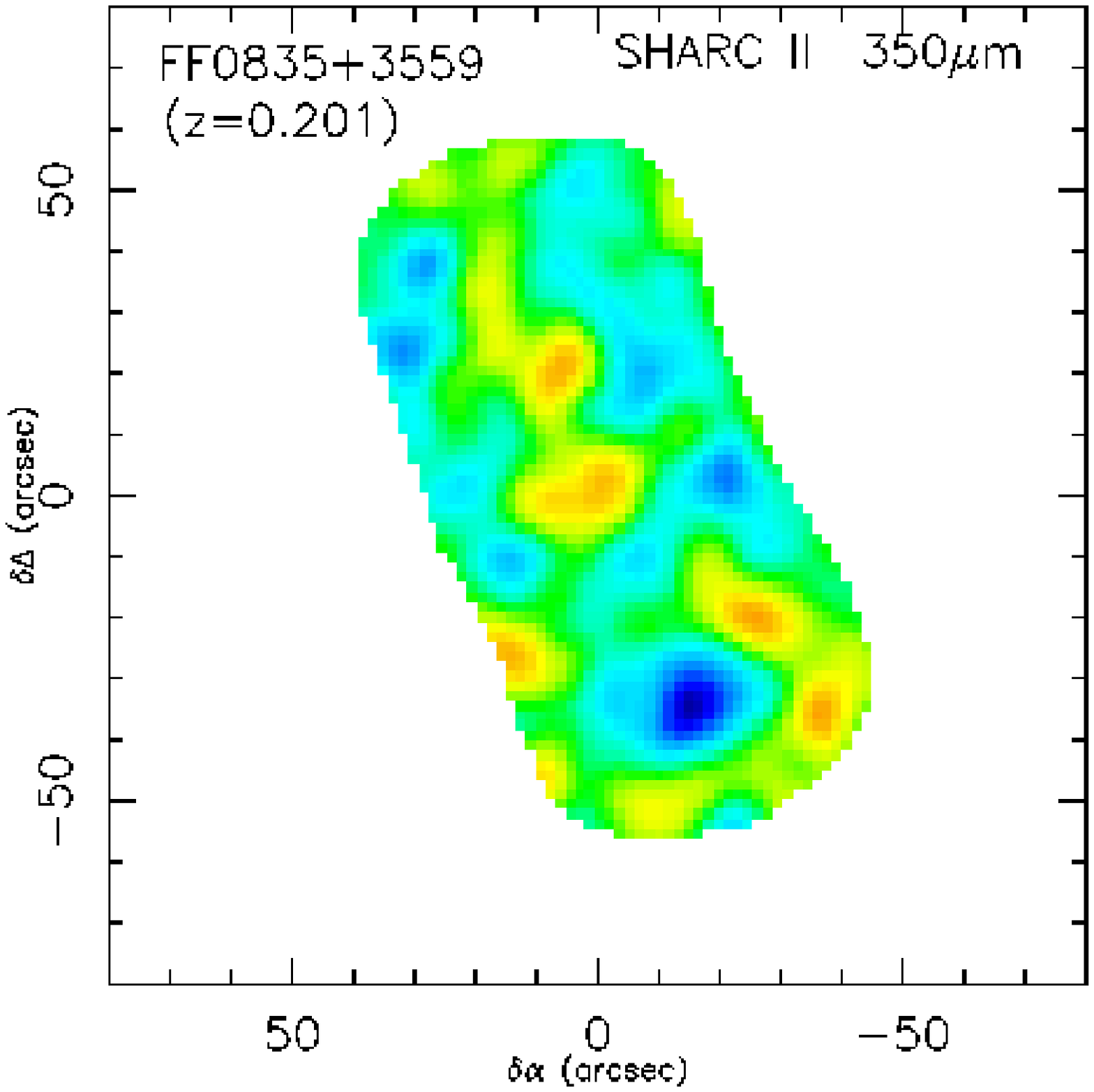}\\
\includegraphics[width=1.5in, angle=270]{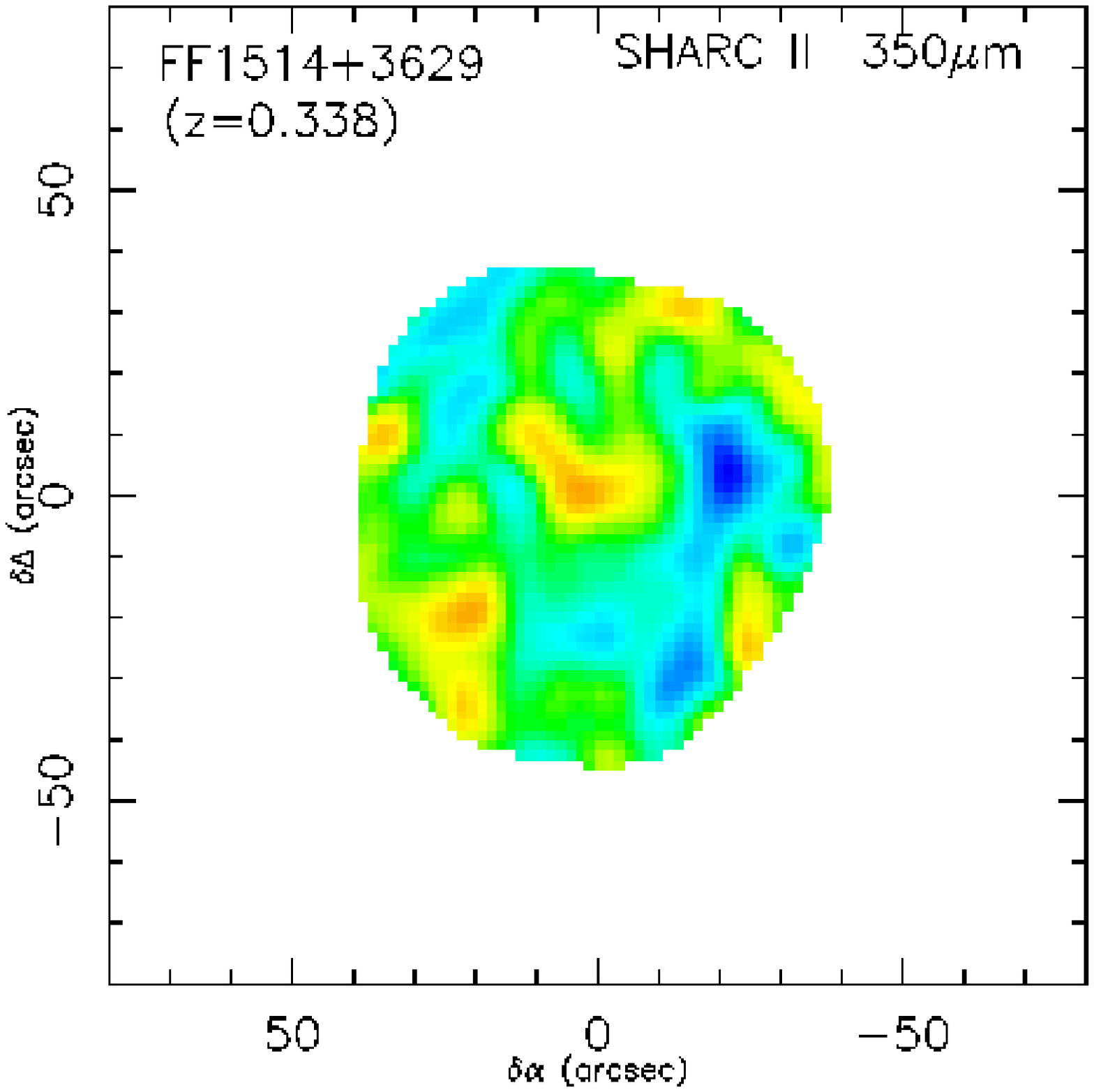}&
\includegraphics[width=1.5in, angle=270]{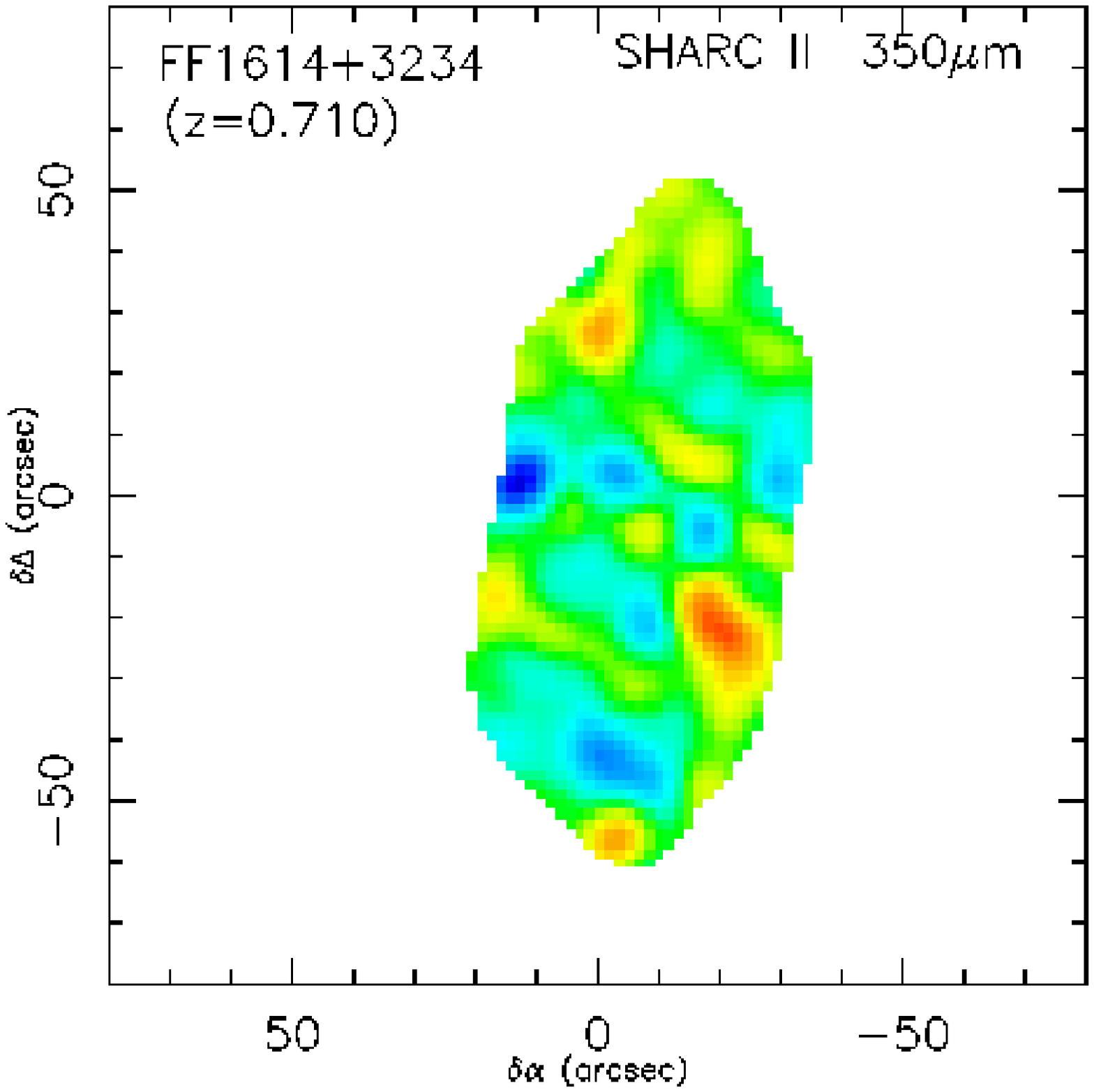}&
\includegraphics[width=1.5in, angle=270]{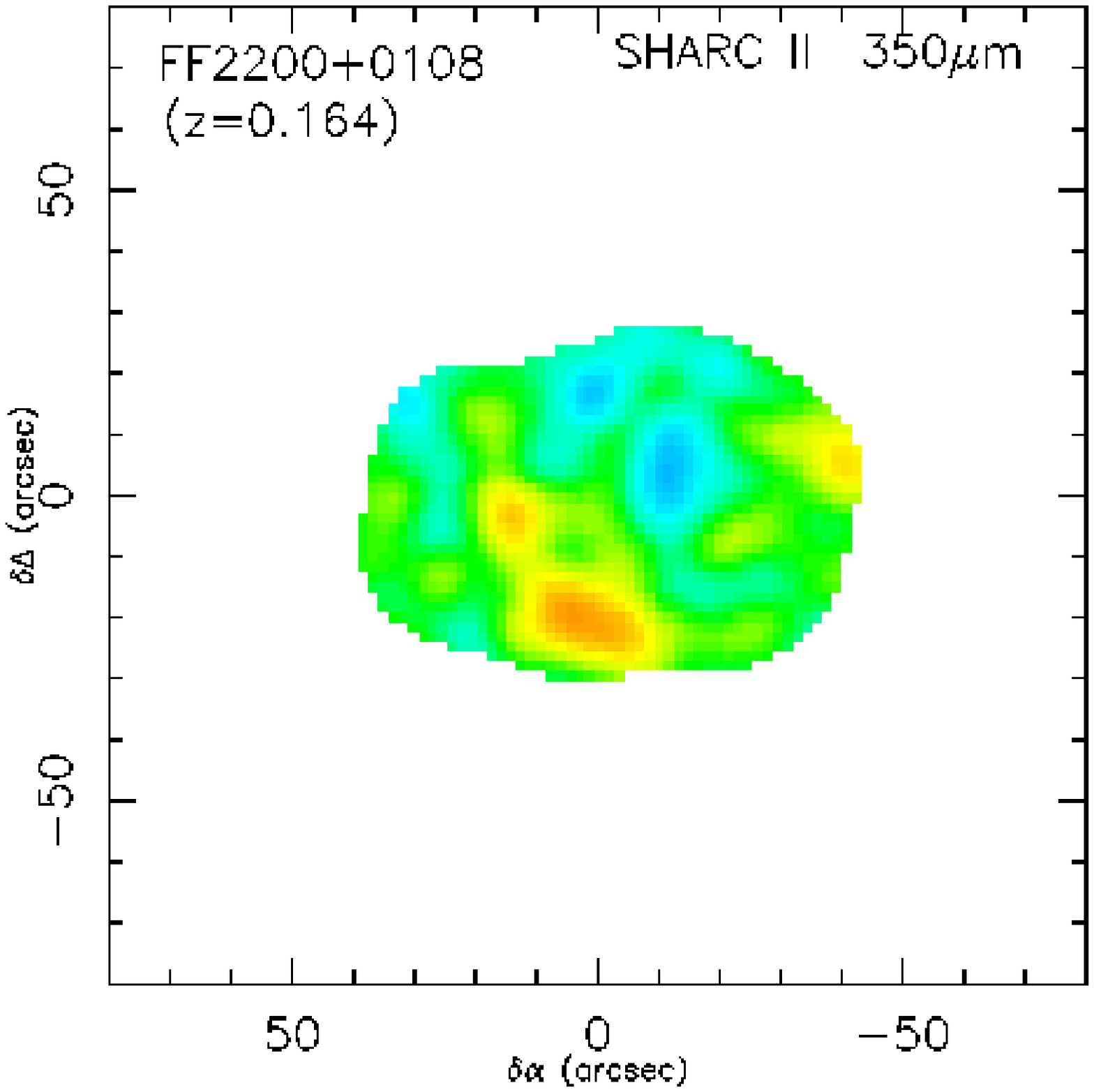}&
\includegraphics[width=1.5in, angle=270]{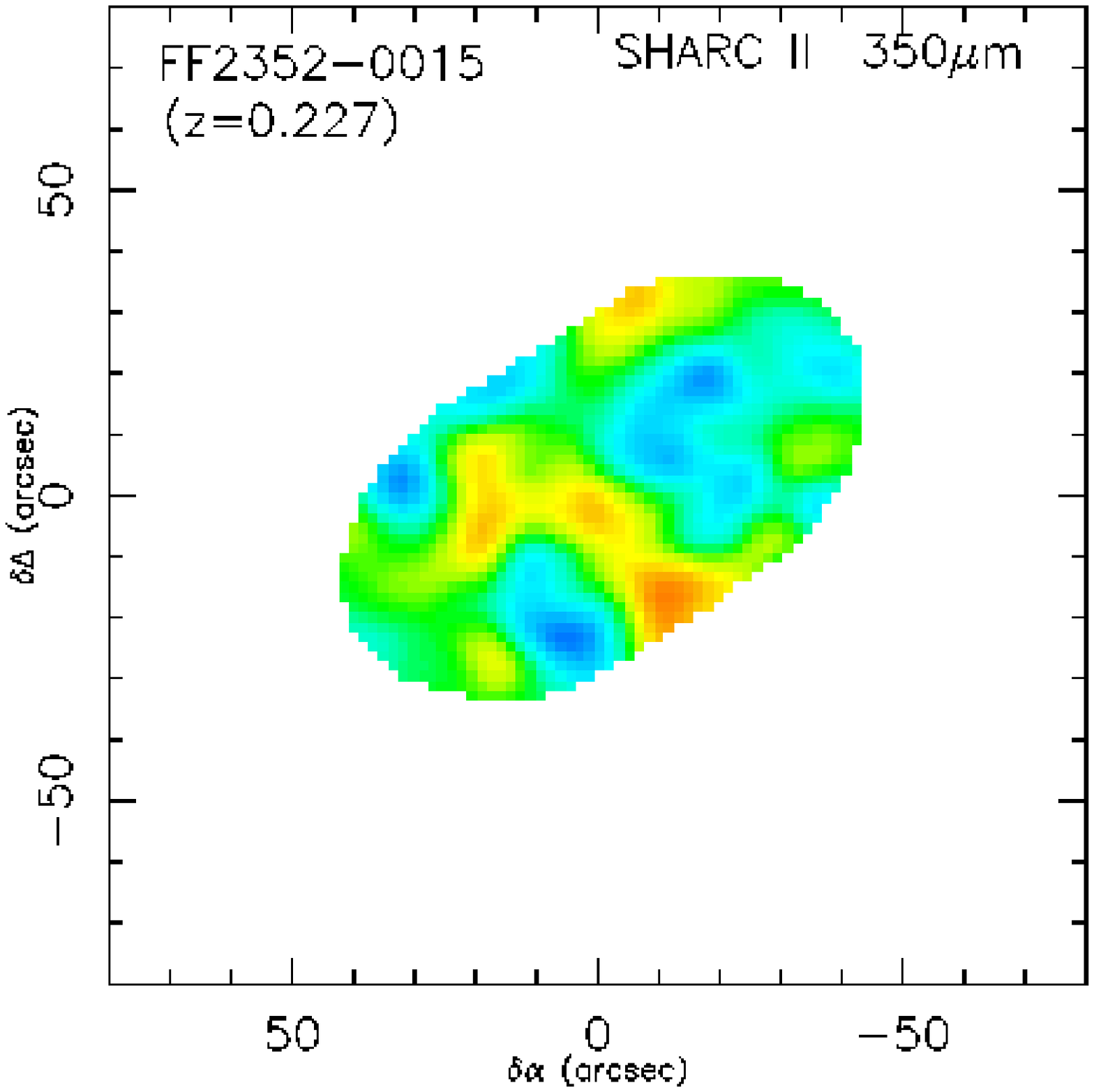}\\
\end{tabular}
\caption{350$\,\mu$m sigma-to-noise maps of the eight FF sources not detected by SHARC-II.}\label{350map_stanford_nondetection}
\end{center}
\end{figure*}

In total, 28 out of the 36 sources in our sample were detected with signal-to-noise ratios $\rm S/N \geq 3$. 
This high detection rate ($\sim$78\%) demonstrates the effectiveness of SHARC-II 
for studying faint 350$\,\mu$m sources at the $50-100$\,mJy level.
Other than the {\it IRAS\,} data, the present results at 350$\,\mu$m are the first reported 
detections of FF sources at submm wavelengths longward of 100$\,\mu$m, except for FF\,1532$+$3242 and FF\,1614$+$3234 
which were previously observed at 90 and 180$\,\mu$m by Verma et al.\ (2002), 
and at 850$\,\mu$m by Farrah et al.\ (2002) respectively. 
Table \ref{target} lists the 350$\,\mu$m flux measurements of the FF sources selected for SHARC-II observations.
The final 350\,$\mu$m signal maps smoothed with a 9\as FWHM beam are shown in 
Figs.\ \ref{350map_stanford_detection} and \ref{350map_stanford_nondetection}. 

\section{Spectral Fits and Derived Quantities}\label{stanford_sed}
We adopt the single-temperature, optically-thin far-IR/submm SED model (Hildebrand 1983)
\begin{equation}
\rm S_\nu = \Omega \, B_\nu(T_d) \, Q_\nu \label{sed_function}
\,,\end{equation}
where \Td is the dust temperature, 
$\rm Q_{\nu} = Q_0 \, (\frac{\nu}{\nu_{0}})^\beta$ is the absorption coefficient
and \bet is the emissivity index. 
A flux error at a uniform 20\% level is assumed at each data point
\footnote{Although the {\it IRAS\,} 100\,$\mu$m measurement are 
probably less reliable than the 60 and 350\,$\mu$m data points.}, 
hence each flux measurement is weighted equally in 
the nonlinear least squares fit.
It is not possible to fit both \Td and \bet as free parameters 
due to the paucity of available data points. 
Additionally, the negative \mbox{\Tds-\bet} degeneracy 
in the SED fitting procedures (Blain et al.\ 2003) often
renders simultaneous estimates of \Td and \bet problematic.
To achieve redundancy, we fit three SED curves for each source 
in our sample with \bet fixed to 1.0, 1.5 and 2.0.
The fitted SEDs for each source are overplotted (Fig.\ \ref{sedplot_stanford}) 
whenever convergence can be achieved by the fitting procedure. 
\bet values favored by the observed photometric data points 
differ from source to source, suggesting real differences 
in emissivity index among the galaxies. 
The negative \mbox{\Tds-\bet} degeneracy is also clearly manifested in
Fig.\ \ref{sedplot_stanford}, which shows that higher estimates of \Td 
are always associated with lower values assumed for \bets.
We assume $\beta=1.5$ - a value most commonly assumed and observed - 
to estimate effective dust temperature and 
derive physical quantities in later analysis and discussion 
(Table\ \ref{target}). 

\begin{figure}
\begin{center}
\includegraphics[width=4.4in, angle=90]{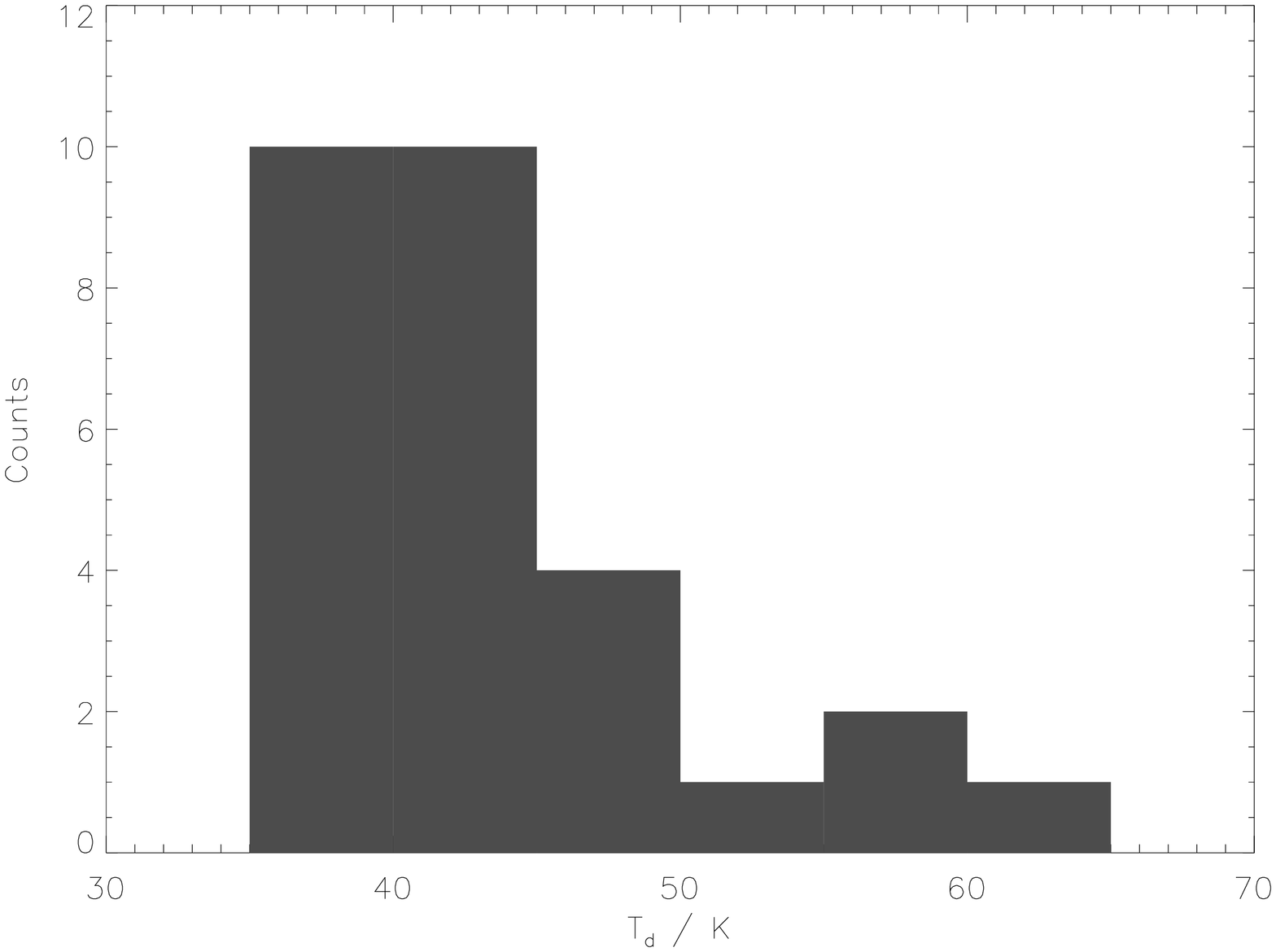}
\caption{Histogram of \Td values derived for our intermediate-redshift ULIRG sample.}\label{stanford-t-histo}
\end{center}
\end{figure}

The best-fitting dust temperatures in the SHARC-II detected FF sample, 
with \bet assumed to be 1.5, are found to range within a wide band of $\rm 35.1 \leq T_d \leq 62.6$\Ks, 
with a sample median and standard deviation $\rm T_d = 42.8 \pm 7.1$\Ks.
A histogram of \Td is shown in Fig.\ \ref{stanford-t-histo}.
All SHARC-II detected FF galaxies have \Td values in the range of 35 to 50\Ks, except for
four sources - FF\,0050$-$0039 ($z$=0.727), FF\,1042$+$3231 ($z$=0.633), FF\,1106$+$3201 ($z$=0.900), 
FF\,1532$+$3242 ($z$=0.926) - that have $\rm T_d \geq 50$\Ks.
The apparent absence of low-\Tds-high-$z$ galaxies within the sample could at least 
in part be caused by the selection effects inherent to the FF catalog,
while the lack of high-\Tds-low-$z$ (effectively high-\Tds-low-\Lfirs) galaxies 
in the sample most likely reflects real properties of the ULIRG population.

The global far-IR luminosities ($\rm L_{\mbox{\tiny{FIR}}}$) and dust masses ($\rm M_{d}$) are calculated using
\begin{eqnarray}
\rm L_{\mbox{\tiny{FIR}}} & = & \rm 4 \pi \, D_{\tiny{L}}^{\,2} \, \int_{40  \, \mu m}^{1000  \, \mu m} S_\nu \, d\nu \label{lfir} \\
\rm M_{d} & = & \rm \frac{S_\nu \, D_{\tiny{L}}^{\,2}}{\kappa \, B_\nu(T_d)} \label{md}
\,,\end{eqnarray}
where \DL is the luminosity distance, and $\rm \kappa \equiv \frac{3 \, Q_{abs}}{4 \, a \, \rho} $ 
is the dust mass absorption coefficient, varying as $\rm \propto \nu^{\,\beta}$.
We adopt dust parameters given by Hildebrand (1983): 
$ \rm Q_{125{\mu}m}=7.5\times10^{-4}$, $\rm \rho = 3.0\,g/{cm^3}$ and $a \rm = 0.1\,{\mu}m$, which implies
$\kappa_{125 \mu m} = 1.875 \, \rm (kg/{m^2})^{-1}$.
While there are significant uncertainties associated with $\kappa$, 
due to the lack of accurate knowledge of interstellar dust properties, the
inferred far-IR luminosities are tightly constrained 
as long as a satisfactory SED fit is achieved. 
The $\rm L_{\mbox{\tiny{FIR}}}$ and $\rm M_d$ values derived for our sample
lie in the ranges $\rm 10^{11.4} \leq L_{\mbox{\tiny{FIR}}} \leq 10^{\,13.1}$ \Lsun and  
$\rm 10^{\,7.2} \leq M_d \leq 10^{\,\,8.5} \,M_\odot$, respectively, with sample medians of
$\rm L_{\mbox{\tiny{FIR}}} = 10^{\,12.2 \pm 0.5} \,L_\odot$ and $\rm M_d = 10^{\,8.3 \pm 0.3} \,M_\odot$.
Thus, our sample spans nearly two orders of magnitude in far-IR luminosity, and more than 
an order of magnitude in dust mass.

If the far-IR luminosity originates from warm dust heated by 
massive, short-lived stars, which we argue in \S \ref{stanford_fir_radio} is the case for majority
of our sources, it provides a measure of the instantaneous star formation rate (SFR) in these
galaxies. Adopting the calibration factor given by Kennicutt (1998), we have 
\begin{equation}
\frac{\rm SFR}{\rm M_\odot \, yr^{-1}} = 1.7 \times \frac{\rm L_{\mbox{\tiny{FIR}}}}{\rm 10^{10}\,L_\odot} \,f\,X
\label{sfrffir1}
,\end{equation}
where \Lfir is integrated over the wavelength range of $40-1000\,\mu$m,
$f$ characterizes the percentage energy contribution to far-IR luminosity 
by massive young stars ($f \ls 1$), and $X$ accounts for integrated flux 
shortward of 40$\,\mu$m ($X > 1$). Using the estimated value of 
$ X \sim 1.4$ (from Eqs.\ \ref{equation:full-LFIRa} and \ref{equation:full-LFIRb}),
we find the median star formation rate of our sample is $\rm SFR = 10^{\,2.5 \pm 0.5} \,M_\odot\,yr^{-1}$.
The implied star formation rates are thus $3-4$ orders of magnitude higher than what is typically
observed in in normal quiescent galaxies (Kennicutt 1998), but comparable to that of the local ULIRG Arp\,220
($\rm SFR \simeq 320\,M \odot\,yr^{-1}$ -- Dopita et al.\ 2005). Such intense starbursts
can be induced by strong tidal interactions, a scenario 
consistent with high-resolution radio and near-IR observations of the FF catalog 
that reveal morphological features typical of 
strong interacting/merging systems for the majority of 
the sources (Stanford et al.\ 2000).

\section{Discussion}
\subsection{Far-IR/Radio Correlation}\label{stanford_fir_radio}
A tight far-IR/radio correlation is known to hold for a wide range of star-forming 
galaxies in the local Universe (Condon 1992). This is explained by the fact that radio emission at $\rm 1.4\,GHz$ 
is thought to arise mainly from nonthermal synchrotron emission of cosmic electrons
that are released by supernovae explosions of massive young stars, thus radio emission  
is expected to be proportional to the present star formation rate, as is far-IR emission.  
Whether and how this far-IR/radio correlation might evolve with galaxy 
characteristics and redshift remains an open question at the present, partly due to the 
substantial observational efforts involved to compile the relevant information 
(i.e., multiband far-IR/submm fluxes, 1.4\,GHz radio flux, and precise spectroscopic redshifts) 
for a large number of galaxies at intermediate and high redshifts. 
Our sample of SHARC-II detected FF sources, having the complete information mentioned above, 
constitutes a statistically significant sample that is suitable for a meaningful 
investigation of the far-IR/radio correlation at moderate redshifts $0.1 \ls z \ls 1.0$. 

\begin{figure}
\begin{center}
\includegraphics[width=5.4in, angle=0]{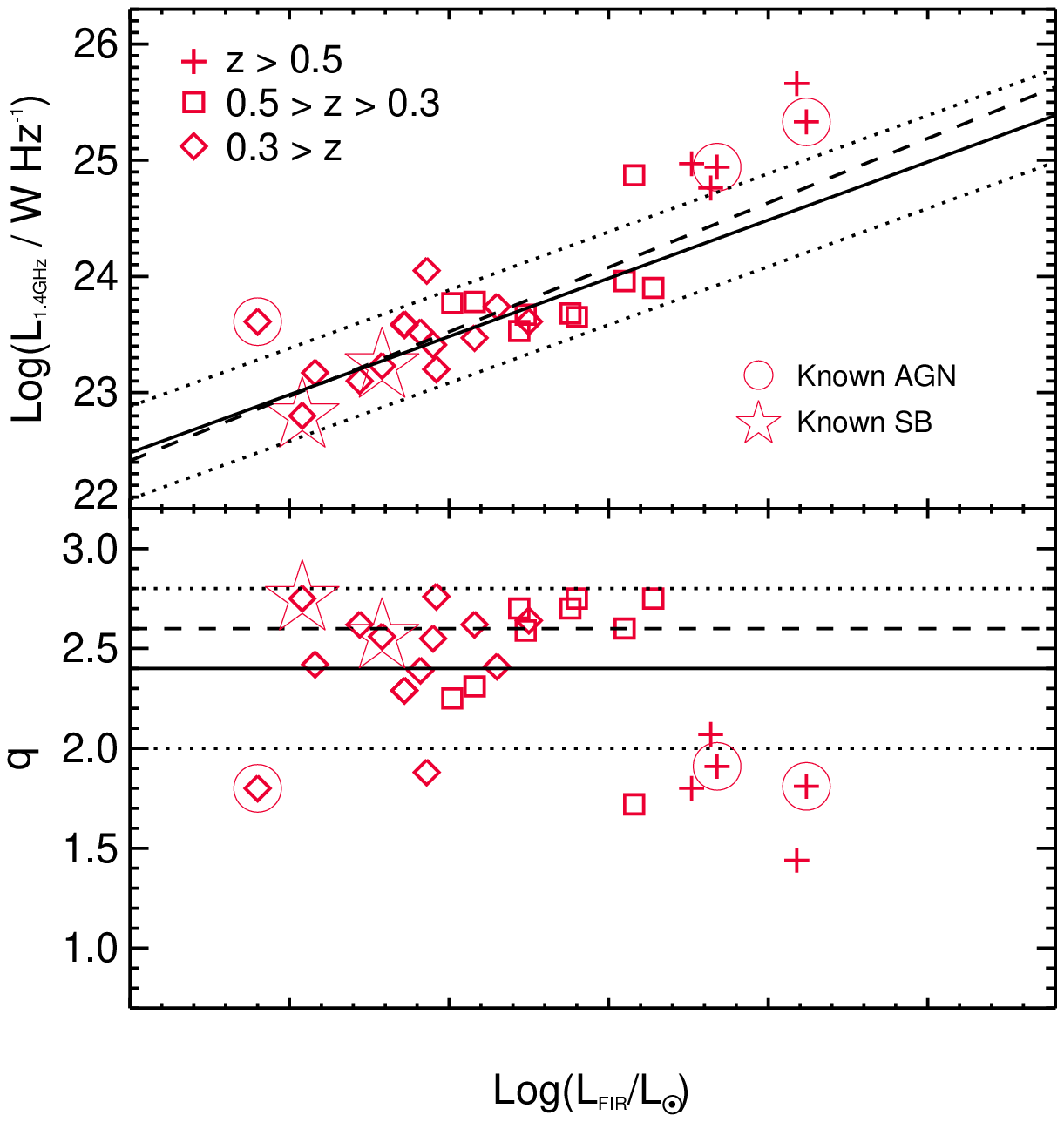}
\caption{{\bf a):} The far-IR/radio correlation for our intermediate-redshift ULIRG sample.
The five sources which have been robustly identified as either AGN or starburst (SB) 
dominated systems have been highlighted. 
The solid line represents a linear fit to the 21 sources 
with $q \geq 2.0$, and the dotted lines mark the $\pm 0.2$ scatter in $q$. The dashed line
is the local far-IR/radio correlation given by Condon et al.\ (1991).
{\bf b):} The distribution of $q$ values as a function of the 
far-IR luminosity. The solid line shows the sample median 
for all 28 sources, and the dotted lines mark the $\pm 0.4$ scatter. 
The dashed line shows the median of the 21 sources with $q \geq 2.0$.}
\label{figure:fir-radio}
\end{center}
\end{figure}

The far-IR/radio correlation for our sample is plotted in Fig.\ \ref{figure:fir-radio}a. 
Note that the correlation is investigated over the full far-IR/submm wavelength range of $40-1000\,\mu$m. 
The far-IR/radio correlation is often expressed in terms of the logarithmic 
far-IR/radio flux ratio $q$ (Helou et al.\ 1985), defined as
\begin{equation}
   q = \rm log (\frac{L_{\mbox{\tiny{FIR}}}}{3.75\times 10^{12}W})-log(\frac{L_{1.4GHz}}{W Hz^{-1}}) \label{q-2}
\,,\end{equation}
where $\rm L_{\tiny{1.4GHz}}$ is the 1.4\,GHz (rest-frame) radio luminosity. 
Accounting for both the $K$-correction, through $(1+z)^{\alpha}$, 
and the bandwidth compression by a factor of $(1+z)^{-1}$, we have
\begin{equation}
\rm log (\frac{L_{\,1.4\,GHz}}{W\,Hz^{-1}})  =  17.08 + 2\,log\,(\frac{D_{\tiny{L}}}{Mpc}) + (\alpha-1)\,log\,(1+z) + log (\frac{S_{\,1.4\,GHz}}{mJy}) \label{lrad}
\,,\end{equation}
where $\alpha$, defined by $\rm S_\nu \propto \nu^{- \alpha}$, is the radio spectral index. 
Realistic constraints on $\alpha$ are not currently available for the FF sample
due to the lack of multi-wavelength radio measurements. 
We therefore assume a constant value of $\rm \alpha \sim 0.7$ in the spectral domain near 1.4\,GHz - 
a value that is appropriate when synchrotron emission is dominant (Condon 1992).
Following Eqs.\ (\ref{q-2}) and (\ref{lrad}), and using rest-frame $\rm L_{40 - 1000\,\mu m}$, 
we have derived $q$ for each source in our sample of SHARC-II detected ULIRGs (Table \ref{target}).
A broad range in $q$ values are found ($ 1.44 \leq q \leq 2.76$), with a sample median of $q = 2.4 \pm 0.4$. 
The distribution of $q$ values for our sample is shown in Fig.\ \ref{figure:fir-radio}b. 
Of the entire sample, 21 sources (75\%) have $q$ values within $\pm 1\sigma$ of the median value, while
the seven remaining outliers all have $q < 2.0$, suggesting that they might be radio-loud AGNs. 
We note that four of the seven outliers are at $z>0.5$ and 
five have $\rm L_{\mbox{\tiny{FIR}}} \geq 10^{\,12.5}$\Lsuns, which is most likely 
caused by the selection biases in our sample that favor radio-loud, 
luminous objects with enhanced AGN activity at high redshifts (\S \ref{ffsample}). 
The suspected AGN fraction of $\sim$25\% is also roughtly consistent with 
the $\sim$10\% found for the entire FF sample, and with the trend of 
an increasing AGN fraction with FIR-luminosity as noted by Stanford et al.\ (2000).
We therefore exclude the seven sources in the investigation of the far-IR/radio correlation by setting a lower bound of
$\langle q \rangle - \sigma_q = 2.0$.
To check this approach we have searched the literature for independent AGN/starburst classification
for our sources. In total, classifications are found for six sources (Stanford et al.\ 2000;
Zakamska et al.\ 2004; Best et al.\ 2005; Best private communications),
of which three are classified as AGNs (FF\,0050$-$0039, FF\,0804$+$3919 and FF\,1532$+$3242) 
and three as starbursts (FF\,0030$-$0027, FF\,0907$+$3931 and FF\,2221$-$0042). 
The three known AGNs have $q <2.0$, while the three known starbursts are within 
$\pm 1\sigma$ of our median $q$ value, in agreement with our lower cut in $q$.
However, we wish to note that a systematic change in the radio slope
would have a direct impact on the estimated $q$ values. 
For instance, Ivison et al.\ (private communications) have suggested 
that SMGs might have shallower synchrotron slopes than "non-SMG" radio sources. 
In this scenario, $q$ values estimated by assuming $\rm \alpha \sim 0.7$ 
would be underestimated, with the underestimates in $q$
being more severe at higher redshifts. As such, the very low $q$ values ($q < 2.0$) 
characteristic of the high-$z$ ($z>0.5$) sources in our sample 
could also be caused by actual radio slopes that are flatter 
than the assumed value of 0.7 in these sources.

The remaining 21 sources with $q \ge 2.0$ show a very tight far-IR/radio correlation 
(Fig.\ \ref{figure:fir-radio}b), with a sample median of $q = 2.6 \pm 0.2$. 
We wish to compare the far-IR/radio correlation observed at intermediate redshifts 
to that in the local Universe, for which $q$ values are found to be bound within a narrow range 
$2.3 \pm 0.2$, with the far-IR luminosity defined to cover the wavelength range of 
$40-120\,\mu$m (Condon et al.\ 1991).
To account for the difference in the wavelength integration range, 
we adopt estimates given by Soifer \& Neugebauer (1991) 
\begin{eqnarray}
\label{equation:full-LFIRa}
\rm L_{8 - 40\,\mu m} & \sim & \rm  40\% \times L_{40 - 1000\,\mu m}\\
\rm L_{8 - 1000\,\mu m} & \sim & \rm (1\,+\,40\%) \times L_{8 - 120\,\mu m},
\label{equation:full-LFIRb}
\end{eqnarray}
from which we derive
\begin{equation}
\rm L_{40 - 1000\,\mu m} \sim \frac{1}{0.6} \times L_{40 - 120\,\mu m}. 
\end{equation}
The local far-IR/radio correlation as given by Condon et al.\ (1991) can thus 
be recast into $q$ values with a median $\langle q \rangle = 2.3 + {\rm log}\,(\frac{1}{0.6}) = 2.5$
and scatter $\sigma_q = 0.2$,
in excellent agreement with the $q$ values calculated for our sample of intermediate redshift ULIRGs.
Furthermore, a fit to log\,(\Lfirs) and log\,($\rm L_{\tiny{1.4GHz}}$)
inferred for our sample yields a slope of $\alpha = 1.00\pm 0.14$, which is also 
consistent with that of the local correlation ($\alpha = 1.11\pm 0.02$ - Condon et al.\ 1991). 

Thus, we conclude that our derived far-IR/radio correlation in the redshift range $0.1 \ls z \ls 0.5$ 
matches very well with that of the present day, suggesting that the correlation does not evolve significantly
from $z\sim 0$ to $z\sim 0.5$. This supports the notion that dust heating in the intermediate-redshift 
ULIRGs originates predominantly from star formation, with modest contribution to the
far-IR luminosity from old stellar populations and AGNs.
Unfortunately, we cannot confidently extend this conclusion to $z\sim 1$, since
four out of the five sources in our sample at $z> 0.5$ were discarded on suspicion 
of being radio-loud AGNs, albeit the possibility of flatter radio slopes in the high-$z$ sources. 
Second, we find no evidence of an increase in the scatter of the correlation with redshift. 
Our study validates the approach by Appleton et al.\ (2004), in which {\it Spitzer} $70\,\mu$m 
data are $K$-corrected in a model-dependent manner to derive the full far-IR luminosities 
for sources out to $z\sim 1$. Such confirmation is important,
allowing further speculation with a single flux density measurement.


\subsection{Luminosity-Temperature Relation}
\begin{figure}
\begin{center}
\includegraphics[width=4in, angle=90]{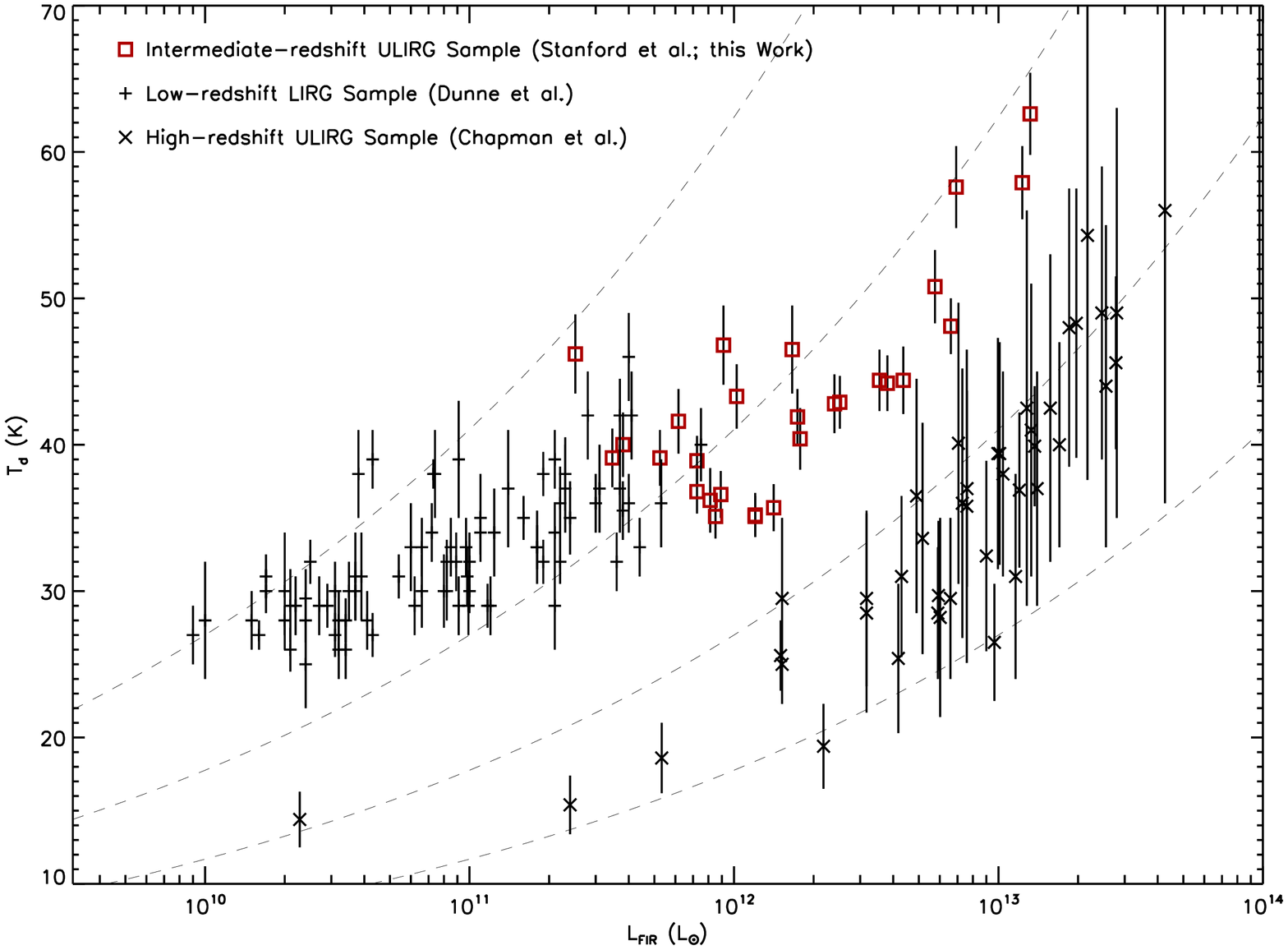}
\caption{\Lfir and \Td derived for luminous dusty galaxy samples at distinct redshifts. 
The dashed lines correspond to \mbox{\Lfirs-\Td} relations given by Eq.~(\ref{lfirtd}), 
assuming \bet=1.5 and $\kappa_{125 \mu m} = 1.875 \, \rm (kg/{m^2})^{-1}$, for dust mass 
spanning over three orders of magnitude $\rm M_d = 10^{\,7 - 10}\,M_\odot$ 
(from top to bottom).}
\label{figure:LT}
\end{center}
\end{figure}

The far-IR luminosities and dust temperatures inferred for our sample are plotted
against each other in Fig.\ \ref{figure:LT}. For comparison, we have also plotted \Lfir and \Td values 
compiled by Blain et al.\ (2004) for samples of luminous infrared galaxies at 
low and high redshifts. In the case of the low-$z$ sample, 60, 100 and 850$\,\mu$m data 
are available and utilized in the SED fitting (Dunne et al.\ 2000). 
By contrast, \Lfir and \Td values of the high-redshift sample 
are derived by assuming the far-IR/radio correlation (Chapman et al.\ 2003; Chapman et al.\ 2005), 
which has been subsequently confirmed by Kov\'{a}cs et al.\ (2006). 

Taken at face value, each sample exhibits an overall trend between \Lfir and $\rm T_d$, 
where sources with large far-IR luminosities having, in general, higher dust temperatures.
This trend can be understood in terms of the approximate relation
\begin{equation}
\rm L_{\mbox{\tiny{FIR}}}
\propto  \kappa_0\,M_d\,T_d^{4+\beta} \label{lfirtd}
\,,\end{equation}
where $\kappa_0$ is the dust mass absorption coefficient normalized at 
some reference frequency $\nu_0$\footnote{
$\rm L_{\mbox{\tiny{FIR}}} = \frac{8\,\pi\,h\,\nu_0^4}{c^2}\,\kappa_0\,M_d\,({k\,T_d}/{h\,\nu_0})^{4+\beta}\,\Gamma(4+\beta)\,\zeta(4+\beta)$, 
where \mbox{$\rm \Gamma(z) = \int_0^\infty t^{z-1}e^{-t}\,dt$} 
and \mbox{$\rm \zeta(s) = \frac{1}{\Gamma(s)} \int_0^\infty \frac{t^{s-1}}{e^t-1}\,dt$} 
are the Gamma and Riemann $\zeta$ functions respectively. This formula 
is equivalent to that given by De Breuck et al.\ (2003). 
With the normal range of $1.0 \leq \beta \leq 2.0$, 
$\zeta(4+\beta)$ is nearly flat while $\Gamma(4+\beta)$ varies rapidly.
Therefore the approximation as stated in Eq.\ (\ref{lfirtd}) is valid 
when variations in \bet are negligible.}.
However, there is significant scatter in the \mbox{\Lfirs-\Td} relation,
due to variations in the dust emissivity, 
as well as in the amount of dust in each galaxy. 

Although significant scatter in \Td for a given \Lfir exists for 
galaxies within each sample, Fig.\ \ref{figure:LT} reveals systematic shifts 
between the three samples shown. 
In terms of dust temperature, the low-$z$ sample spans a fairly narrow range ($\rm T_d\simeq 25-45$\,K)
with a median temperature of $\rm T_d = 35.6\pm4.9$\,K (Dunne et al.\ 2000). The latter is 
consistent with the median temperatures found for our sample ($\rm T_d = 42.8\pm 7.1\,$K) 
and the SMGs ($\rm T_d = 36.0\pm 7.0\,$K -- Chapman et al.\ 2005), although the 
latter two span a larger range of temperatures. In particular, we notice that all sources in our
sample have $\rm T_d\gs 30\,$K, whereas $\sim$30\% of the SMGs have dust temperatures below
$30\,$K. Furthermore, there are four sources in our sample for which we can confidently 
say they have dust temperatures above $50\,$K, while all SMGs are consistent
with having $\rm T_d<50\,$K given their large error bars. Thus, it seems that our sample
overall consists of sources with somewhat warmer dust than both the local and high-$z$ sample.
Assuming the dust emissivity properties in starburst galaxies do not change substantially 
as a function of redshift, and by Eq.\ (\ref{lfirtd}), SMGs thus 
appear to contain cooler but considerably larger amounts of dust than (U)LIRGs at 
low and moderate redshifts (Fig.\ \ref{figure:LT}). 
This is consistent with observations of CO rotational line emission in bright
SMGs, which have revealed that they are very massive systems, 
having at least four time more molecular gas than that in 
local ULIRGs (Neri et al.\ 2003; Greve et al.\ 2005). 
We argue the observed difference can be explained as the combined effects of 
selection effects and intrinsic evolution in the ULIRG population 
as a function of redshift.

First, the very high far-IR luminosities characteristic of 
the high-$z$ SMG sample can be caused by the flux cut-offs 
of present-day submm and radio surveys, which limit the detections of
high-$z$ sources with $\rm L_{\mbox{\tiny{FIR}}} \ls 10^{12}\,$\Lsuns.
Second, by virtue of their selection at submm wavelengths, the SMGs 
are naturally biased in favor of cooler dust temperatures. 
This has been confirmed by recent {\it Spitzer} imaging, which 
shows that the vast majority ($\sim 80-90$\%) of the {\it Spitzer} 
detected SMGs have near-IR/mid-IR properties typical of cold, 
starburst-like ULIRGs (Egami et al.\ 2004). However,
it has been suggested that a significant population of optically faint
radio galaxies (OFRGs), which so far have remained undetected by current 850\,$\mu$m surveys
due to their warmer dust ($\rm T_d\gs 50\,$K), might exist and constitute an extension of SMGs to
warmer temperatures (Chapman et al.\ 2004b). 
OFRGs are luminous in the radio and the 24\,$\mu$m wavebands, and it seems our intermediate redshift sample,
being select by {\it IRAS}, might be more representative of OFRGs than SMGs. 
Finally, the existence of low-to-intermediate redshift (U)LIRGs with cool 
dust temperatures 
cannot be ruled out, since such sources are unlikely to have made it into the 60 and 100$\,\mu$m 
{\it IRAS}-selected samples of local (U)LIRGs. The selection biases are more
severe for the intermediate-redshift sample than for the local sample, as 
predicted by the shape of the $K$-correction at these wavebands.
Unbiased wide surveys with SCUBA-II and Herschel at $200-850\,\mu$m are
expected to uncover cool sources at $z\sim 0-1$ if they exist, thus
allowing for a more direct comparison with the high-$z$ population.

While at least part of the observed differences in dust masses and temperatures between the various
samples can be attributed to the selection effects specific to each sample,
real cosmological evolution of the ULIRG population with redshift appears to play
a significant role.  This is strongly evidenced by the lack of very luminous
($\rm L_{\mbox{\tiny{FIR}}} \gs 10^{12}\,$\Lsuns) sources in the low-$z$ sample, which reflects
the well-known decrease in the abundance of such extreme systems by about three orders of magnitude
from $z\sim 2-3$ to the present day. At moderate redshifts, such systems are more common
as suggested by the overlap in far-IR luminosity between our sample and the high-$z$ SMGs.

\subsection{Dust Temperature and Physical Scale of Star Formation}
 
Finally, we wish to argue that the higher dust temperatures observed in our sample of ULIRGs
at moderate redshifts, as compared to those of the high-$z$ SMGs, may be linked to variations 
in the characteristic physical scale on which star formation dominates
at these two different epochs. Eq.\ (\ref{lfirtd}) suggests that the observed $\rm T_d$, characterizing 
the large-scale SED of a galaxy, is determined by the total 
dust heating (and by implication the global star formation rate) per unit 
dust mass within the galaxy. The dust temperature is therefore linked to the 
global star formation efficiency, subject to uncertainties in 
dust emissivity, additional dust heating, and gas-to-dust ratio. 
At the same time, the star formation efficiency is known to be primarily regulated by 
gas density in star-forming regions (Schmidt law) across enormous ranges 
spanning $5-6$ orders of magnitude in gas densities and star formation rates (Kennicutt 1998). 
Calculations by Yang\ (2006) have shown that a general relationship can be 
expected to exist between the observed \Td and the spatial 
extent of star formation, suggesting higher observed \Td are generally 
indicative of star formation that occurs over more concentrated regions. 
This proposed link is in good agreement with observational evidence;
high-resolution radio observations of (cooler) high-redshift SMGs 
show spatially extended starbursts on scales of $\sim$10\,kpc (Chapman et al.\ 2004a), 
while the few local ULIRGs with far-IR luminosities comparable to that of SMGs in general
have higher dust temperatures with the dust and molecular gas being concentrated
in compact regions with scales on the order of 0.1$\sim$1\,kpc (Young \& Scoville 1991; Sanders \& Mirabel 1996). 
Today's mm interferometers such as the IRAM Plateau de Bure Interferometer
(PdBI) have just about the sufficient sensitivity as well as spatial resolution 
(FWHM$\simeq 0.6"$ at 1.3mm for PdBI in its extended B and A configuration) 
to allow us to detect and marginally resolve the dust emission in sources such as the ones in our sample.
However, such observations are extremely challenging and time consuming, and
will at best provide us with only a rough gauge, or even perhaps only an upper limit,  of
the physical extent of the dust.
Of course, once ALMA comes on-line the situation will improve dramatically,
resulting in detailed
maps of the dust emission that will not only pinpoint the hot dust, which
presumably originates from compact regions
of intense star formation and/or AGN-activity, but also outline the
morphology and scale of any underlying, cooler dust
reservoir that may exist.
 
\section{Conclusion}
This paper reports 350$\,\mu$m observations of 36 intermediate-redshift 
ULIRGs ($0.089 \leq z \leq 0.926$), out of which 28 are detected. 
The newly acquired 350$\,\mu$m data are the first reported 
flux measurements longward of 100$\,\mu$m for these galaxies, 
and lead to meaningful estimates of their dust 
temperatures ($\rm T_d = 42.8 \pm 7.1$\Ks) and far-IR luminosities 
($\rm L_{\mbox{\tiny{FIR}}} = 10^{\,12.2 \pm 0.5} \,L_\odot$). 
These intermediate-redshift ULIRGs appear to host large amount of 
warm dust ($\rm M_d = 10^{\,8.3 \pm 0.3} \,M_\odot$) and 
are experiencing intense starburst activity ($\rm SFR = 10^{\,2.5 \pm 0.5} \,L_\odot\,yr^{-1}$). 
The far-IR/radio correlation observed in star-forming galaxies in the local Universe 
remains valid for ULIRGs in the more distant Universe 
over the redshift range of $0.1 \ls z \ls 0.5$, where
dust heating seems predominantly powered by ongoing star formation, 
with modest contribution to the far-IR luminosity from older stars and AGNs.
Far-IR luminosities and dust temperatures derived for dusty galaxies 
over a wide range of redshifts reveal a general positive relation between 
theses two quantities. However, there is significant scatter in the 
\mbox{\Lfirs-\Td} relation, due to differing selection biases 
as well as variations in dust mass and grain properties.
We argue that the somewhat higher dust temperatures of our sample
compared to the high-$z$ SMGs are possibly tied to the more compact 
spatial scales of the ongoing star formation.

\acknowledgments
We thank Attila Kov\'{a}cs for help with data reduction and taking data on our behalf; 
Andrew Blain for useful discussions and providing part of the \mbox{\Lfirs-\Td} data; 
and Philip Best for providing us with AGN/starburst classifications for
some of our sources. We are deeply grateful to the referee for 
insightful and thorough comments. The CSO is supported by the NSF fund 
under contract AST 02-29008.

\begin{singlespace}
\bibliographystyle{apj}

\end{singlespace}

\begin{deluxetable}{lccccccccccc}
\tabletypesize{\scriptsize}
\rotate
\tablecaption{FF sources selected for SHARC-II observations (36).}
\tablewidth{0pt}
\tablehead{
           \colhead{Source Name} & \colhead{RA}     & \colhead{Dec}   & \colhead{$z$} & \colhead{$\rm S_{60\mu m}$\tablenotemark{a}} & \colhead{$\rm S_{100\mu m}$\tablenotemark{b}} & \colhead{$\rm S_{350\mu m}$} &\colhead{$\rm S_{1.4GHz}$} & \colhead{$\rm T_d$}    & \colhead{log\, \Lfir}   &  \colhead{log\, \Md}   & \colhead{$q$}\\
\colhead{}                       & \colhead{J2000}  & \colhead{J2000} & \colhead{}    & \colhead{Jy}                                 & \colhead{Jy}                                  & \colhead{mJy}                & \colhead{mJy}             & \colhead{K}            & \colhead{\Lsun}         &  \colhead{\Msun}       & \colhead{}\\
}
\startdata
FF\,0030$-$0027                  &  00 30 09.099    & $-$00 27 44.40  &     0.242     &   0.59                                       &  1.11                                         & $134 \pm 16.8$                     &  2.52         & $40.4\pm2.1$ & 12.25 & 8.28 & 2.64\\              
FF\,0050$-$0039                  &  00 50 09.806    & $-$00 39 00.96  &     0.727     &   0.22                                       &  0.49                                         & $43.0 \pm 10.4$                    &  4.32         & $57.6\pm2.8$ & 12.84 & 8.15 & 1.91\\              
FF\,0123$+$0114                  &  01 23 06.973    & $+$01 14 10.03  &     0.089     &   0.23                                       &  0.55                                         & (\nodata$\pm 9.75$)                &  1.78         & \nodata            & \nodata                 & \nodata                &  \nodata    \\ 
FF\,0240$-$0042                  &  02 40 08.576    & $-$00 42 03.56  &     0.410     &   0.20                                       &  0.69                                         & $66.5 \pm 10.4$                    &  0.84         & $42.9\pm1.8$ & 12.40 & 8.30 & 2.75\\           
FF\,0245$+$0123                  &  02 45 55.355    & $+$01 23 28.40  &     0.798     &   0.17                                       &  0.32                                         & ($32.4 \pm 25.6$)                  &  2.03         & \nodata            & \nodata                 & \nodata                &  \nodata    \\ 
FF\,0312$+$0058                  &  03 12 38.445    & $+$00 58 33.86  &     0.130     &   0.26                                       &  0.61                                         & ($19.2 \pm 12.1$)                  &  3.84         & \nodata            & \nodata                 & \nodata                &  \nodata    \\ 
FF\,0317$-$0129                  &  03 17 43.635    & $-$01 29 07.33  &     0.265     &   0.21                                       &  0.34                                         & $106 \pm 9.84$                     &  1.67         & $36.2\pm2.2$ & 11.91 & 8.18 & 2.39\\            
FF\,0738$+$2856                  &  07 38 29.856    & $+$28 56 38.74  &     0.334     &   0.25                                       &  0.64                                         & $68.7 \pm 8.41$                    &  1.38         & $41.9\pm1.9$ & 12.24 & 8.20 & 2.59\\            
FF\,0748$+$3343                  &  07 48 10.591    & $+$33 43 27.13  &     0.356     &   0.64                                       &  1.18                                         & $140 \pm 11.7$                     &  2.03         & $44.4\pm2.3$ & 12.64 & 8.47 & 2.75\\            
FF\,0758$+$2851                  &  07 58 45.956    & $+$28 51 32.76  &     0.126     &   0.62                                       &  0.92                                         & $87.5 \pm 9.52$                    &  3.72         & $40.0\pm2.2$ & 11.58 & 7.64 & 2.42\\          
FF\,0804$+$3919                  &  08 04 07.399    & $+$39 19 27.63  &     0.164     &   0.25                                       &  0.48                                         & $20.7 \pm 5.75$                    &  5.90         & $46.2\pm2.7$ & 11.40 & 7.15 & 1.80\\          
FF\,0823$+$3202                  &  08 23 54.616    & $+$32 02 12.03  &     0.396     &   0.23                                       &  0.52                                         & $73.0 \pm 12.1$                    &  0.97         & $42.8\pm2.0$ & 12.38 & 8.28 & 2.70\\          
FF\,0826$+$3042                  &  08 26 11.644    & $+$30 42 44.17  &     0.248     &   0.31                                       &  0.88                                         & $154 \pm 19.9$                     &  3.19         & $35.7\pm1.6$ & 12.15 & 8.46 & 2.41\\            
FF\,0835$+$3559                  &  08 35 27.440    & $+$35 59 33.07  &     0.201     &   0.33                                       &  0.51                                         & ($38.4 \pm 20.5$)                  &  0.85         & \nodata            & \nodata                 & \nodata                &  \nodata    \\ 
FF\,0856$+$3450                  &  08 56 24.852    & $+$34 50 24.82  &     0.220     &   0.24                                       &  0.75                                         & $114 \pm 20.2$                     &  8.51         & $35.1\pm1.5$ & 11.93 & 8.28 & 1.88\\            
FF\,0907$+$3931                  &  09 07 42.264    & $+$39 31 49.47  &     0.224     &   0.26                                       &  0.47                                         & $46.2 \pm 7.69$                    &  1.24         & $41.6\pm2.2$ & 11.79 & 7.76 & 2.56\\          
FF\,1016$+$3951                  &  10 16 08.616    & $+$39 51 20.46  &     0.307     &   0.21                                       &  0.39                                         & $43.0 \pm 10.5$                    &  2.10         & $43.3\pm2.2$ & 12.01 & 7.90 & 2.25\\          
FF\,1018$+$3649                  &  10 18 34.539    & $+$36 49 51.75  &     0.490     &   0.20                                       &  0.56                                         & $78.6 \pm 10.1$                    &  9.18         & $44.2\pm1.9$ & 12.58 & 8.42 & 1.72\\          
FF\,1042$+$3231                  &  10 42 40.815    & $+$32 31 30.99  &     0.633     &   0.21                                       &  0.44                                         & $61.2 \pm 7.48$                    &  6.34         & $50.8\pm2.5$ & 12.76 & 8.31 & 1.80\\          
FF\,1106$+$3201                  &  11 06 35.716    & $+$32 01 46.39  &     0.900     &   0.19                                       &  0.64                                         & $62.5 \pm 16.2$                    &  13.7         & $57.9\pm2.5$ & 13.09 & 8.40 & 1.44\\          
FF\,1242$+$2905                  &  12 42 32.497    & $+$29 05 14.75  &     0.260     &   0.19                                       &  0.50                                         & $85.3 \pm 26.5$                    &  1.34         & $36.6\pm1.6$ & 11.95 & 8.21 & 2.55\\          
FF\,1412$+$3014                  &  14 12 24.952    & $+$30 14 09.78  &     0.257     &   0.15                                       &  0.53                                         & $63.3 \pm 14.2$                    &  2.06         & $36.8\pm1.5$ & 11.86 & 8.11 & 2.29\\          
FF\,1456$+$3337                  &  14 56 58.427    & $+$33 37 09.98  &     0.443     &   0.27                                       &  0.60                                         & $84.1 \pm 11.6$                    &  1.43         & $44.4\pm2.1$ & 12.55 & 8.38 & 2.60\\          
FF\,1514$+$3629                  &  15 14 33.118    & $+$36 29 42.36  &     0.338     &   0.24                                       &  0.54                                         & ($22.5 \pm 10.6$)                  &  0.82         & \nodata            & \nodata                 & \nodata                &  \nodata    \\ 
FF\,1532$+$3242                  &  15 32 44.052    & $+$32 42 46.73  &     0.926     &   0.26                                       &  0.50                                         & $52.3 \pm 13.4$                    &  5.89         & $62.6\pm2.8$ & 13.12 & 8.28 & 1.81\\          
FF\,1614$+$3234                  &  16 14 22.105    & $+$32 34 03.66  &     0.710     &   0.15                                       &  0.20                                         & ($29.8 \pm 28.5$)                  &  1.19         & \nodata            & \nodata                 & \nodata                &  \nodata    \\
FF\,1659$+$3549                  &  16 59 24.669    & $+$35 49 01.74  &     0.371     &   0.29                                       &  0.33                                         & $53.0 \pm 9.78$                           &  0.79  & $46.5\pm3.0$ & 12.22 & 7.95 & 2.70\\                 
FF\,1707$+$3725                  &  17 07 11.795    & $+$37 25 55.32  &     0.311     &   0.13                                       &  0.45                                         & $101 \pm 16.0$                          &  2.11    & $35.1\pm1.4$ & 12.08 & 8.44 & 2.31\\                 
FF\,1713$+$3843                  &  17 13 46.085    & $+$38 43 04.77  &     0.171     &   0.37                                       &  0.79                                         & $72.1 \pm 21.2$                           &  1.66  & $39.1\pm1.9$ & 11.72 & 7.82 & 2.62\\                 

FF\,2131$-$0141                  &  21 31 53.490    & $-$01 41 43.35  &     0.730     &   0.11                                       &  0.47                                         & $64.9 \pm 21.2$                           &  2.79  & $48.1\pm1.9$ & 12.82 & 8.48 & 2.07\\

FF\,2136$-$0112                  &  21 36 34.229    & $-$01 12 08.38  &     0.210     &   0.29                                       &  1.11                                         & $69.5 \pm 25.8$                           &  3.20  & $38.9\pm1.7$ & 11.86 & 7.98 & 2.29\\                 
FF\,2200$+$0108                  &  22 00 51.859    & $+$01 08 27.08  &     0.164     &   0.20                                       &  0.72                                         &  (\nodata$\pm 41.9$)                     &  2.82   & \nodata            & \nodata                 & \nodata                &  \nodata    \\ 
FF\,2216$+$0058                  &  22 16 02.721    & $+$00 58 10.65  &     0.212     &   0.51                                       &  0.85                                         & $48.0 \pm 16.4$                           &  1.31  & $46.8\pm2.7$ & 11.96 & 7.68 & 2.76\\                 
FF\,2221$-$0042                  &  22 21 26.066    & $-$00 42 39.08  &     0.189     &   0.20                                       &  0.37                                         & $42.8 \pm 12.8$                           &  0.66  & $39.1\pm2.0$ & 11.54 & 7.65 & 2.75\\                 
FF\,2330$-$0025                  &  23 30 34.920    & $-$00 25 03.98  &     0.252     &   0.25                                       &  0.72                                         & $137 \pm 33.8$                          &  1.65    & $35.2\pm1.5$ & 12.08 & 8.42 & 2.62\\                 
FF\,2352$-$0015                  &  23 52 53.171    & $-$00 15 24.69  &     0.227     &   0.19                                       &  0.61                                         & ($88.8 \pm 48.6$)                      &  0.59     & \nodata            & \nodata                 & \nodata                &  \nodata    \\ 
\enddata
\tablenotetext{a}{The template amplitude of the noise-weighted mean scan (1003), 
given by {\it IRAS\,} Scan Processing and Integration (SCANPI), Version 5.0.}
\tablenotetext{b}{The template amplitude of the noise-weighted mean scan (1003), 
given by SCANPI, Version 5.0, except for FF\,0123$+$0114 and FF\,2200$+$0108, 
for which the 100$\,\mu$m fluxes are not available from SCANPI and thus values in Stanford et al.\ (2000) are adopted.}
\tablecomments{Three additional far-IR/submm flux measurements are available from the literature - 
FF\,1532$+$3242: $S_{90\mu m}$= 478\,mJy, $S_{180\mu m}$= 397\,mJy (Verma et al.\ 2002); 
FF\,1614$+$3234: \mbox{$S_{850\mu m}$= 8.47\,mJy} (Farrah et al.\ 2002).}
\tablecomments{\bet=1.5 is assumed in the SED fitting.}
\label{target}
\end{deluxetable}

\begin{figure*}
\begin{center}
\begin{tabular}{cccc}
\includegraphics[width=1.5in, angle=270]{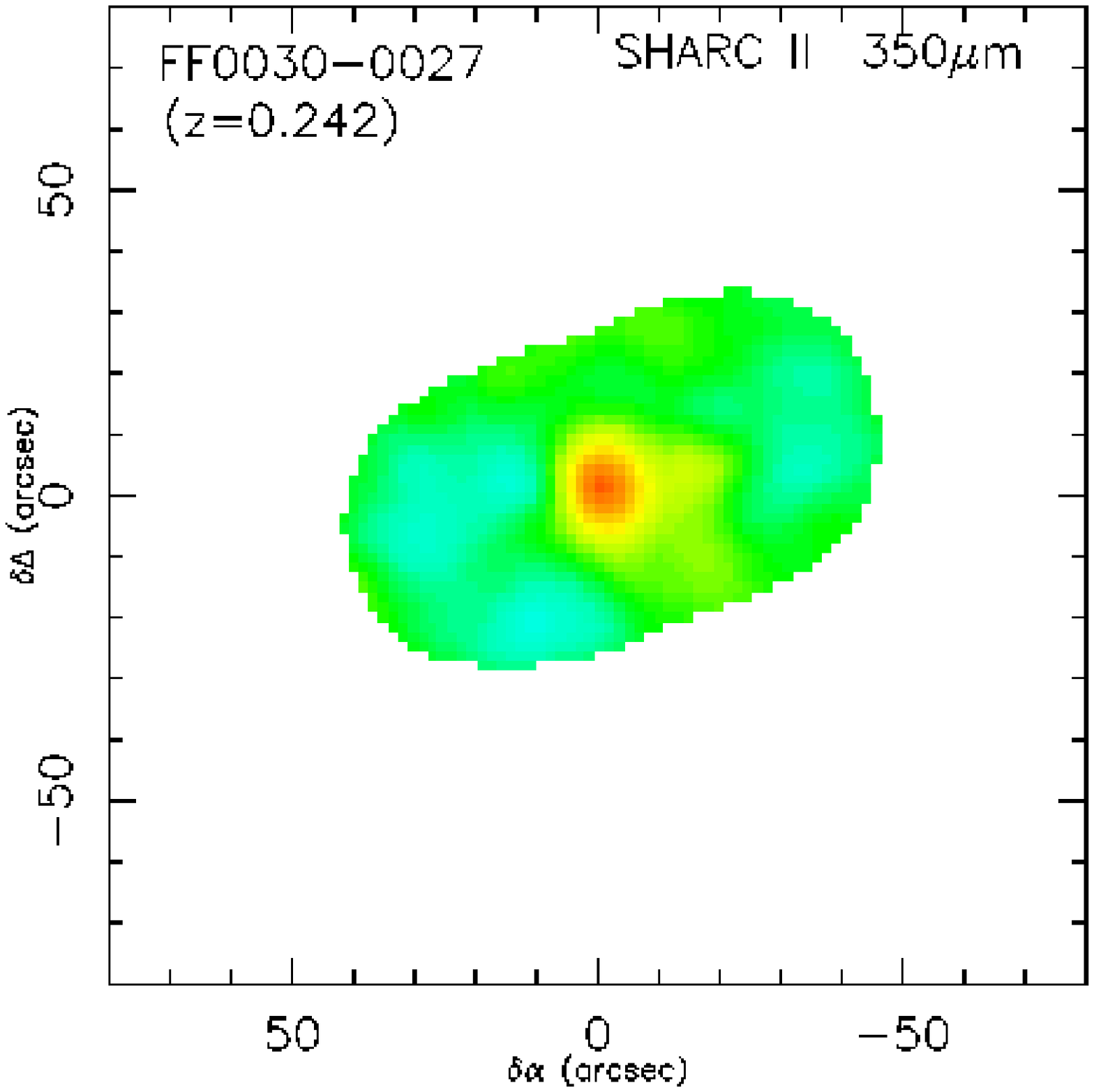}&
\includegraphics[width=1.5in, angle=270]{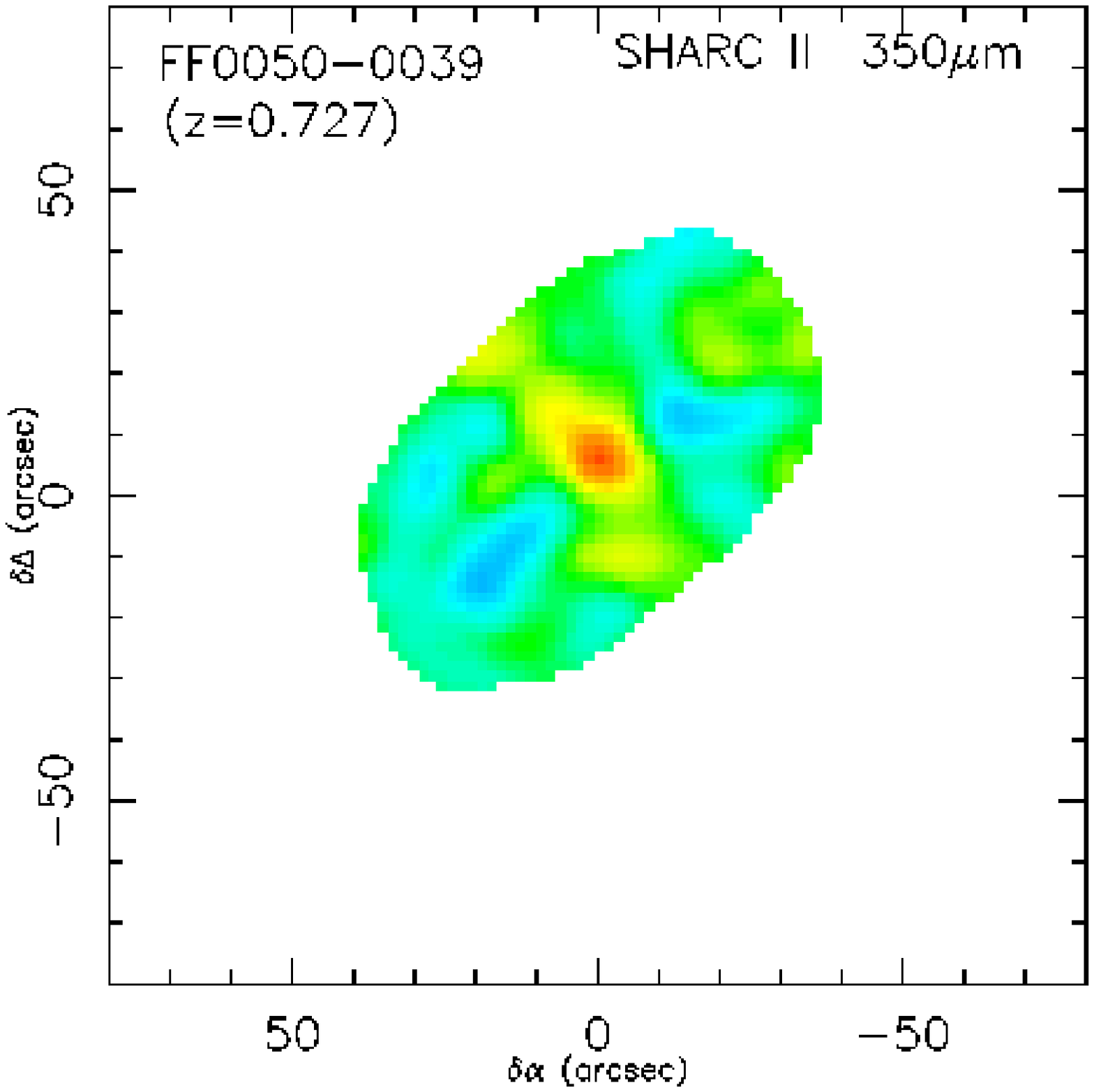}&
\includegraphics[width=1.5in, angle=270]{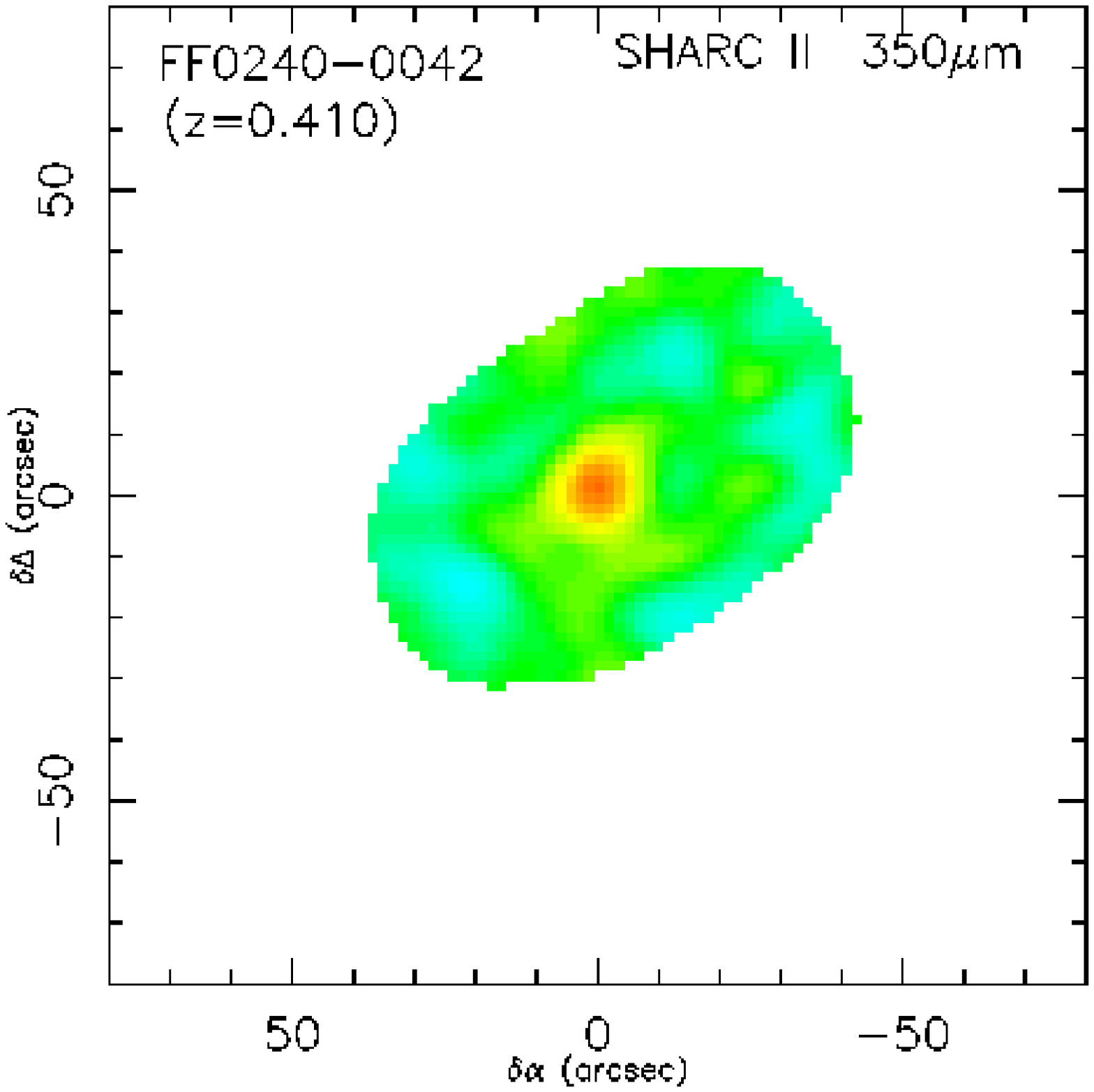}&
\includegraphics[width=1.5in, angle=270]{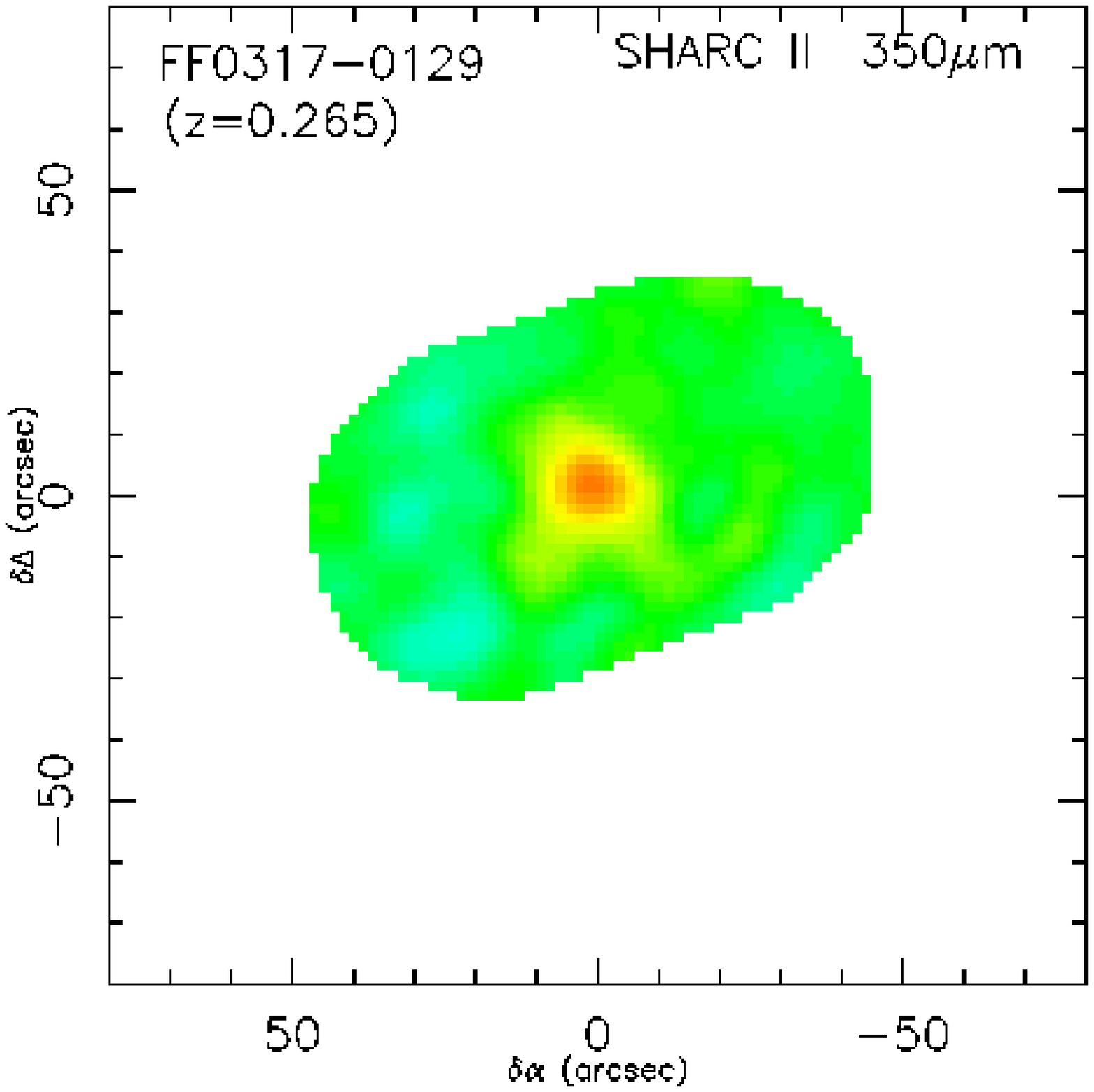}\\
\includegraphics[width=1.5in, angle=270]{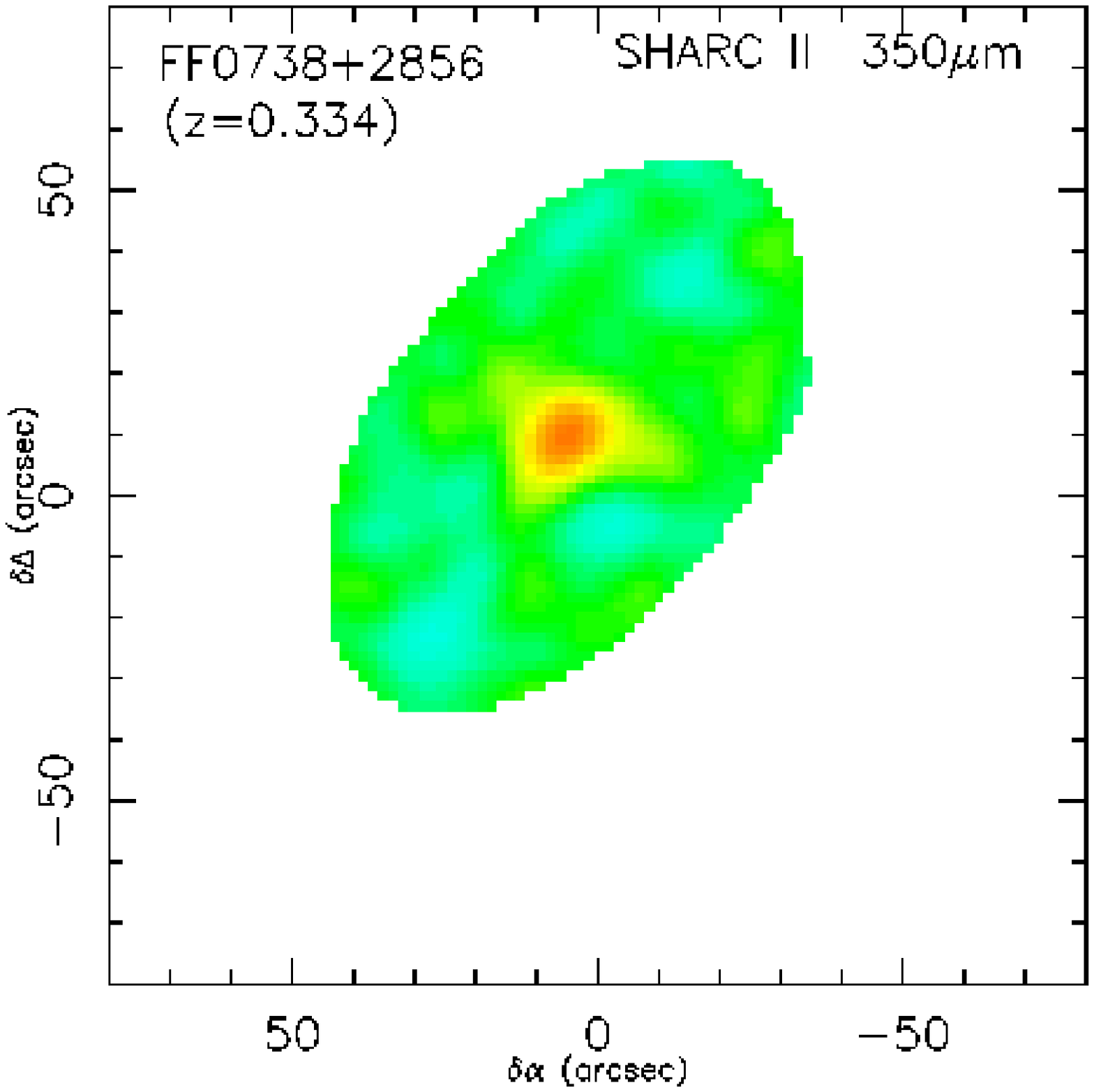}&
\includegraphics[width=1.5in, angle=270]{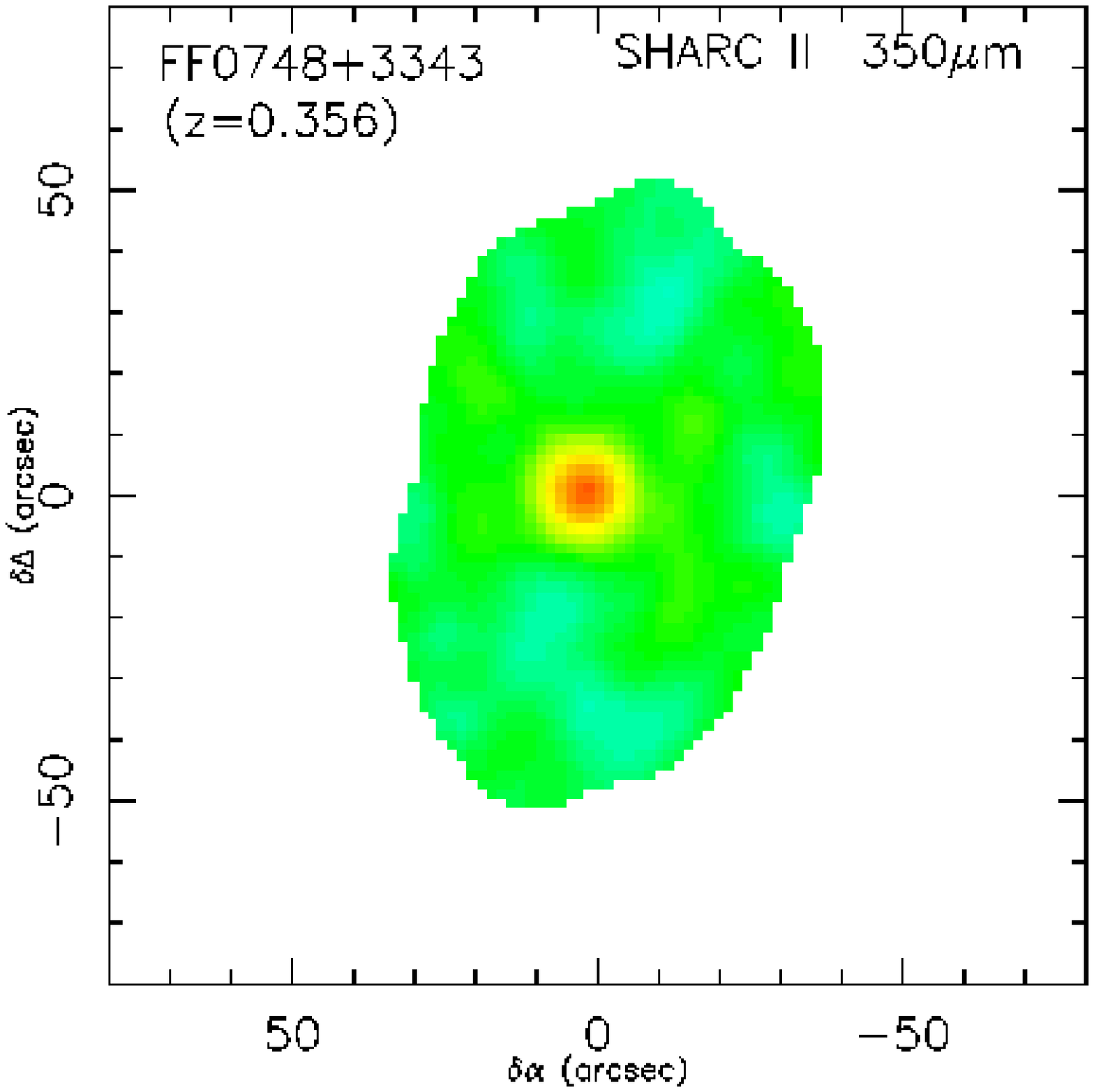}&
\includegraphics[width=1.5in, angle=270]{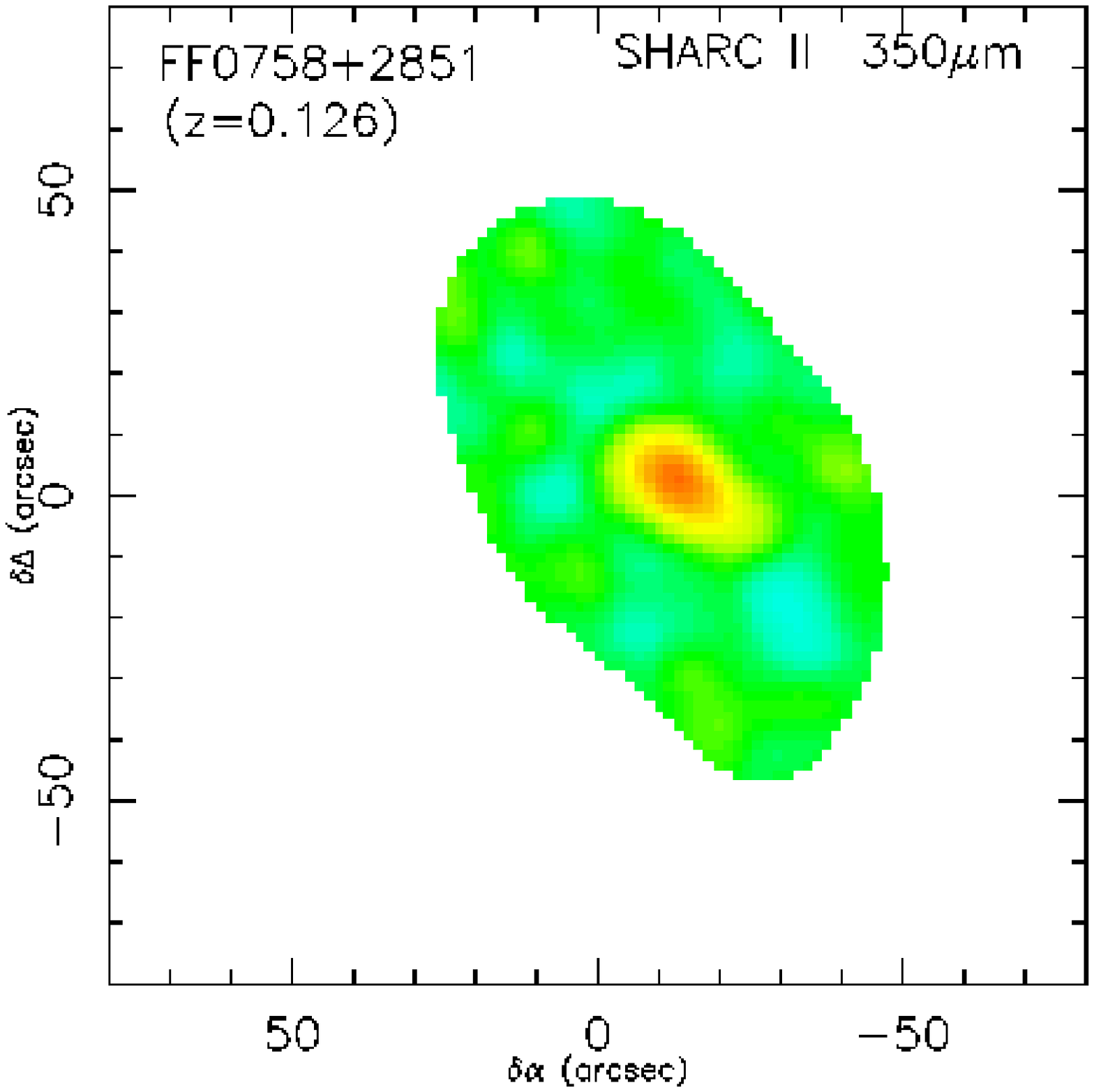}&
\includegraphics[width=1.5in, angle=270]{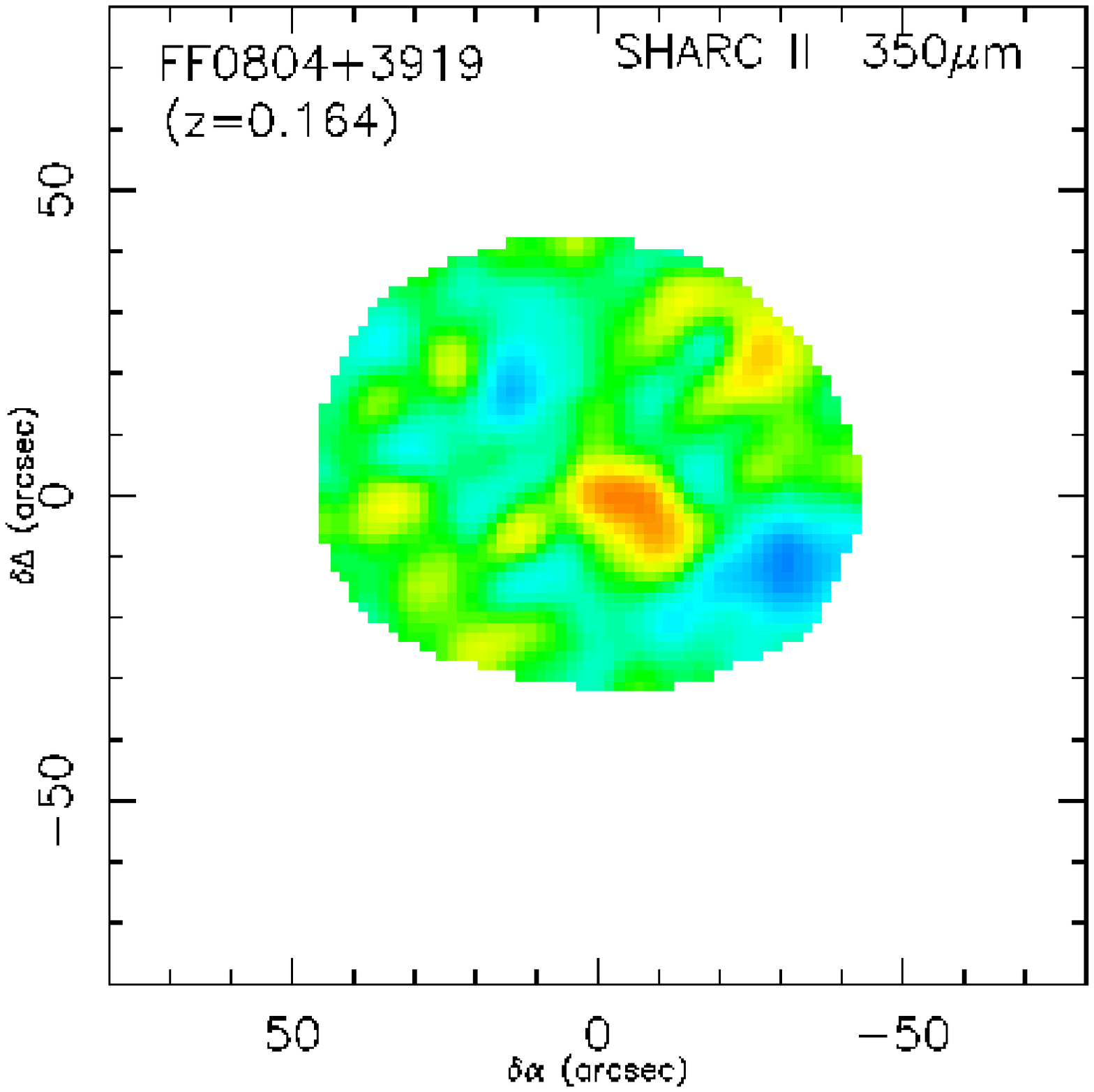}\\
\includegraphics[width=1.5in, angle=270]{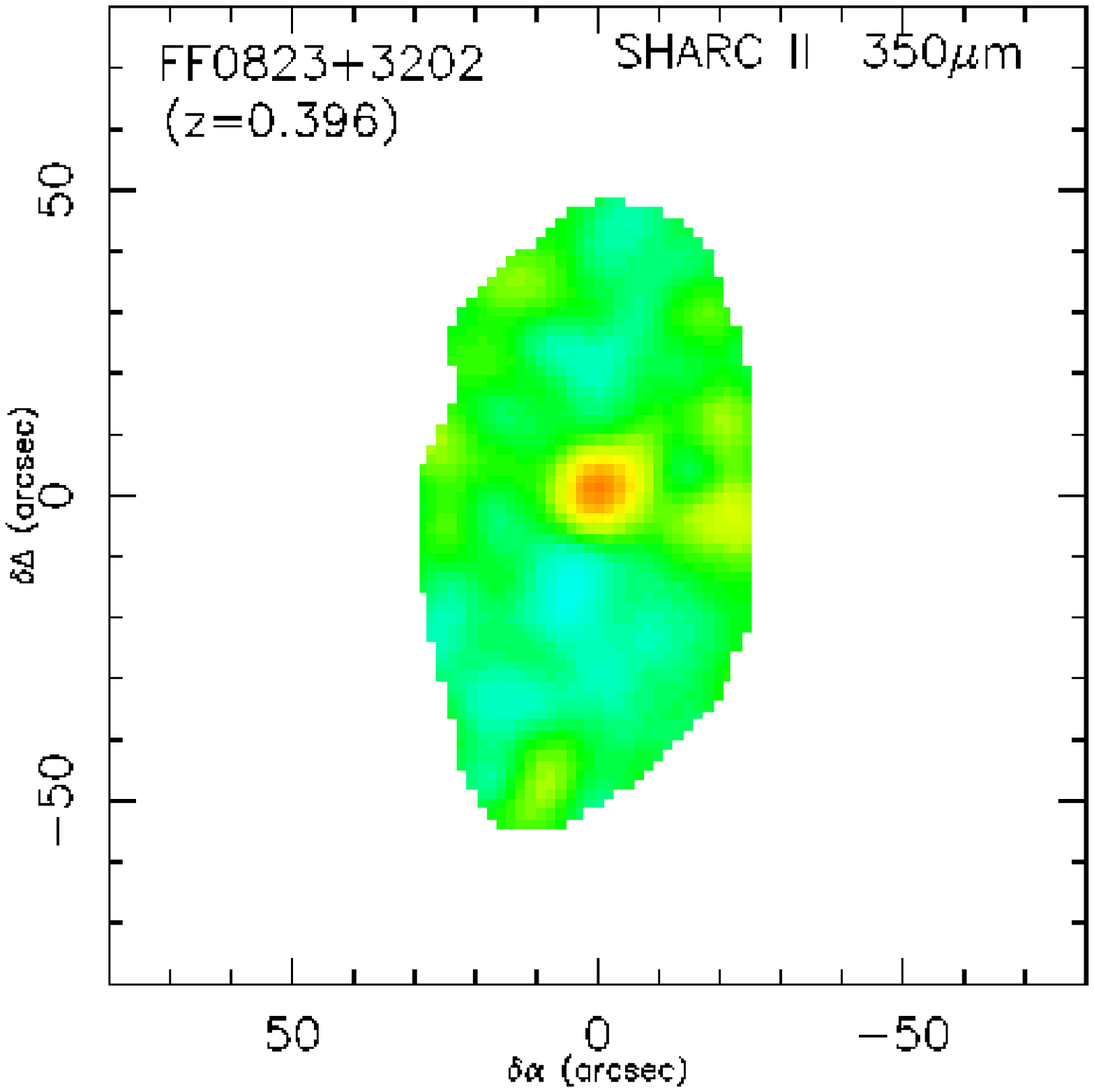}&
\includegraphics[width=1.5in, angle=270]{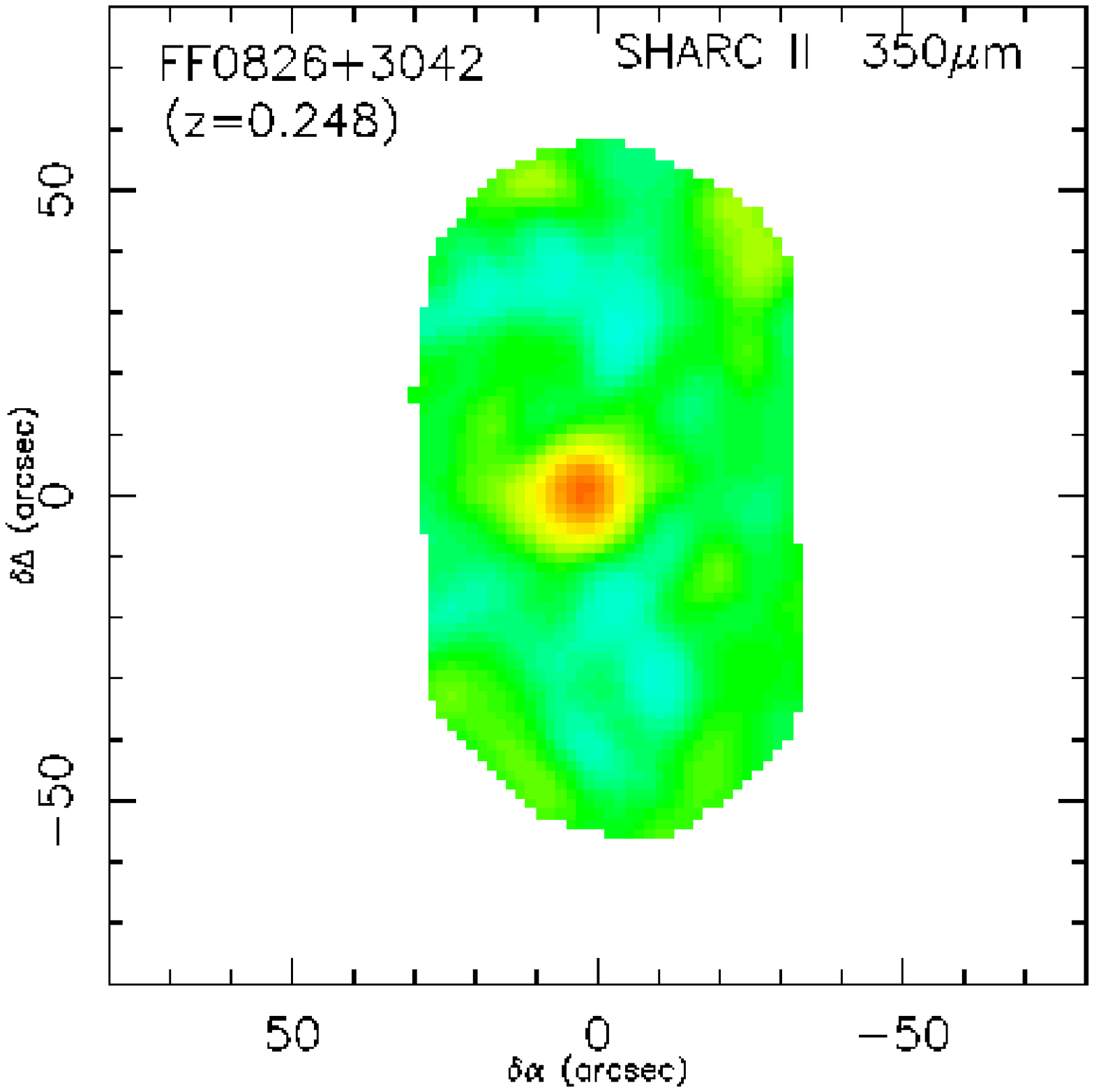}&
\includegraphics[width=1.5in, angle=270]{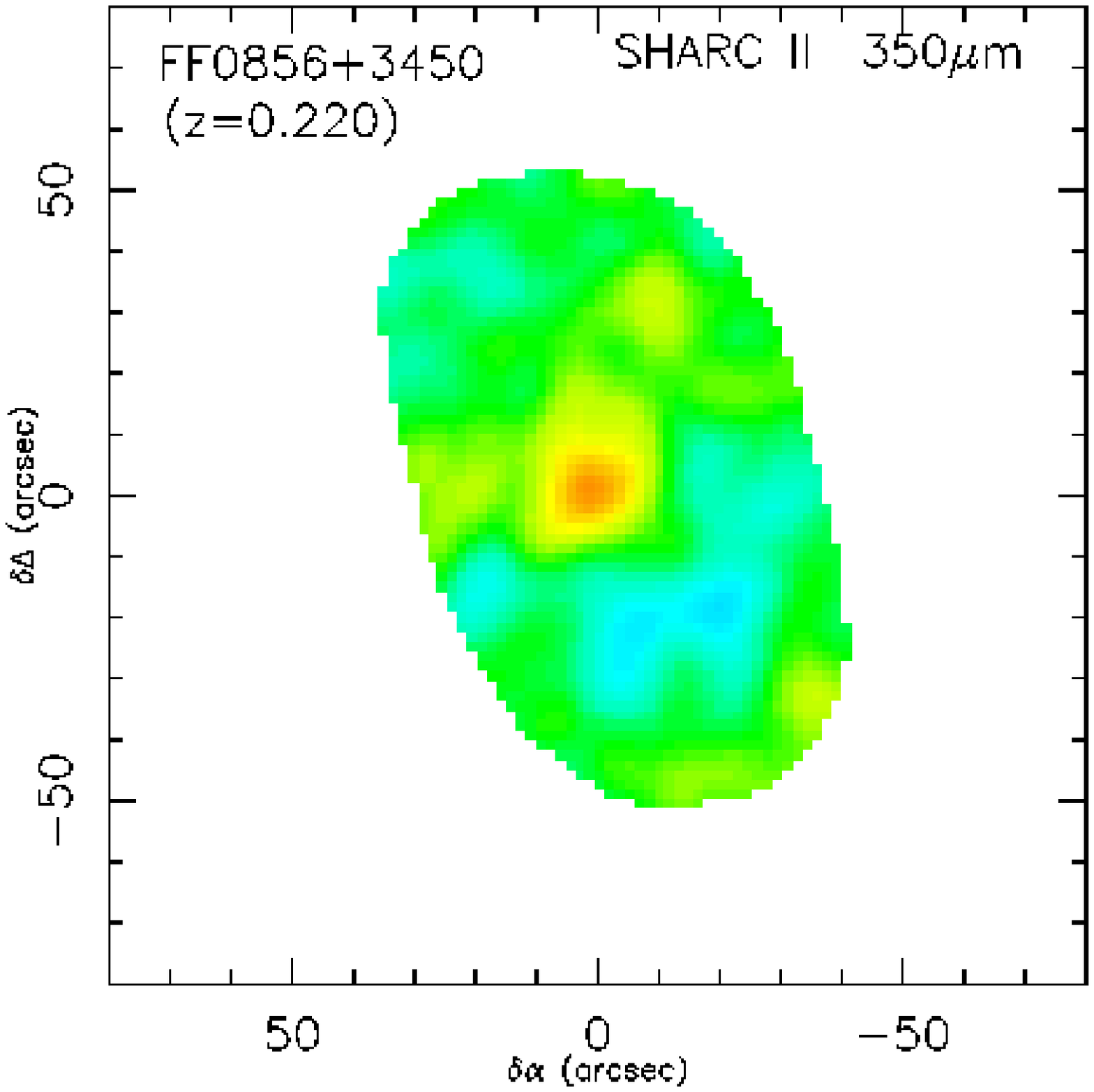}&
\includegraphics[width=1.5in, angle=270]{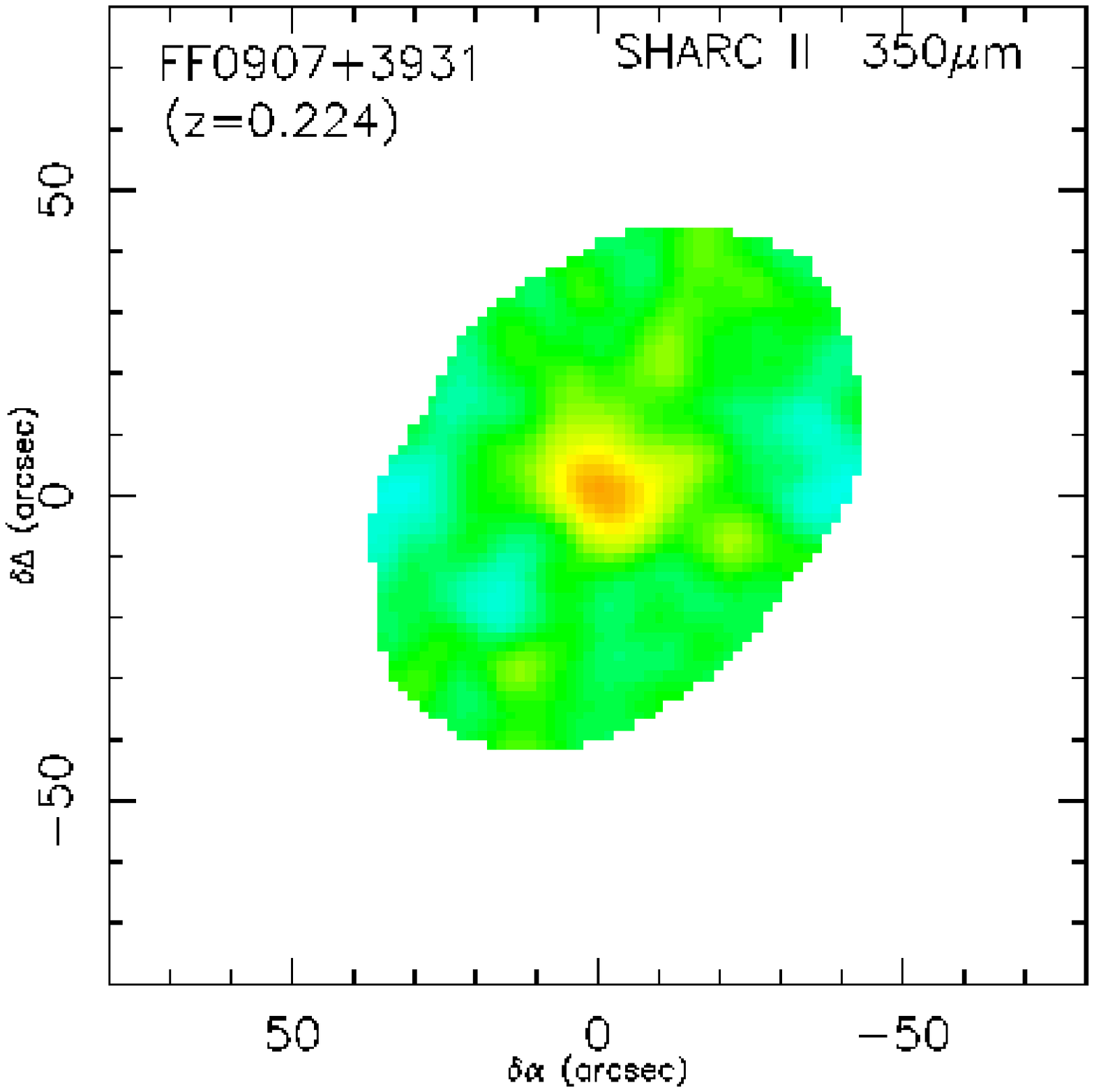}\\
\includegraphics[width=1.5in, angle=270]{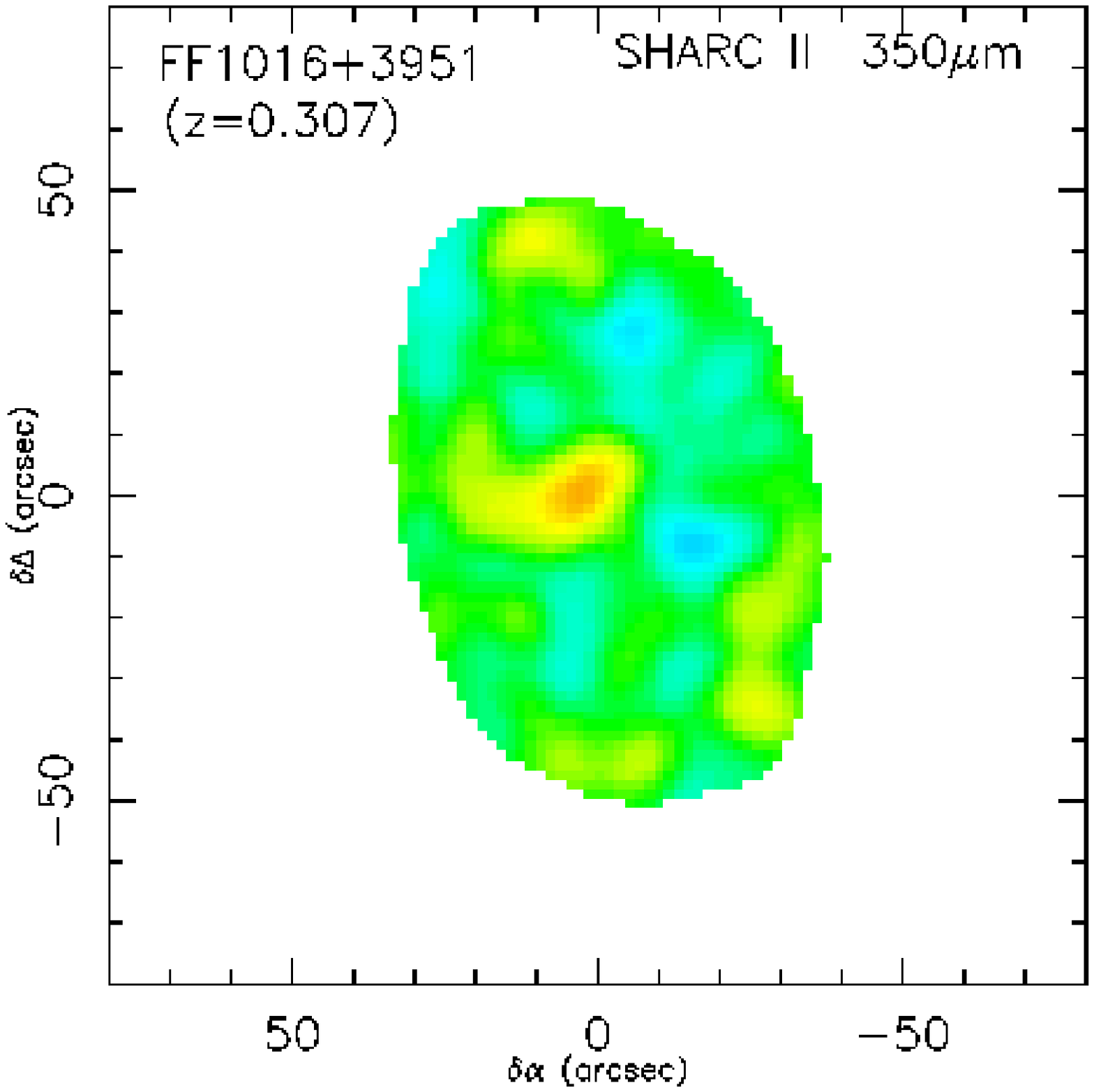}&
\includegraphics[width=1.5in, angle=270]{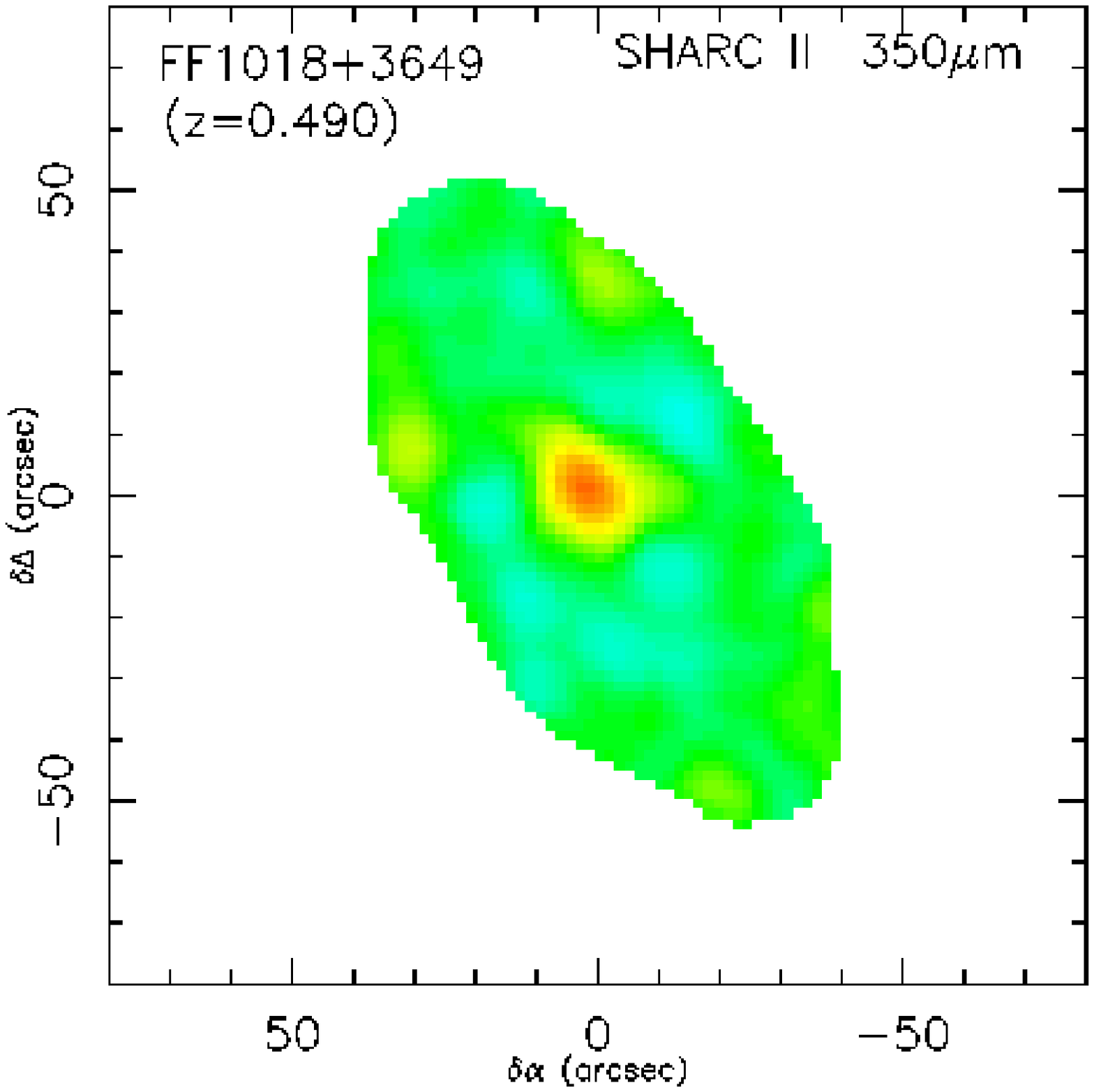}&
\includegraphics[width=1.5in, angle=270]{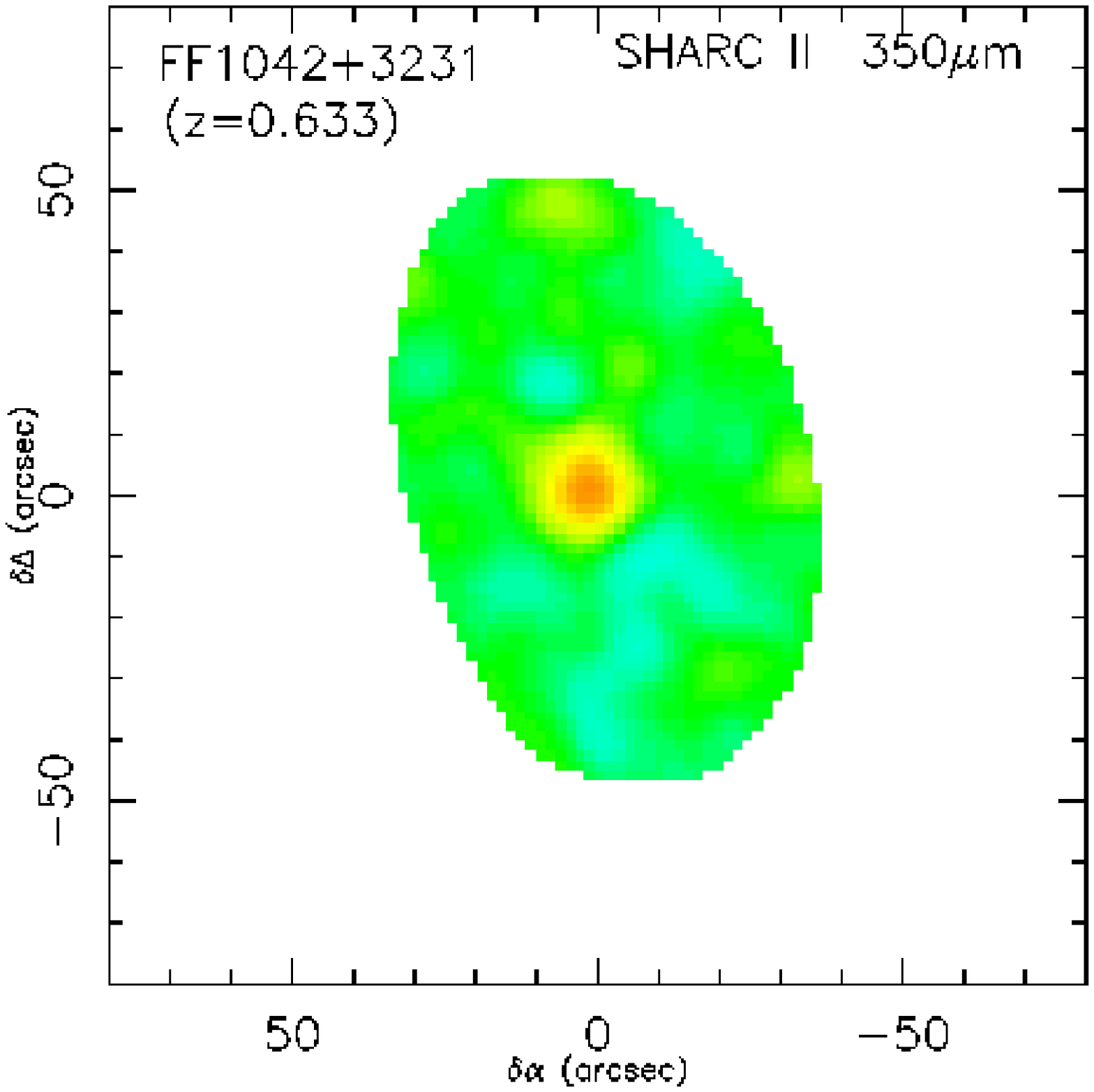}&
\includegraphics[width=1.5in, angle=270]{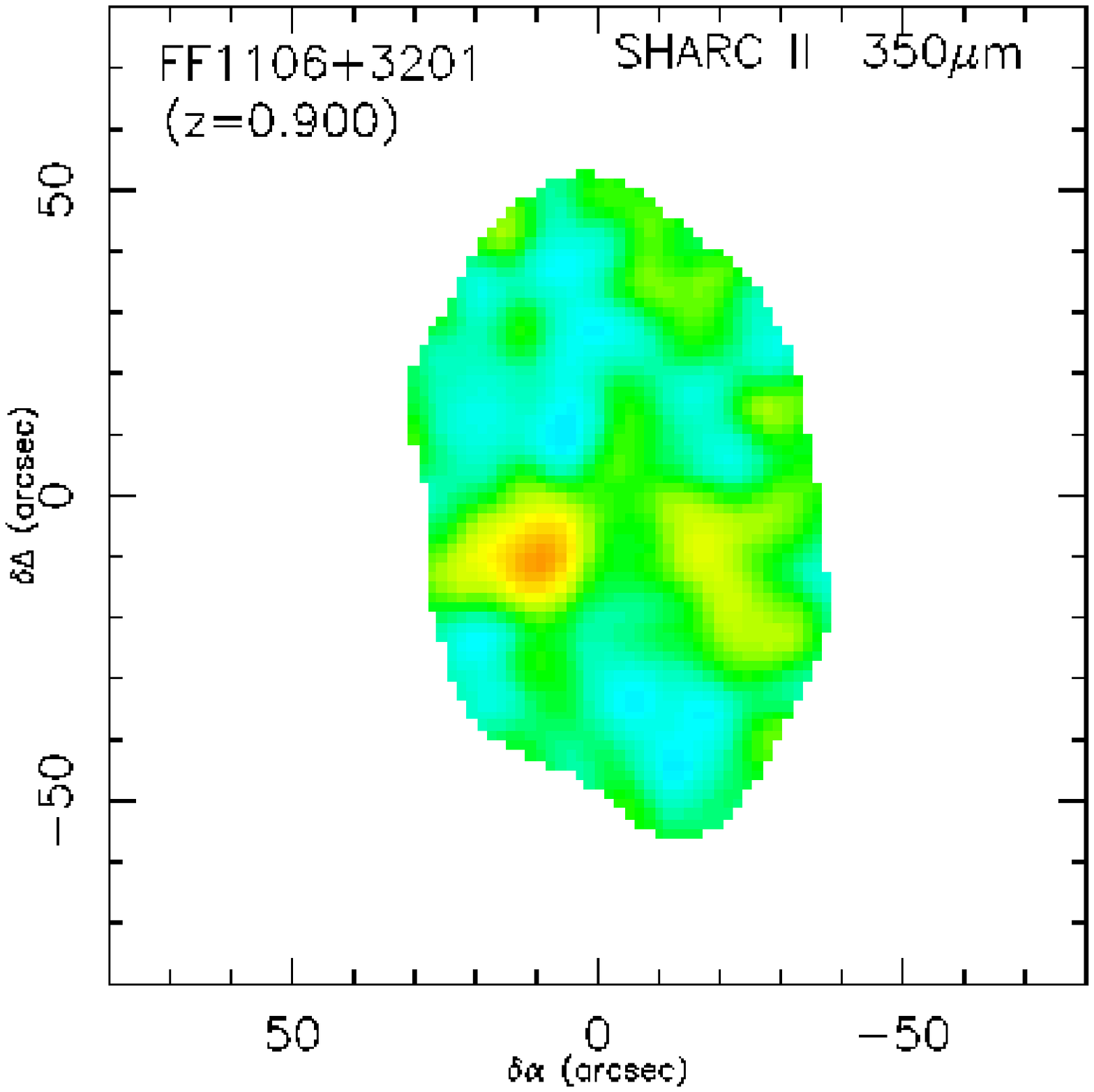}\\
\includegraphics[width=1.5in, angle=270]{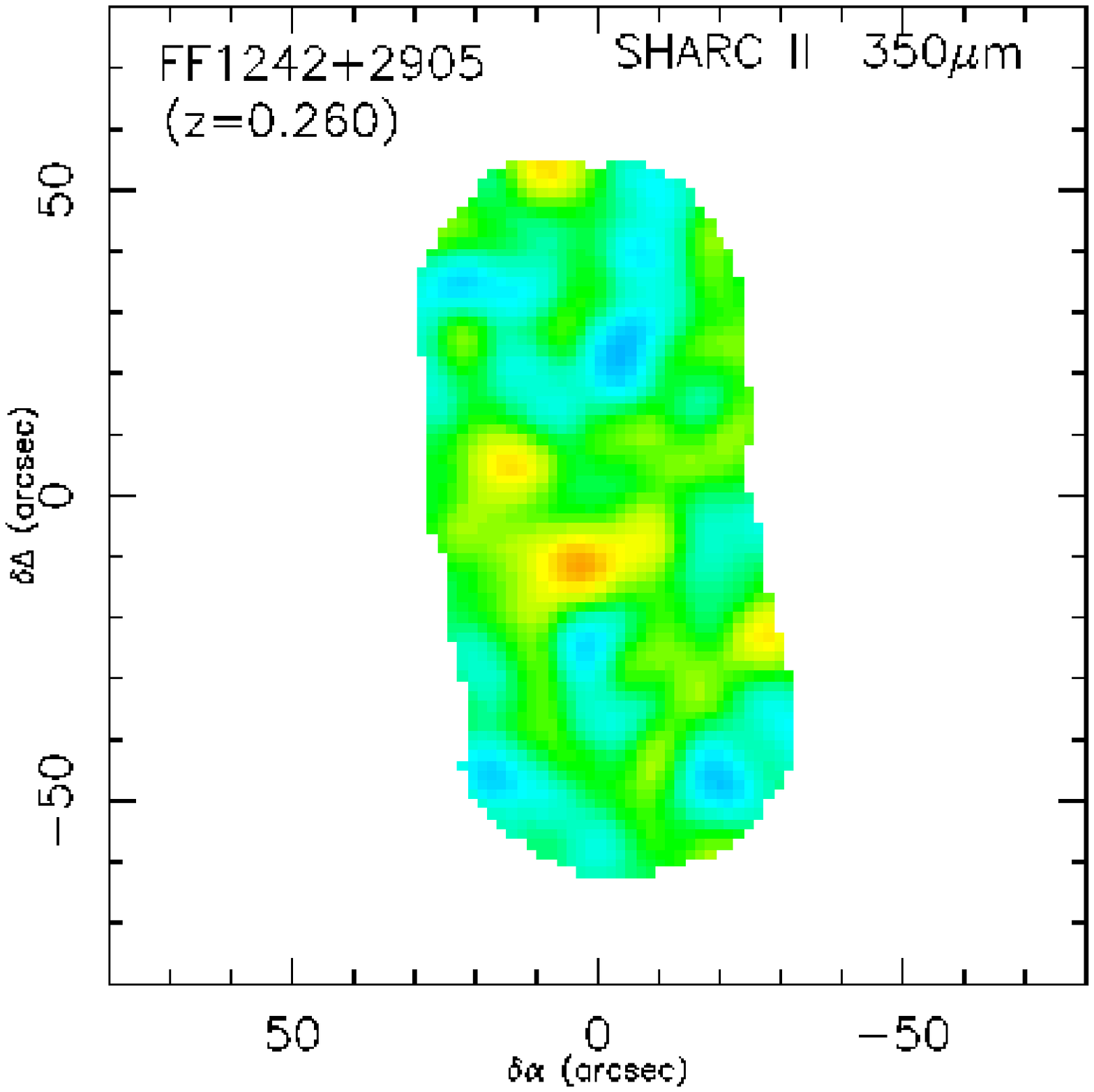}&
\includegraphics[width=1.5in, angle=270]{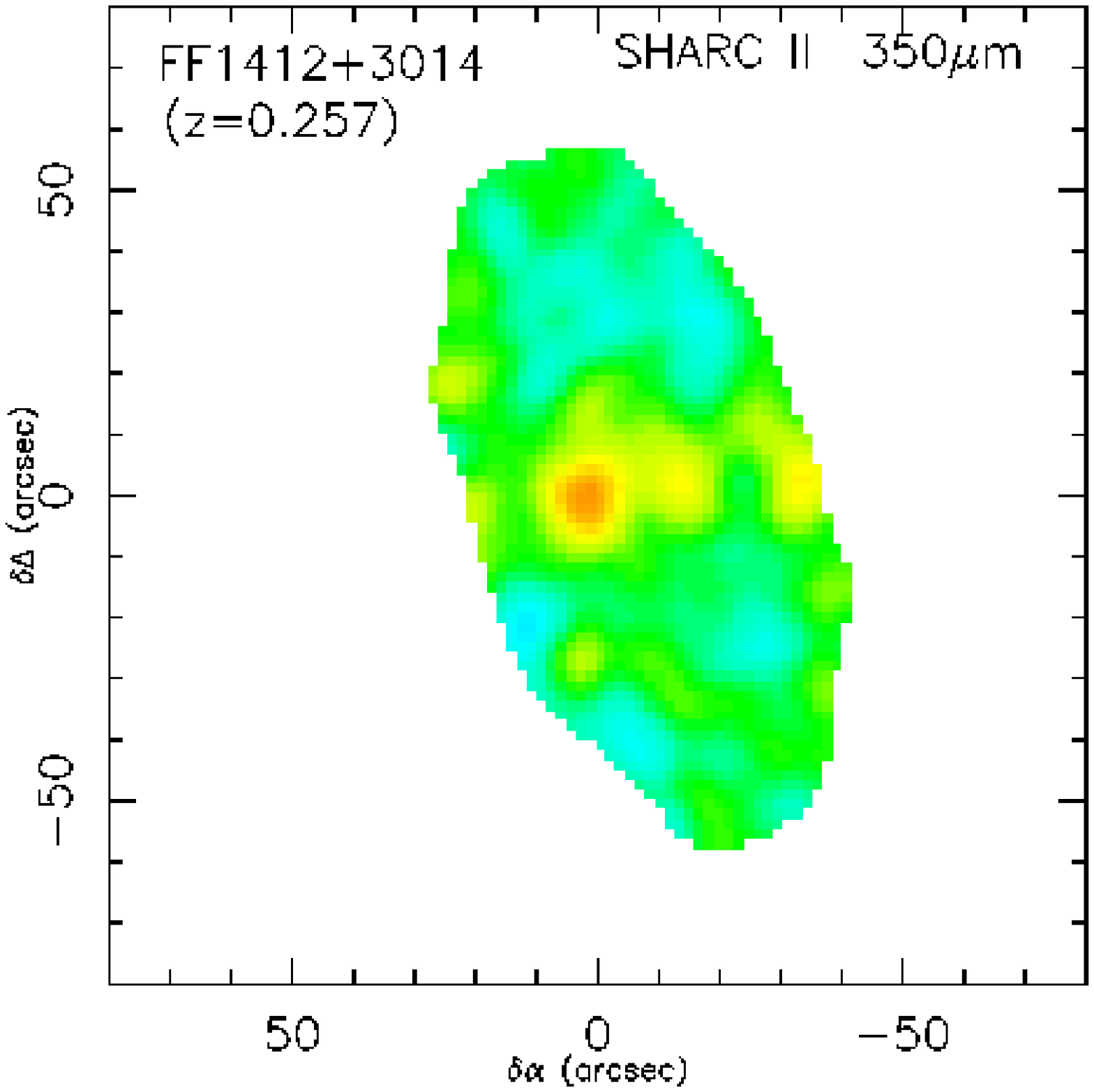}&
\includegraphics[width=1.5in, angle=270]{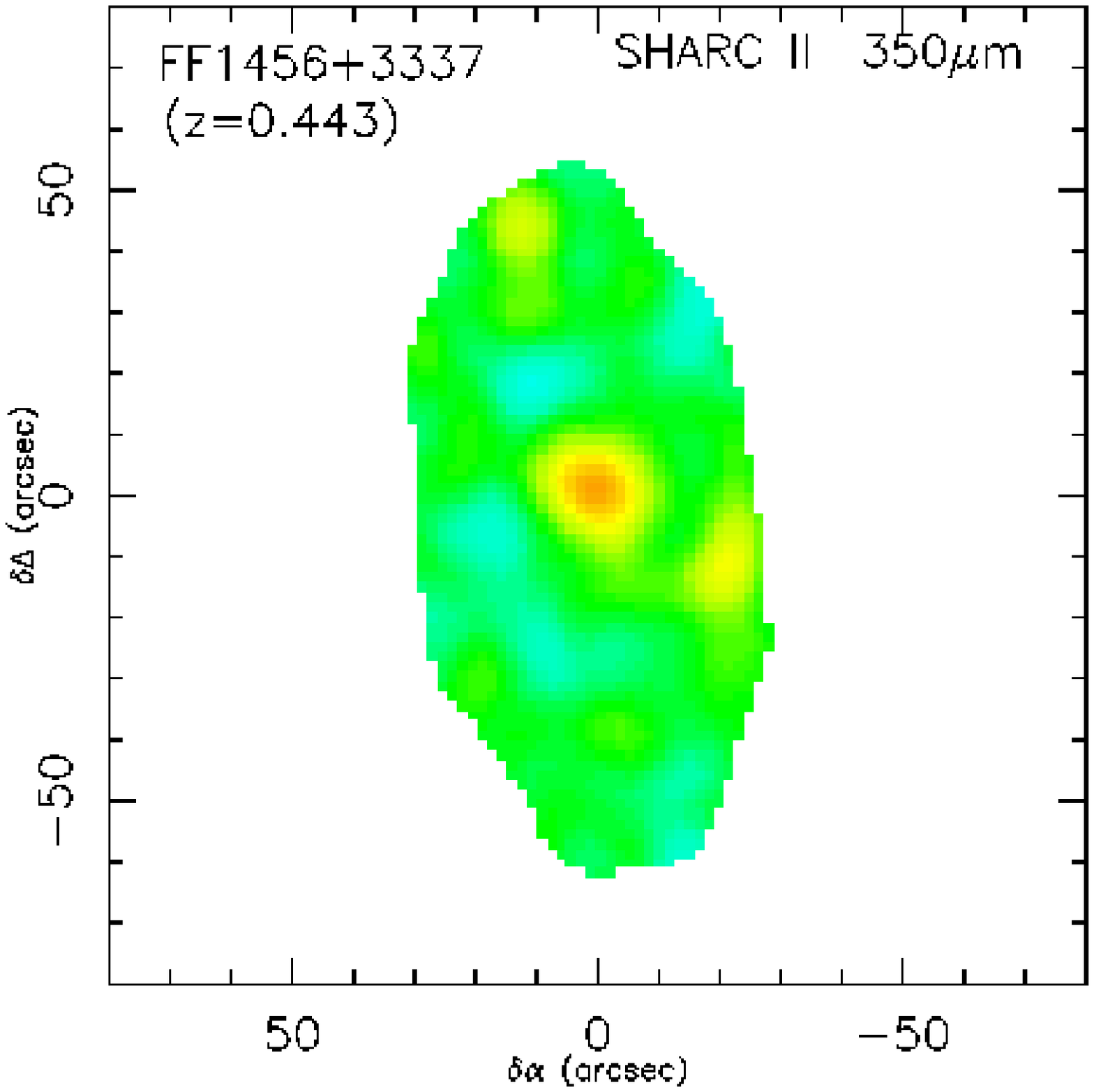}&
\includegraphics[width=1.5in, angle=270]{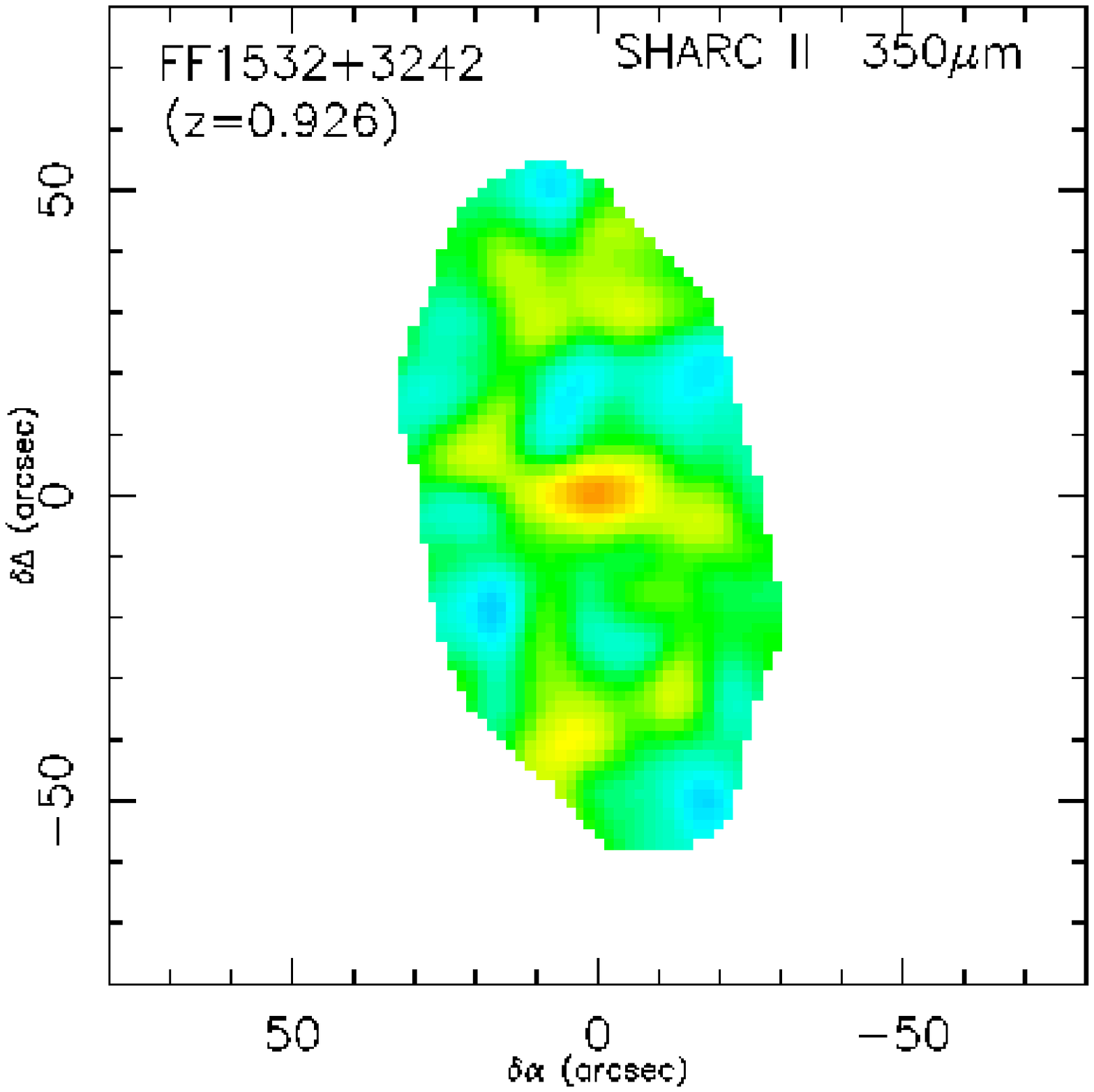}\\
\end{tabular}
\caption{350$\,\mu$m sigma-to-noise maps of the 28 FF sources detected by SHARC-II.}\label{350map_stanford_detection}
\end{center}
\end{figure*}

\begin{figure*}
\setcounter{figure}{1}
\begin{center}
\begin{tabular}{cccc}
\includegraphics[width=1.5in, angle=270]{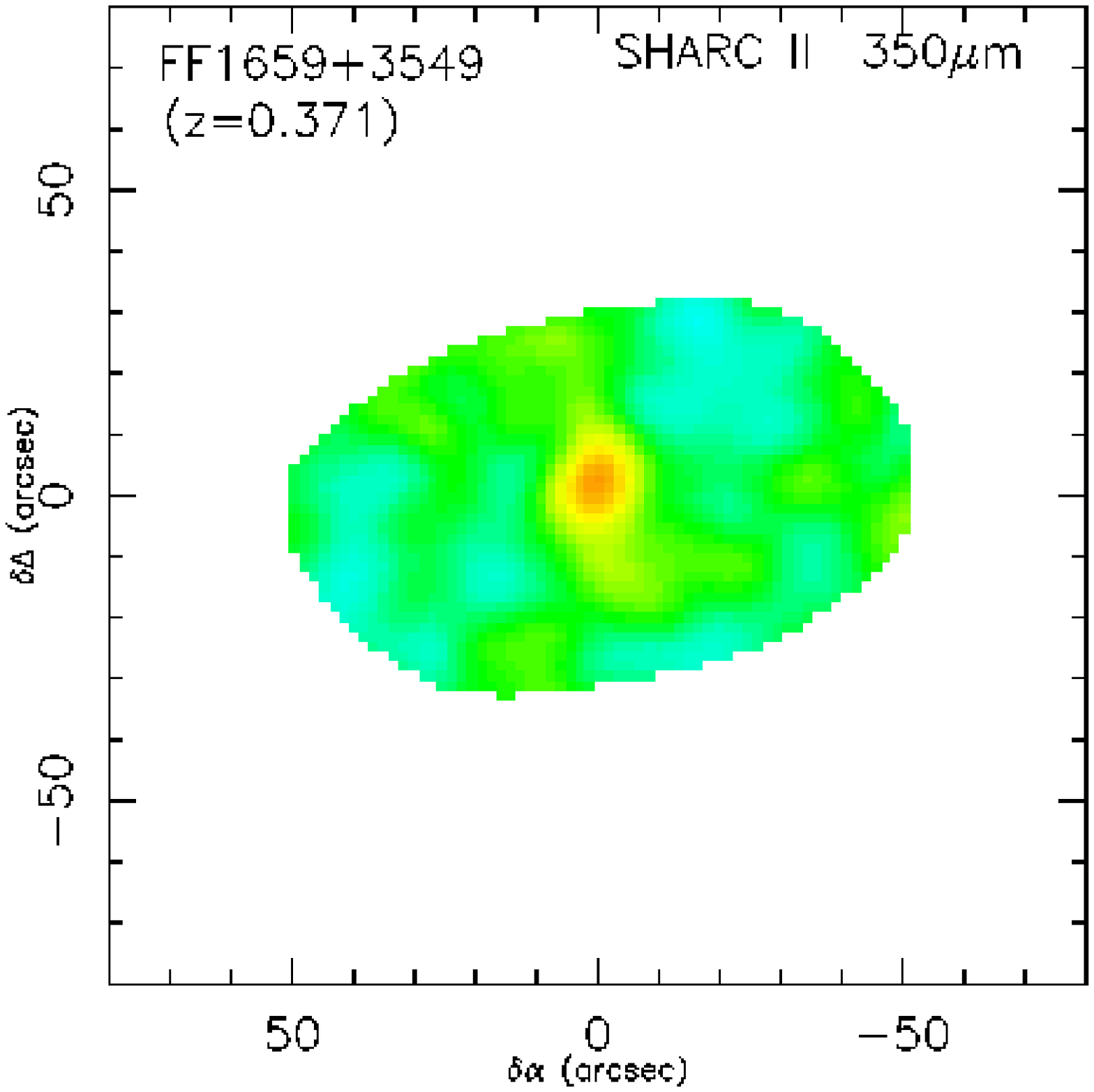}&
\includegraphics[width=1.5in, angle=270]{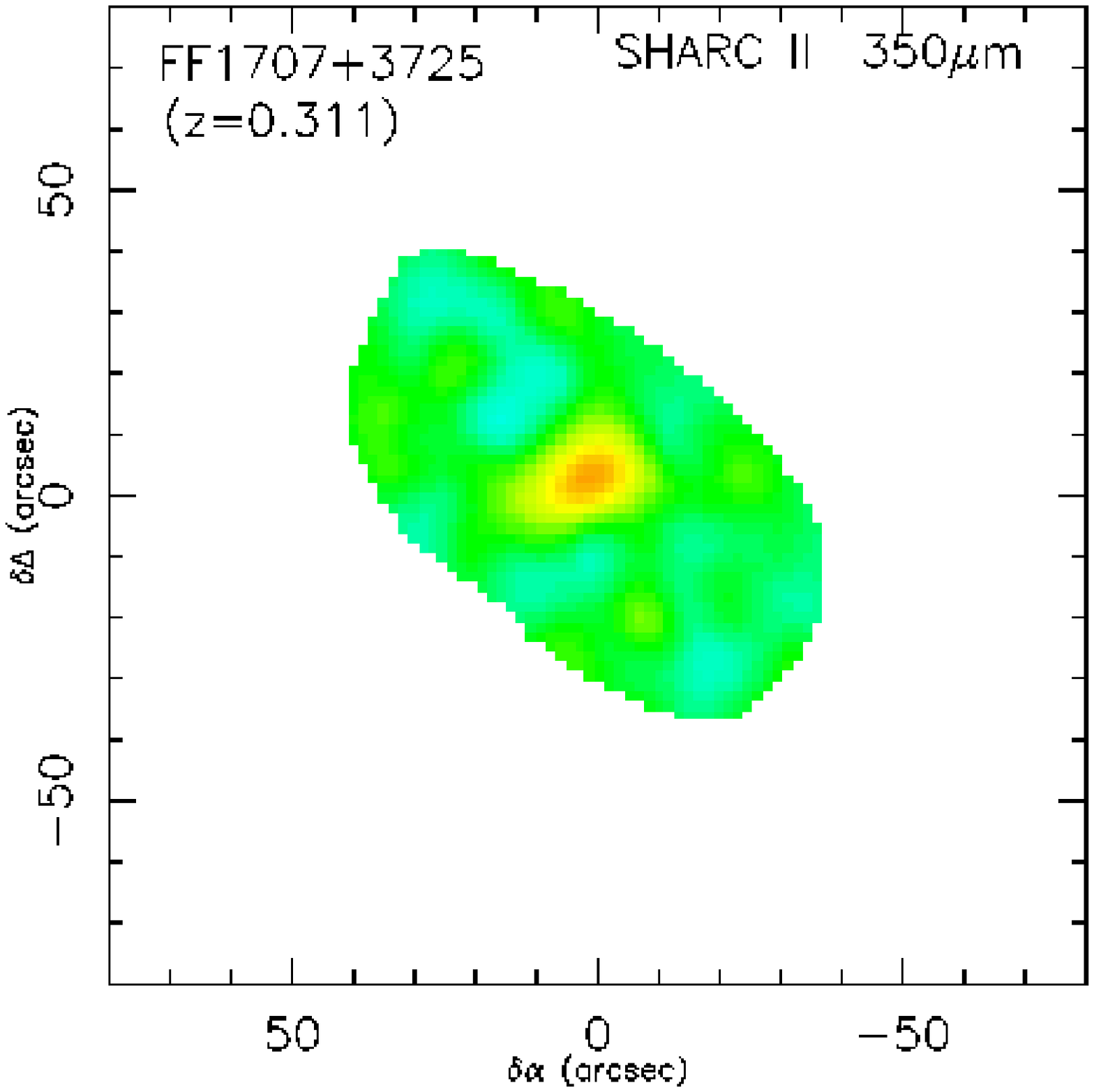}&
\includegraphics[width=1.5in, angle=270]{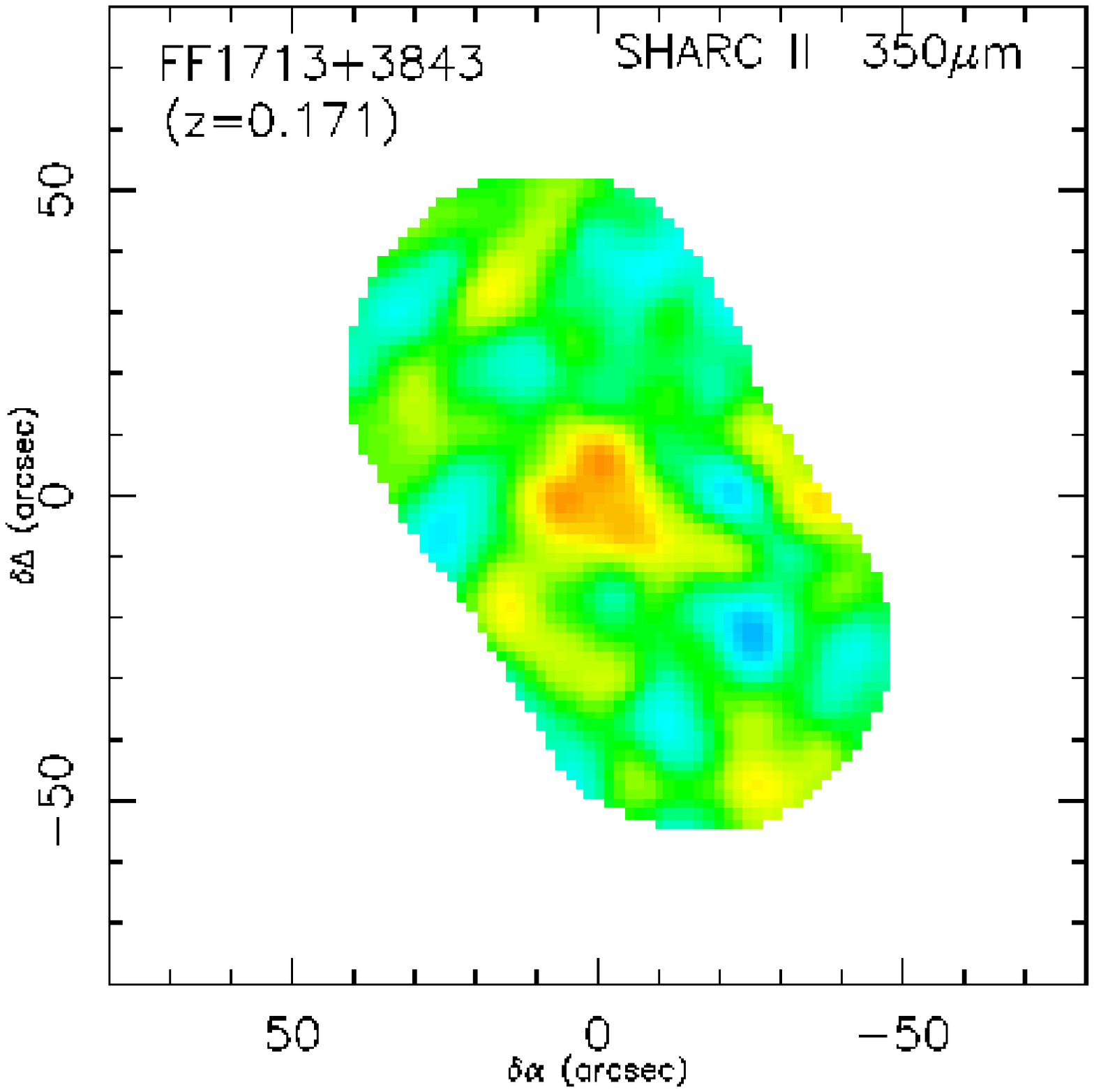}&
\includegraphics[width=1.5in, angle=270]{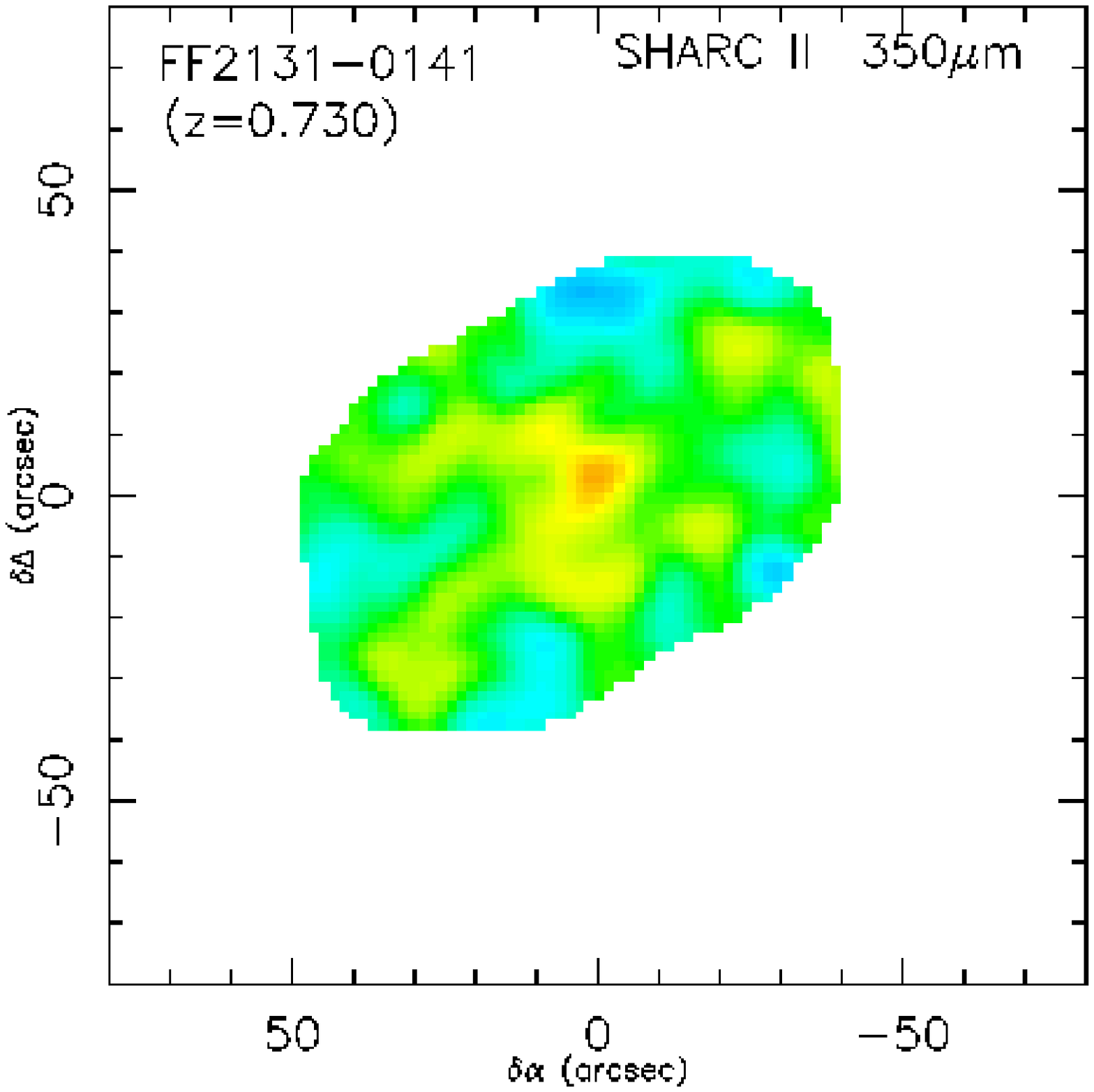}\\
\includegraphics[width=1.5in, angle=270]{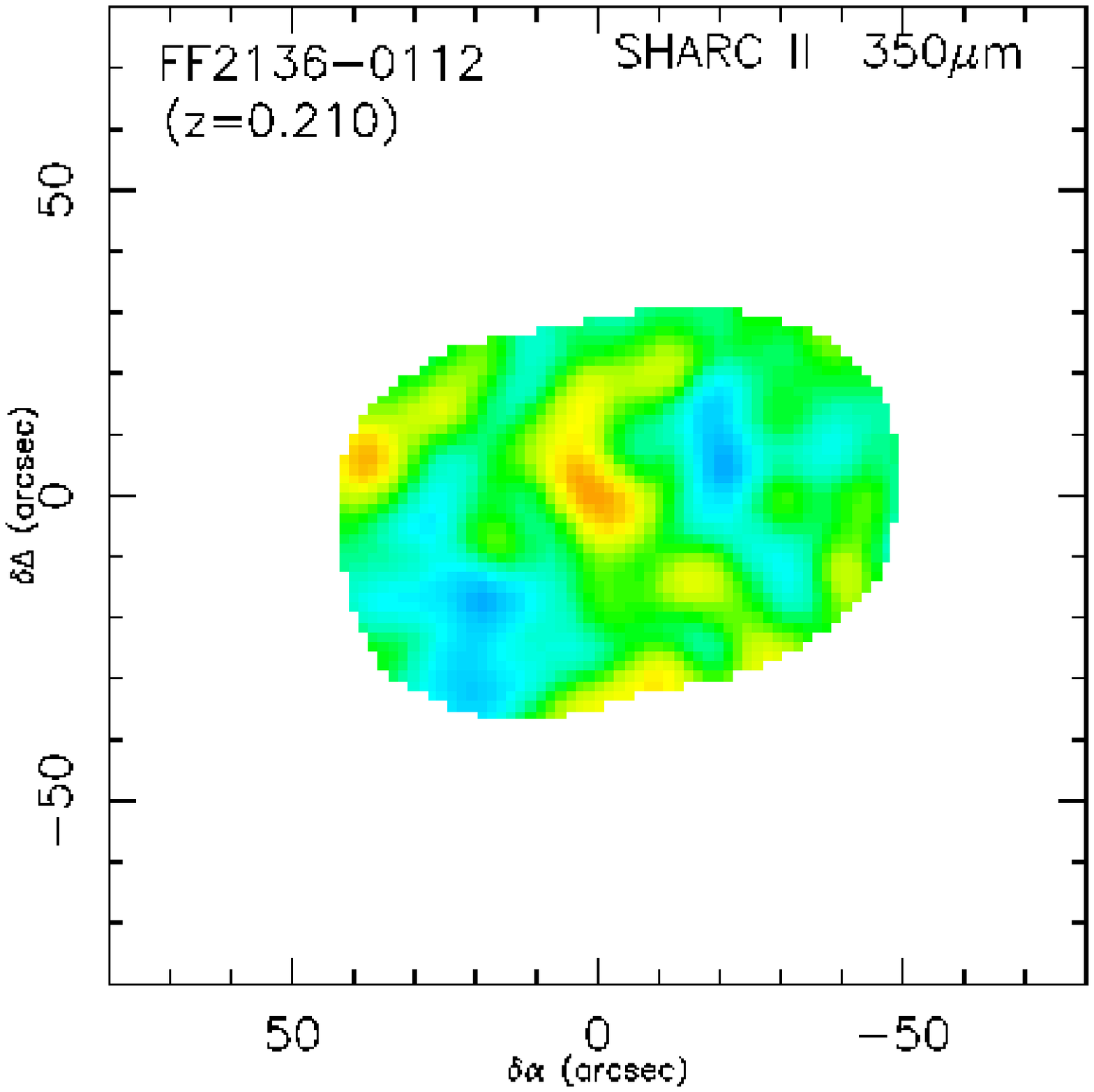}&
\includegraphics[width=1.5in, angle=270]{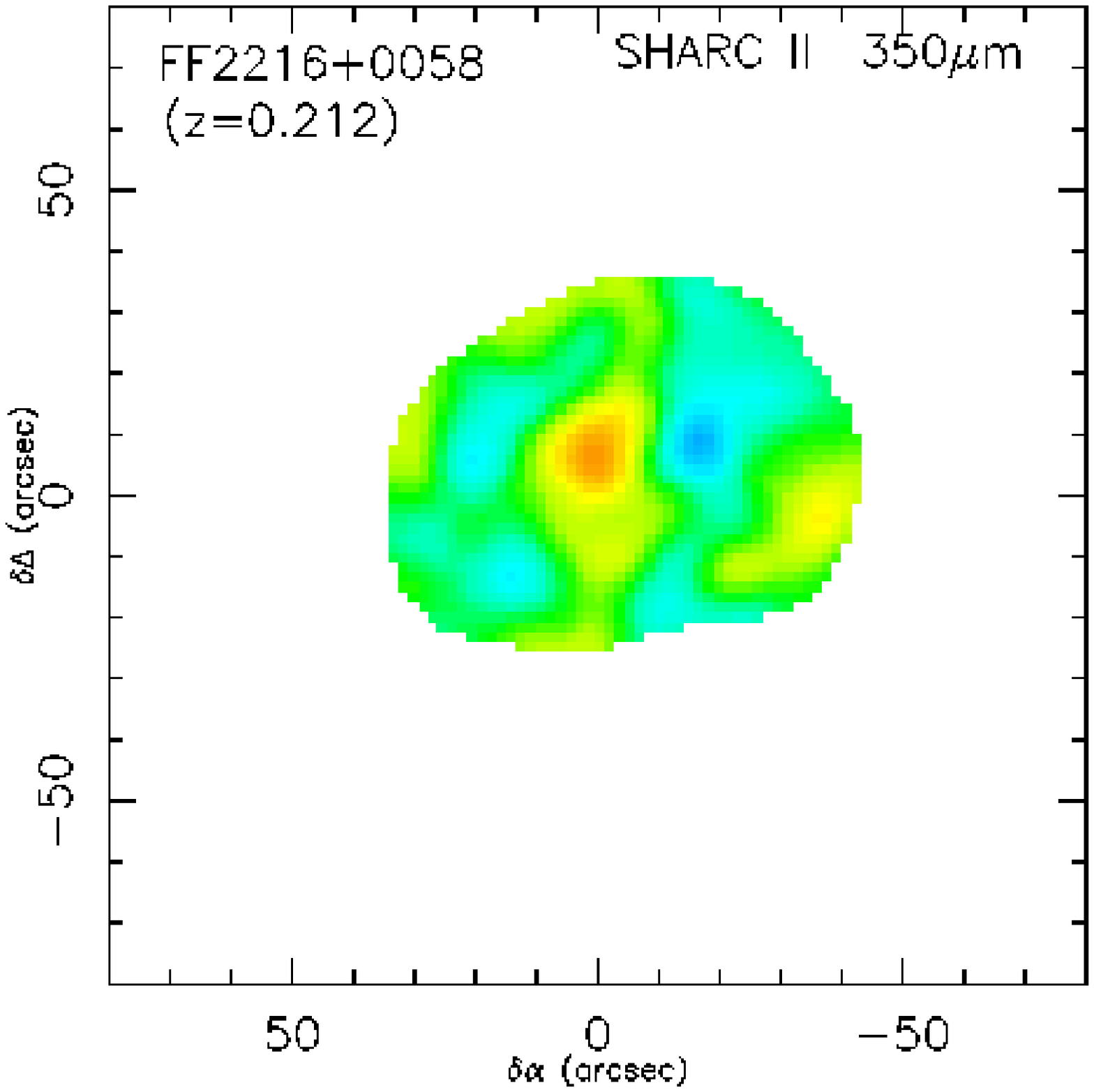}&
\includegraphics[width=1.5in, angle=270]{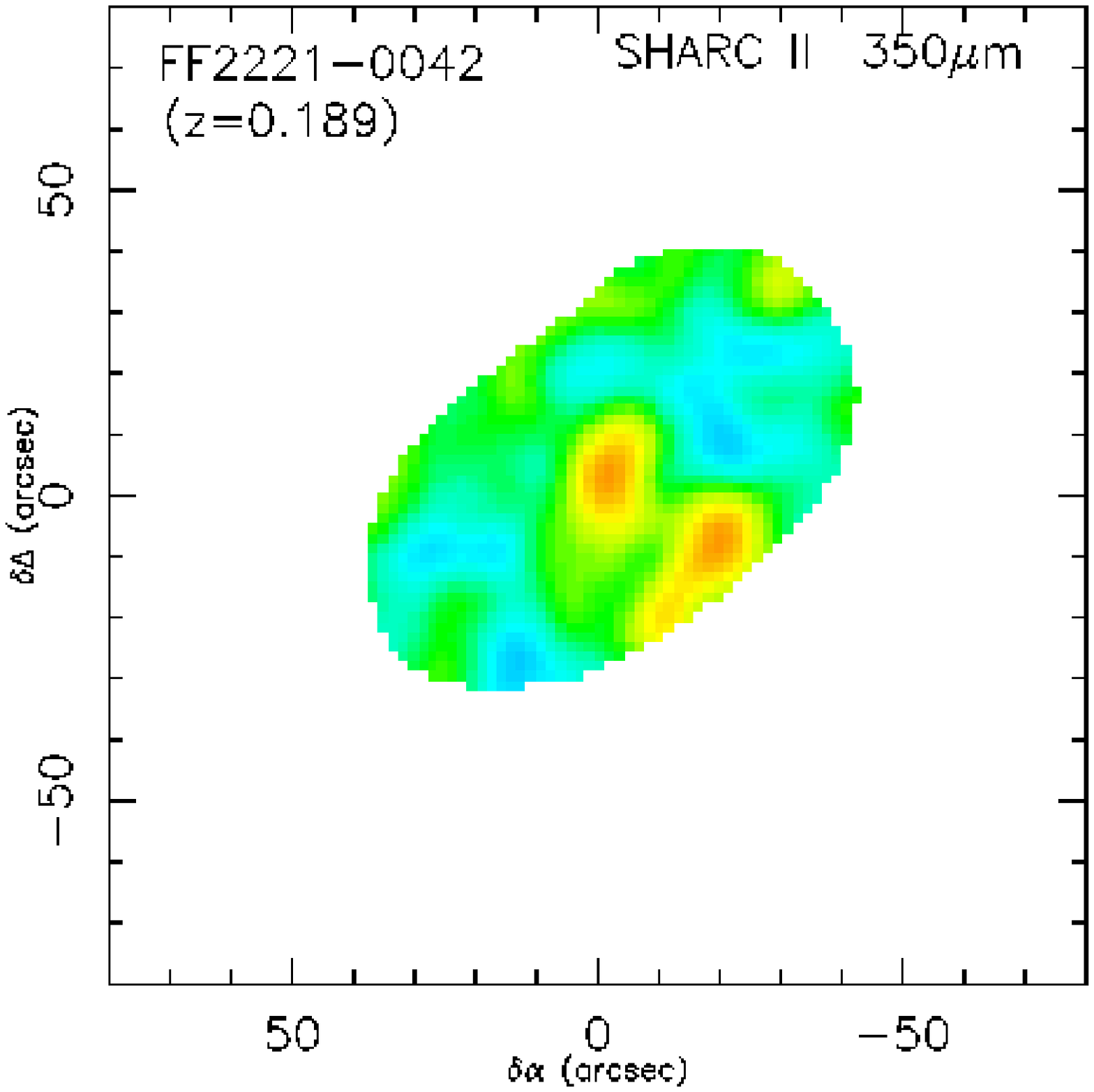}&
\includegraphics[width=1.5in, angle=270]{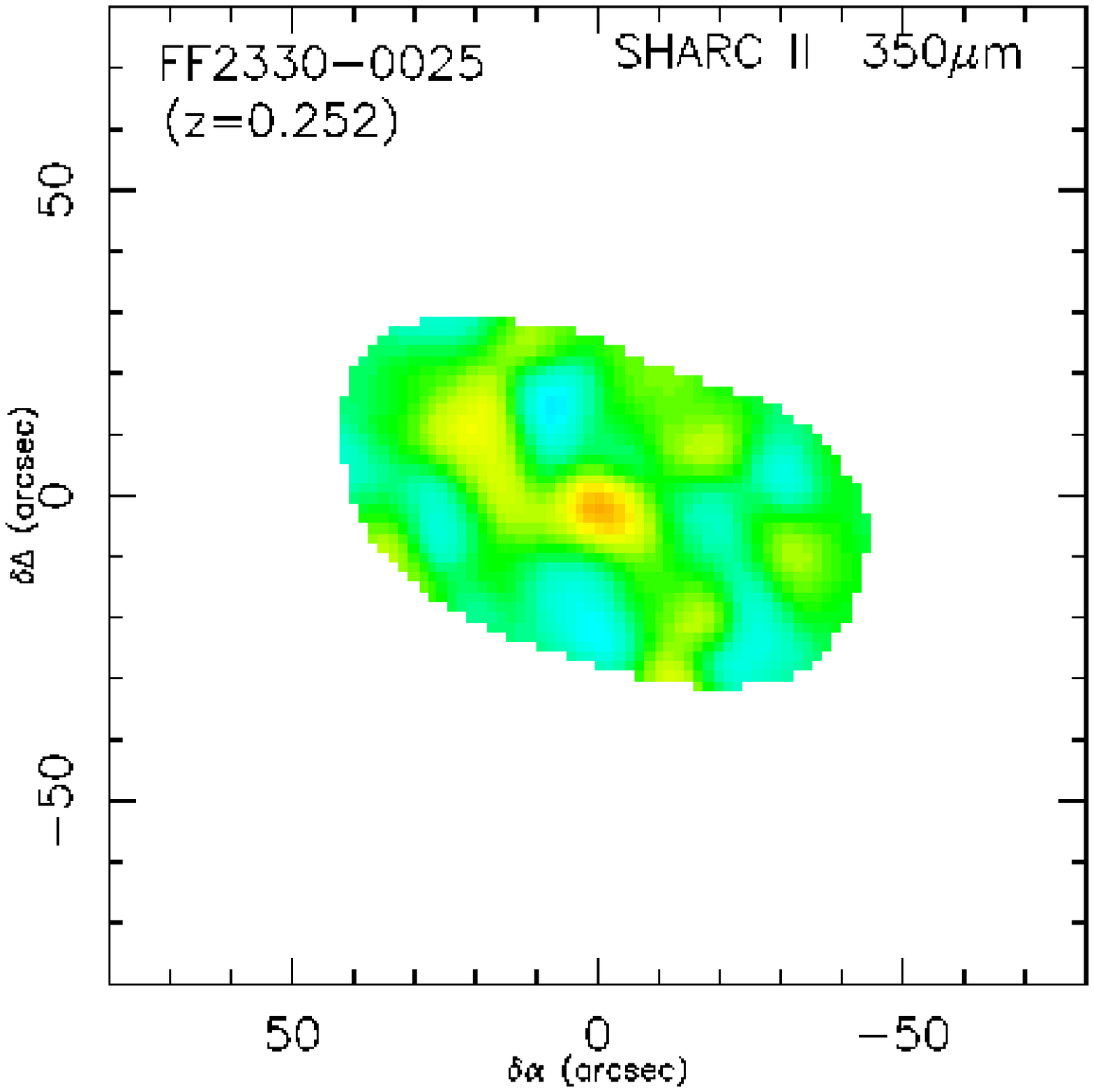}\\
\end{tabular}
\caption{Continued}
\end{center}
\end{figure*}

\begin{figure*}
\begin{center}
\begin{tabular}{ccc}
\includegraphics[width=1.5in, angle=90]{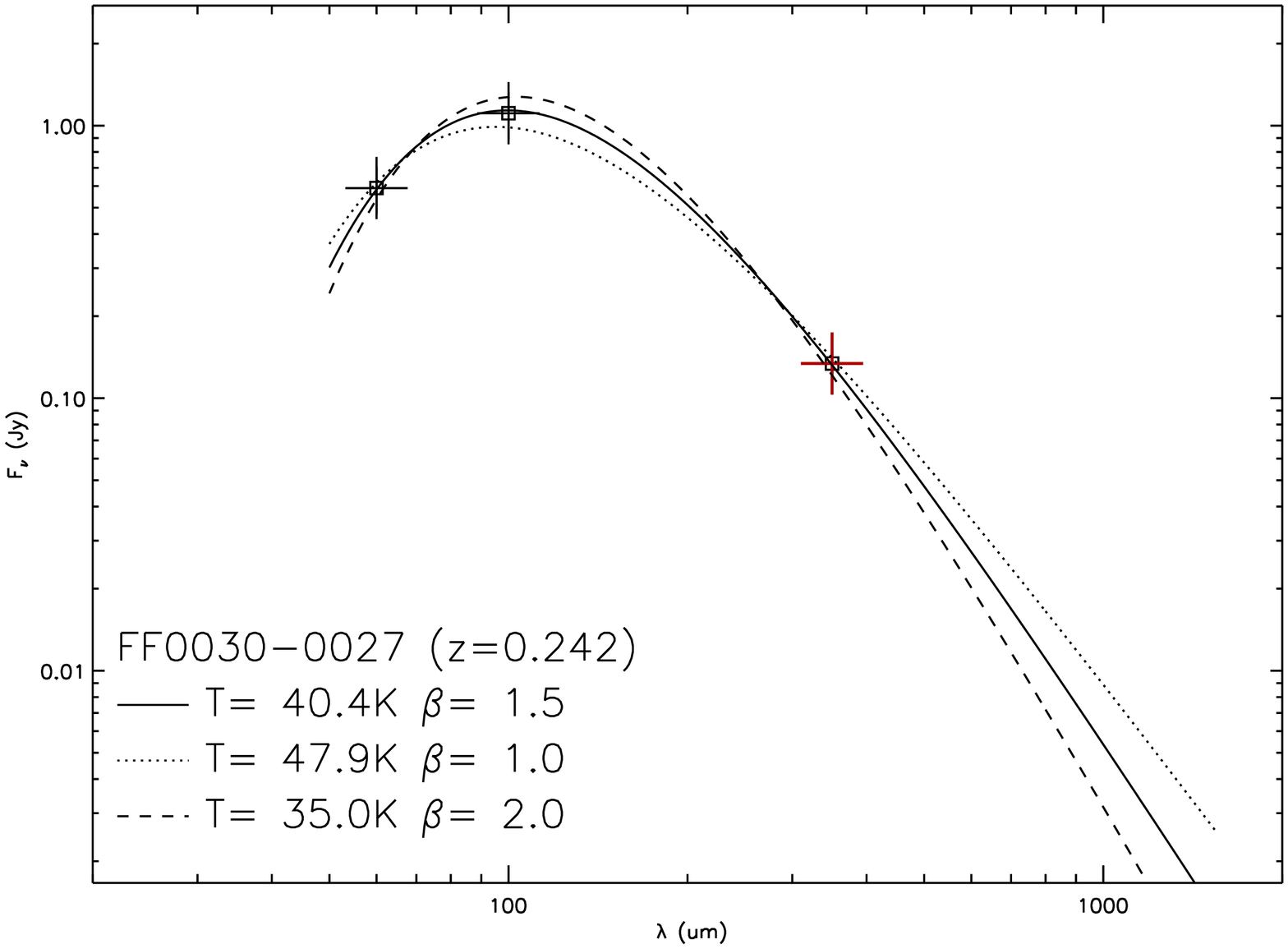}&
\includegraphics[width=1.5in, angle=90]{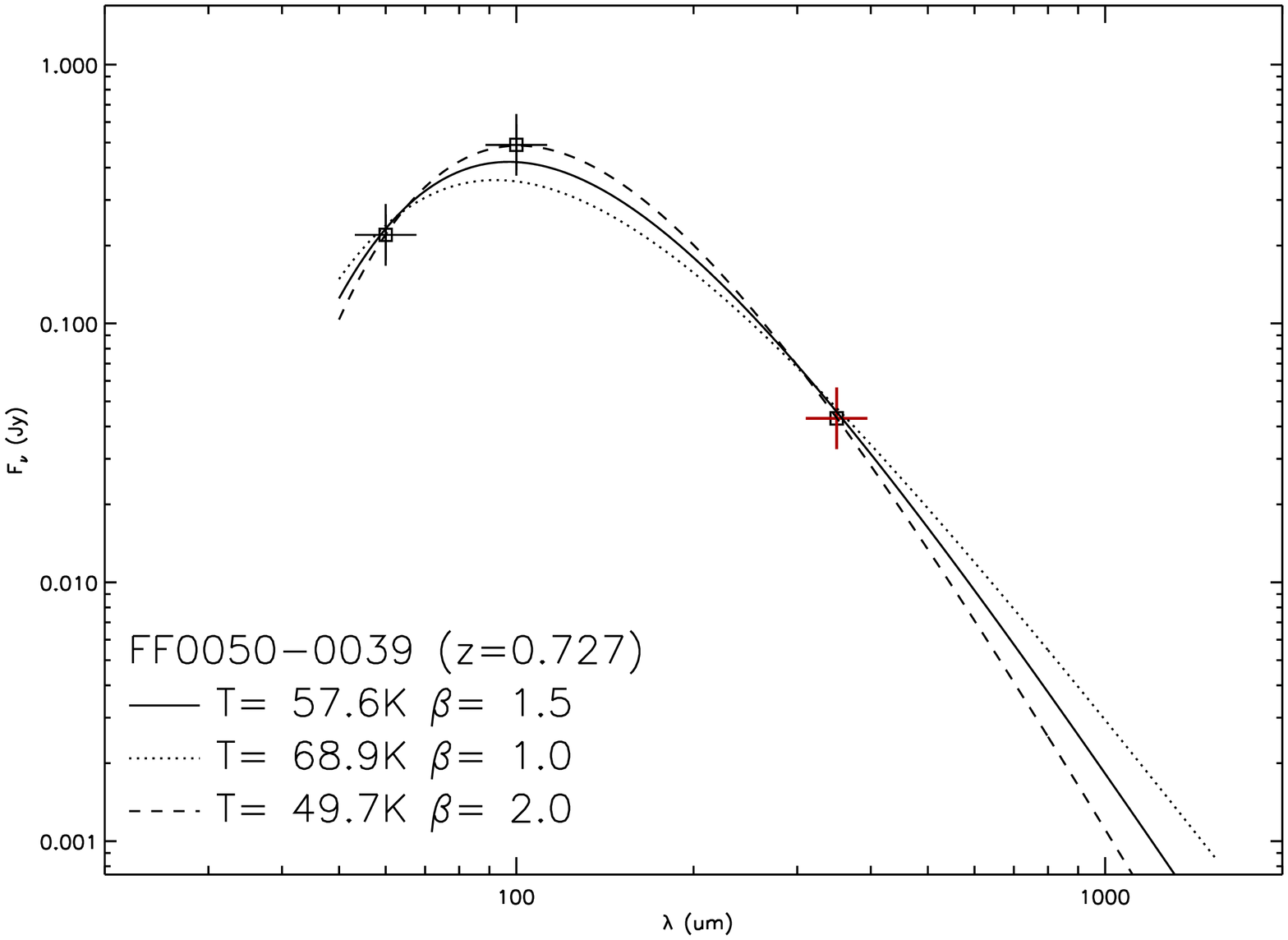}&
\includegraphics[width=1.5in, angle=90]{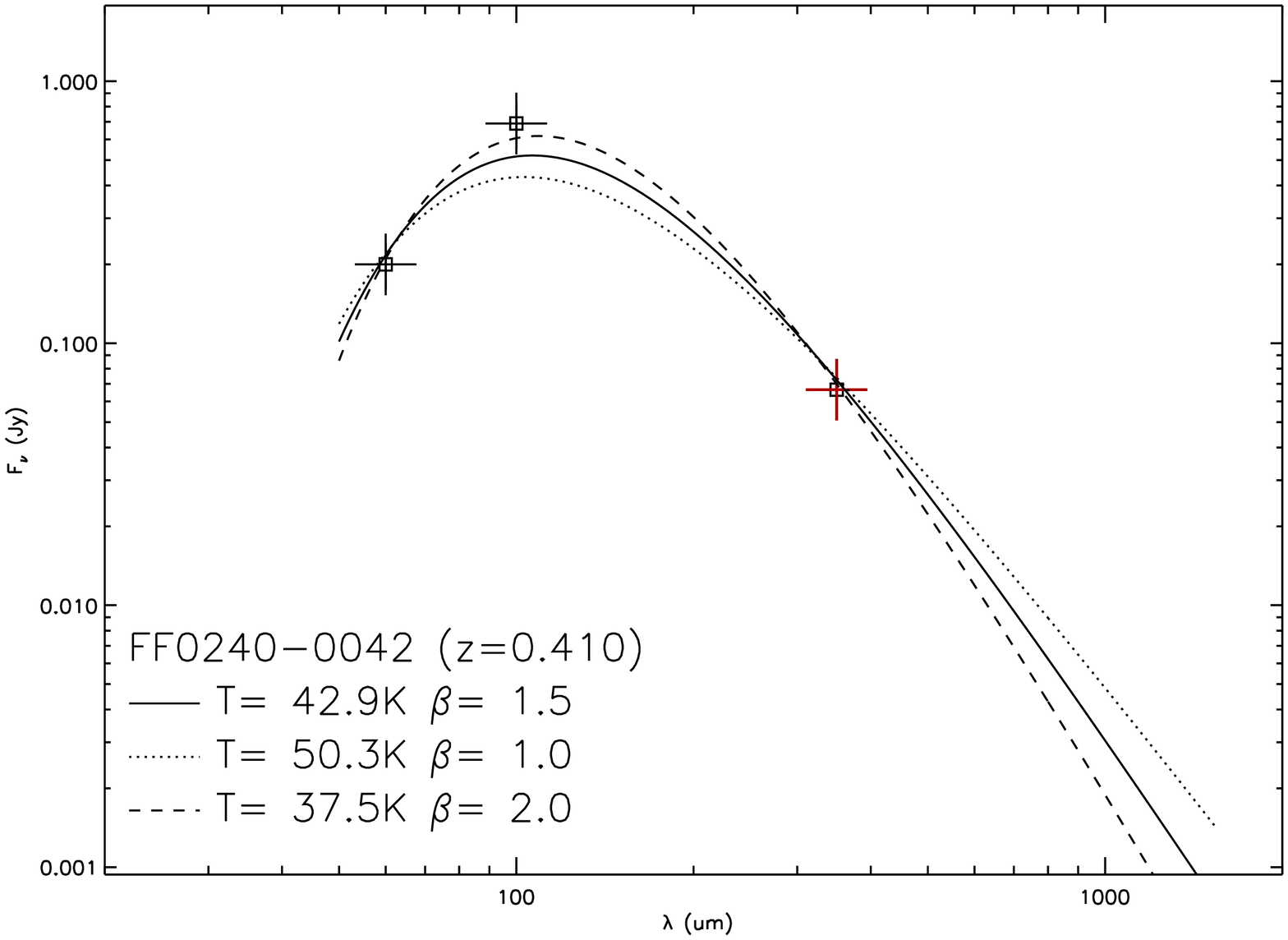}\\
\includegraphics[width=1.5in, angle=90]{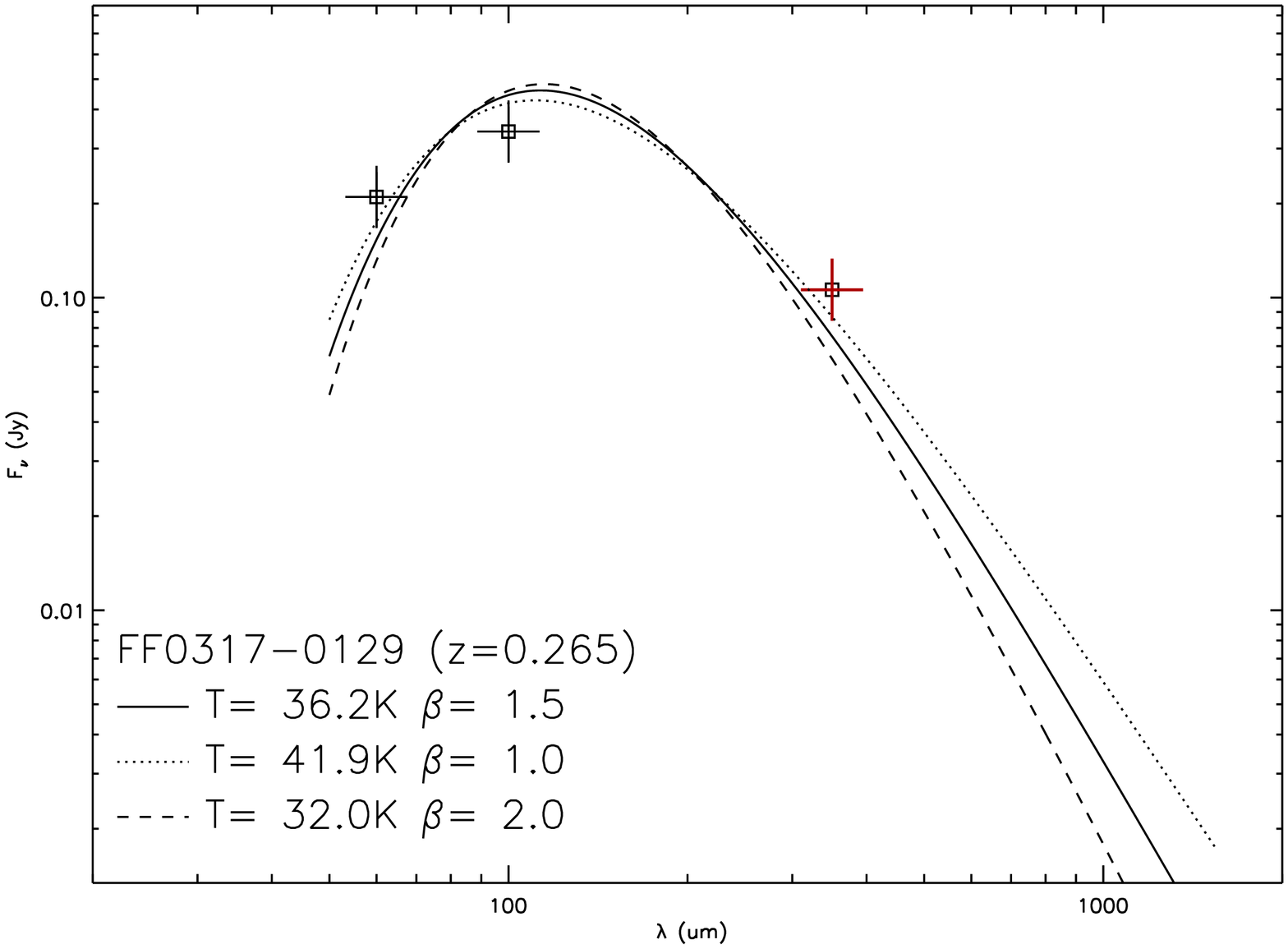}&
\includegraphics[width=1.5in, angle=90]{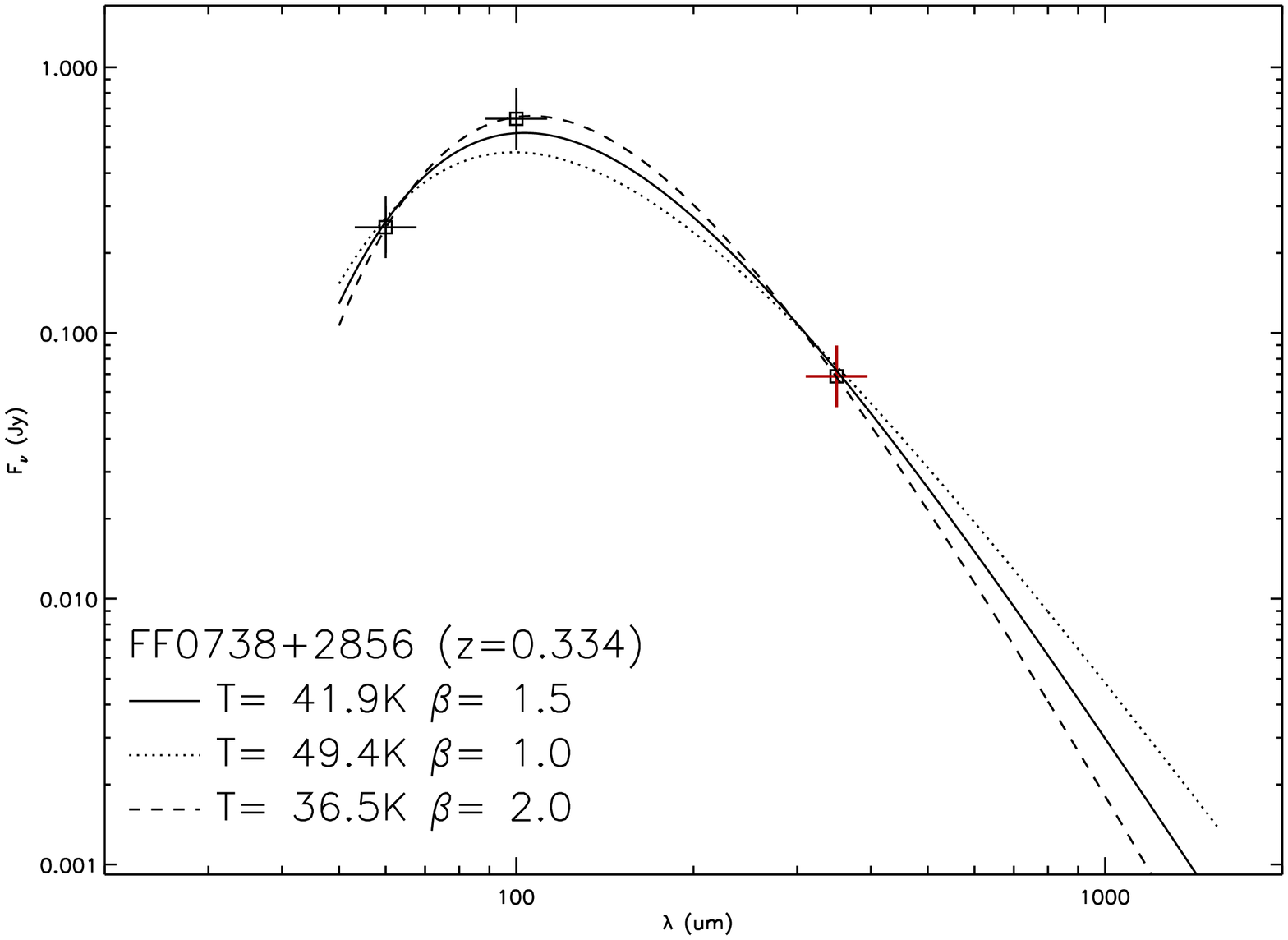}&
\includegraphics[width=1.5in, angle=90]{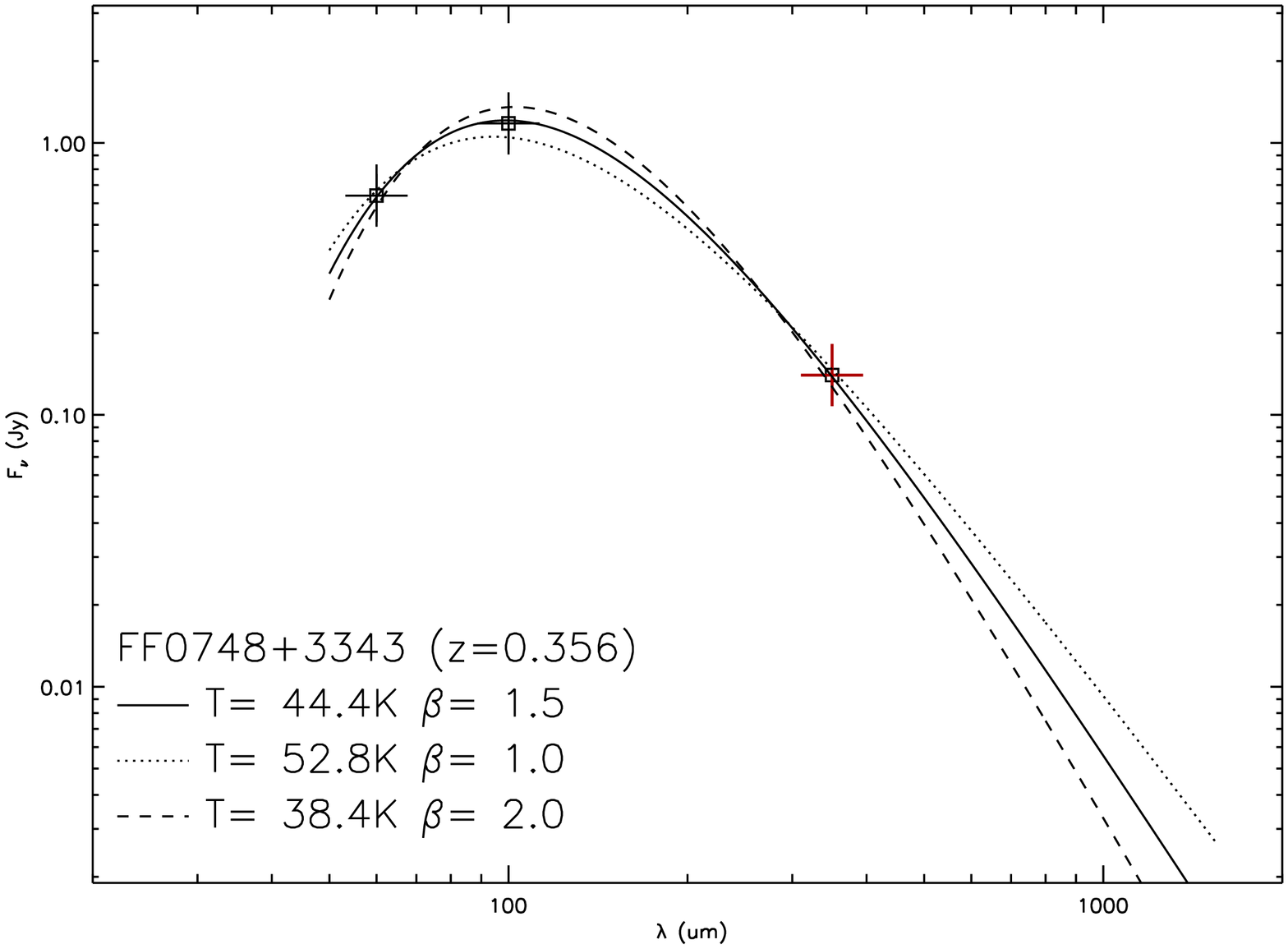}\\
\includegraphics[width=1.5in, angle=90]{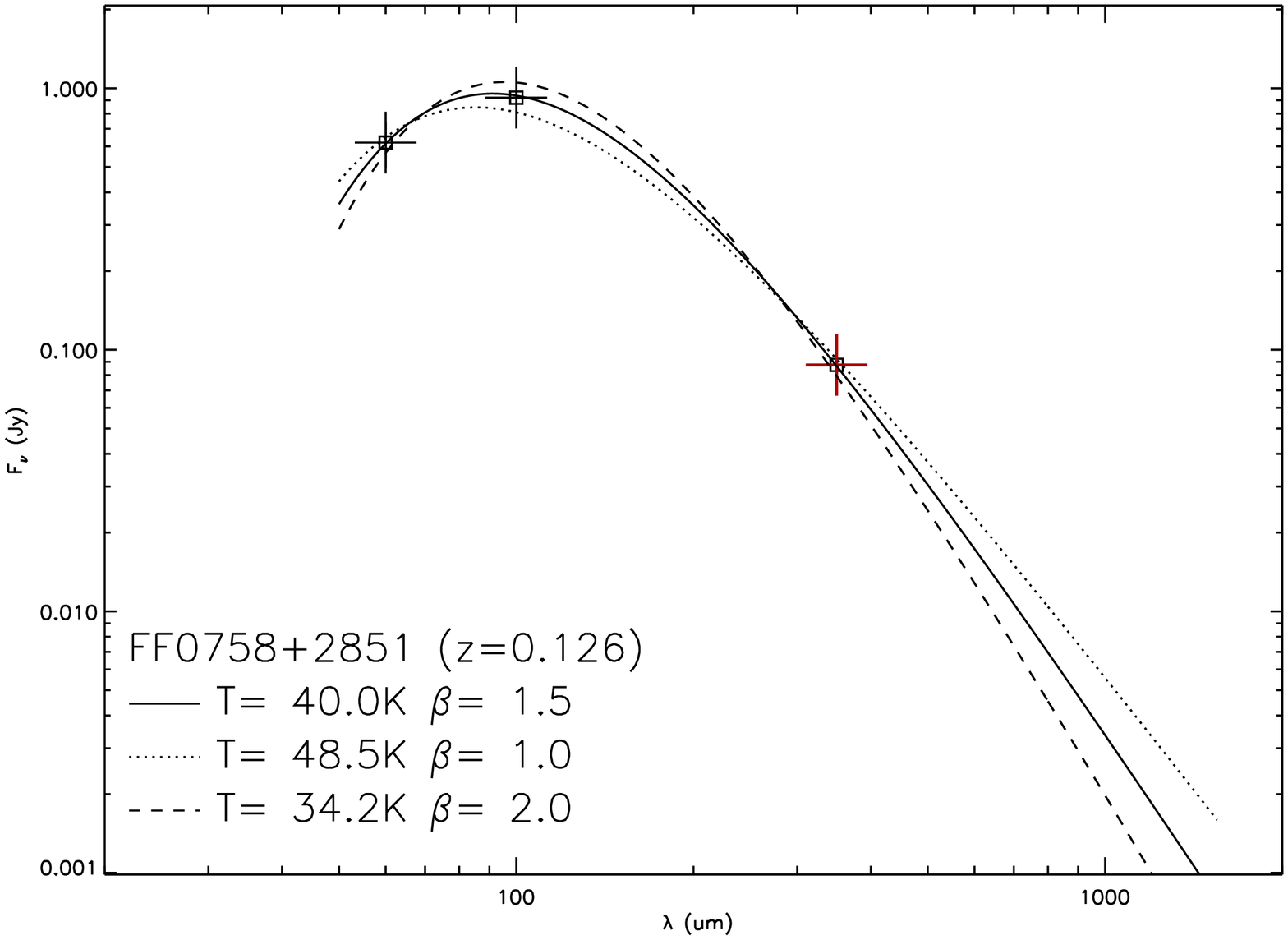}&
\includegraphics[width=1.5in, angle=90]{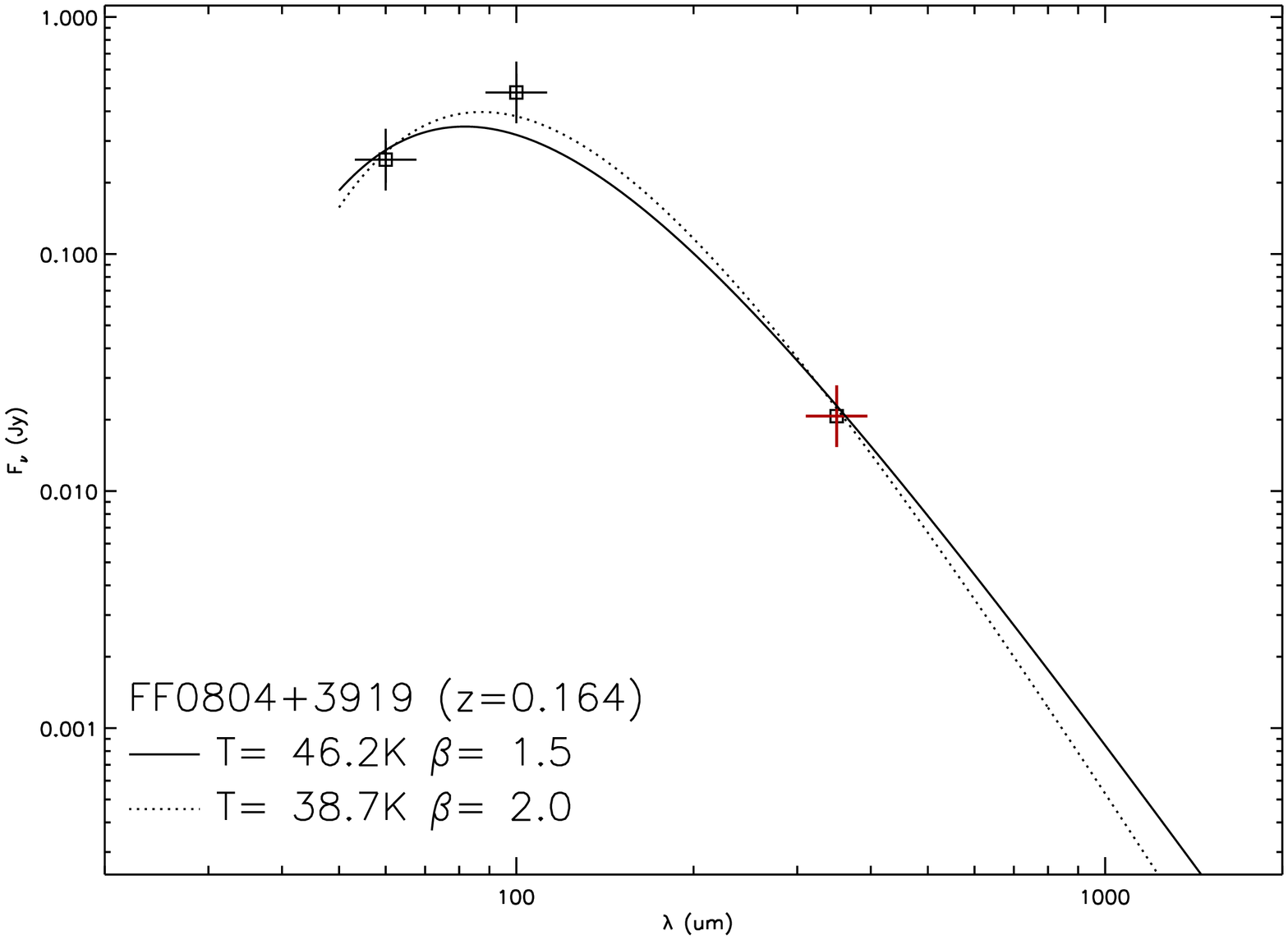}&
\includegraphics[width=1.5in, angle=90]{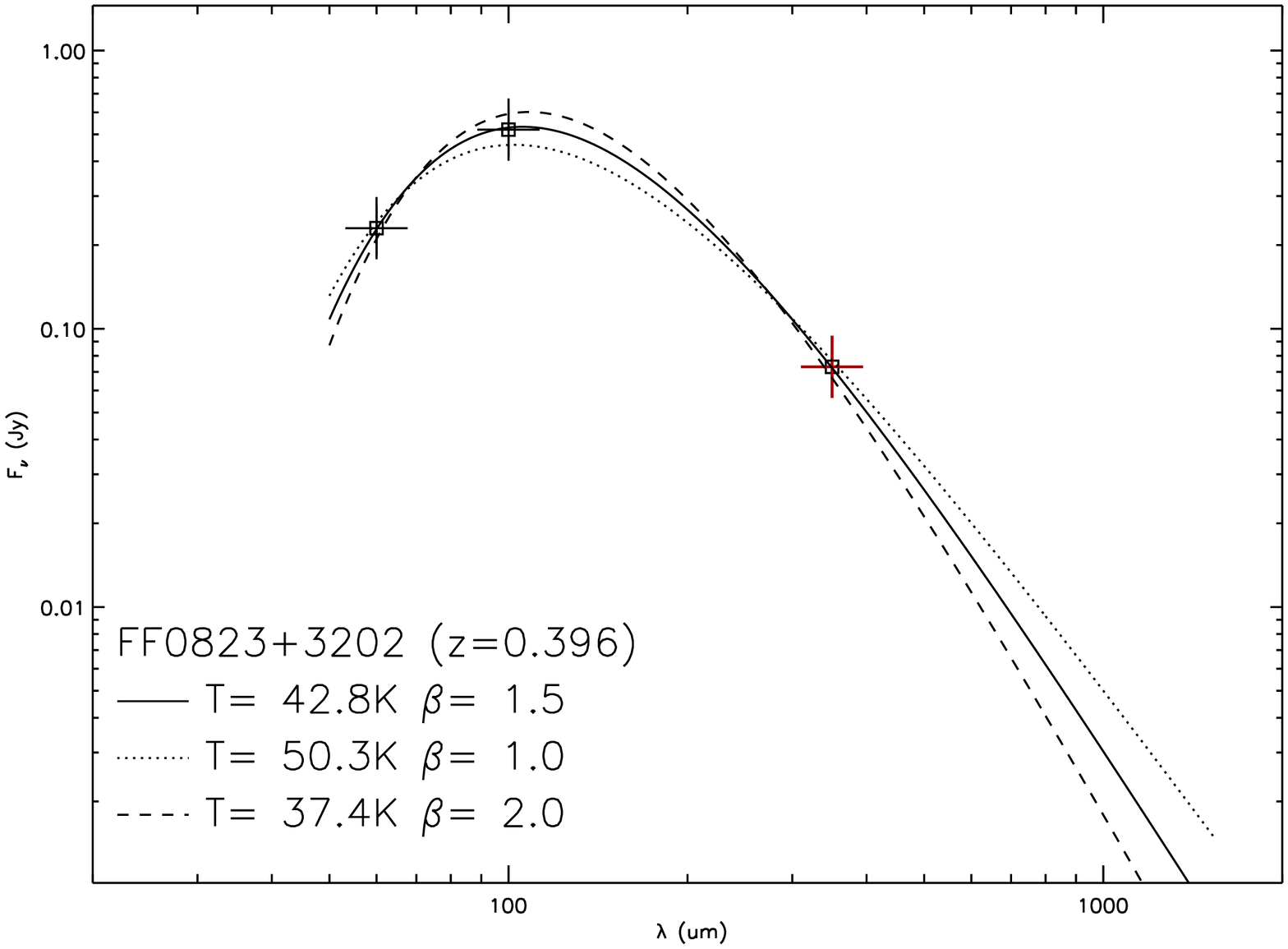}\\
\includegraphics[width=1.5in, angle=90]{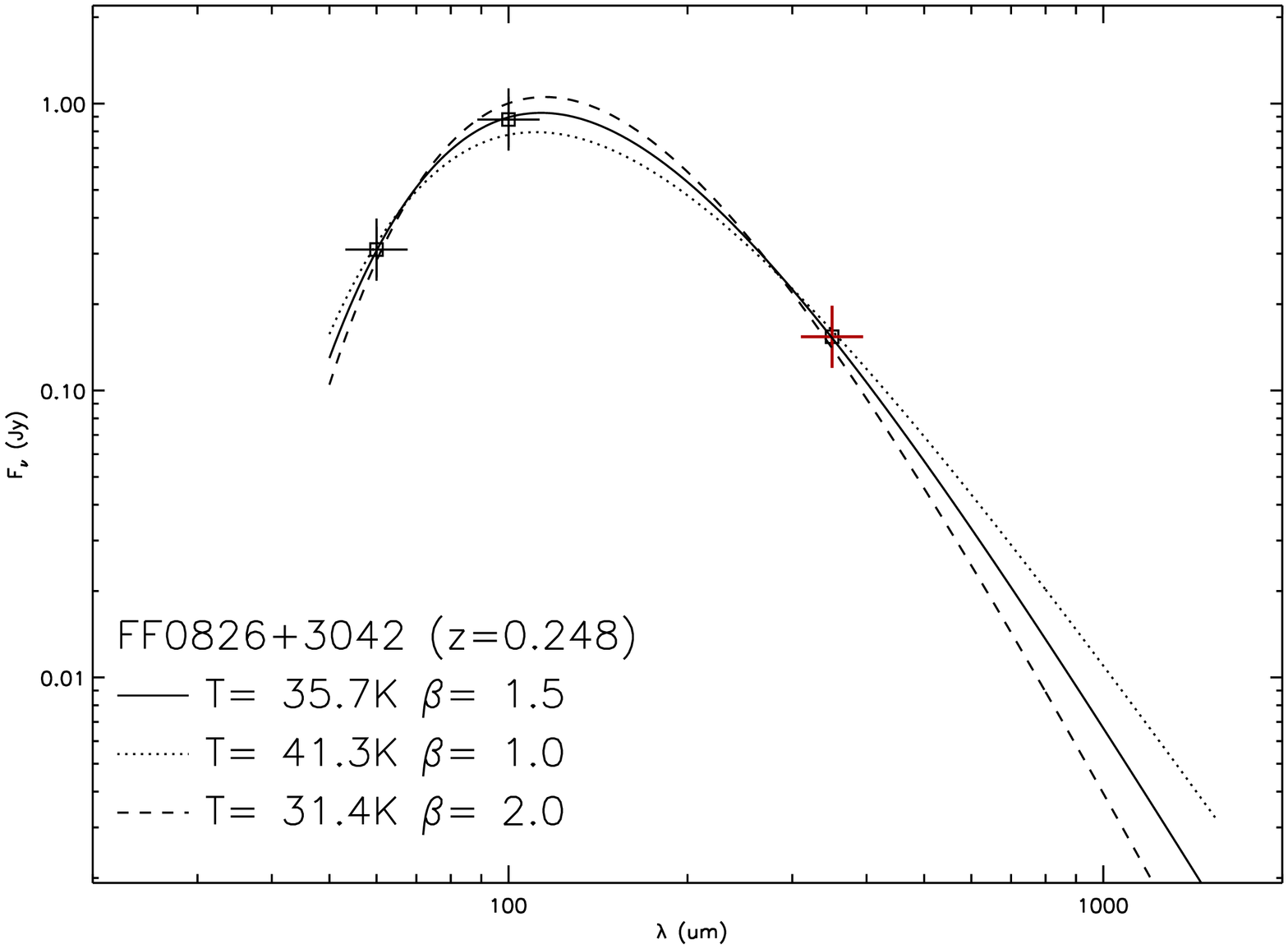}&
\includegraphics[width=1.5in, angle=90]{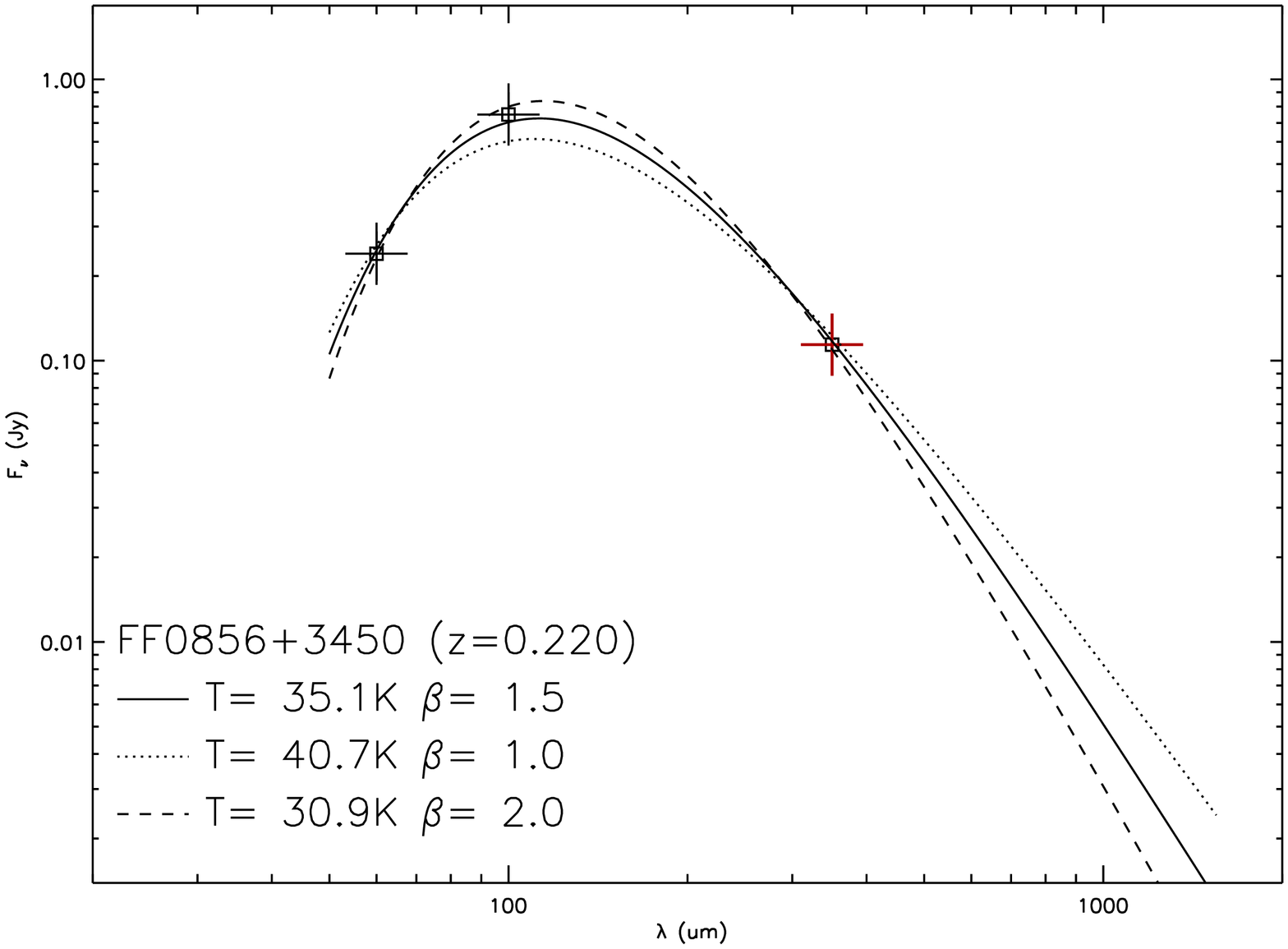}&
\includegraphics[width=1.5in, angle=90]{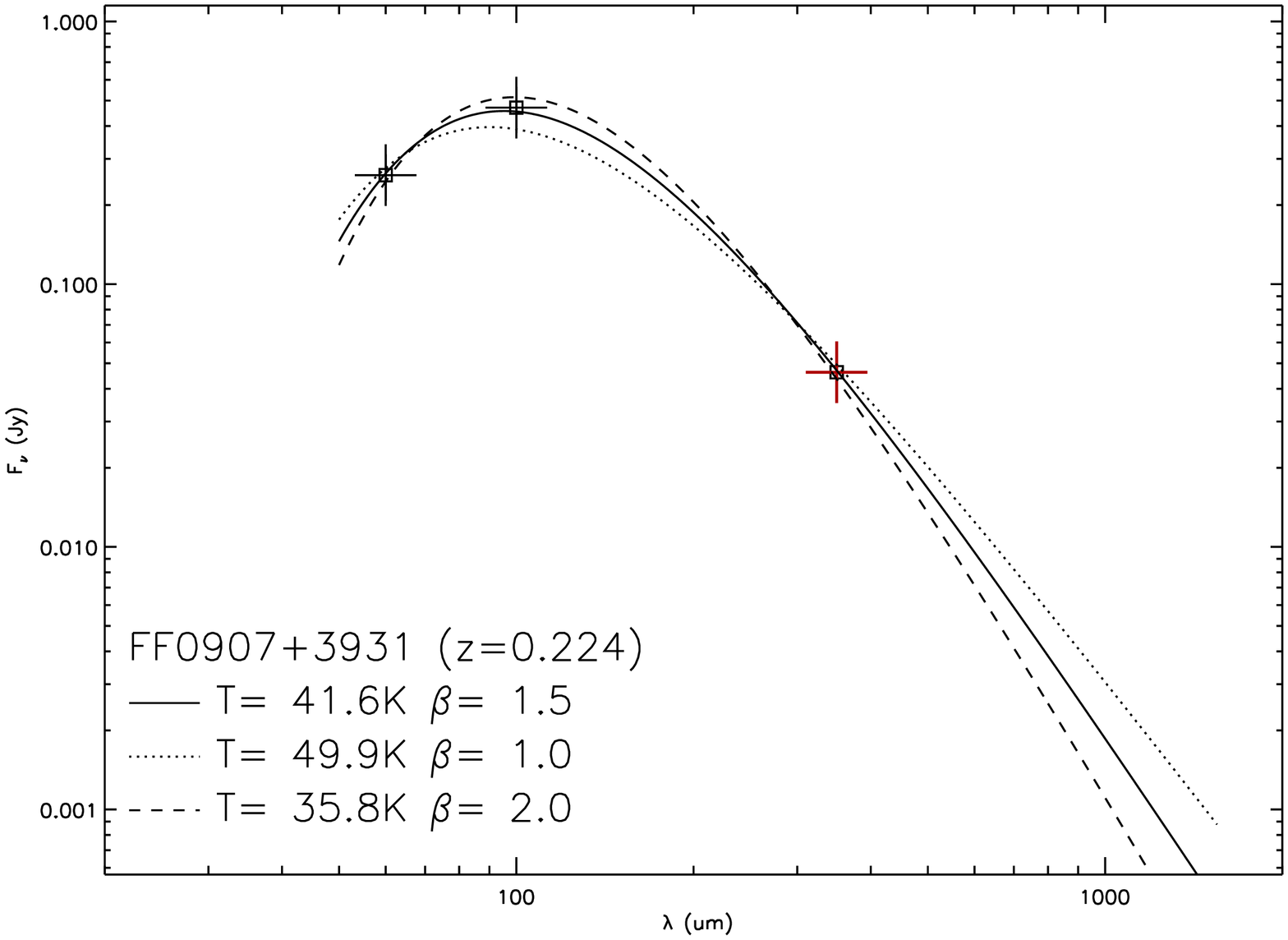}\\
\includegraphics[width=1.5in, angle=90]{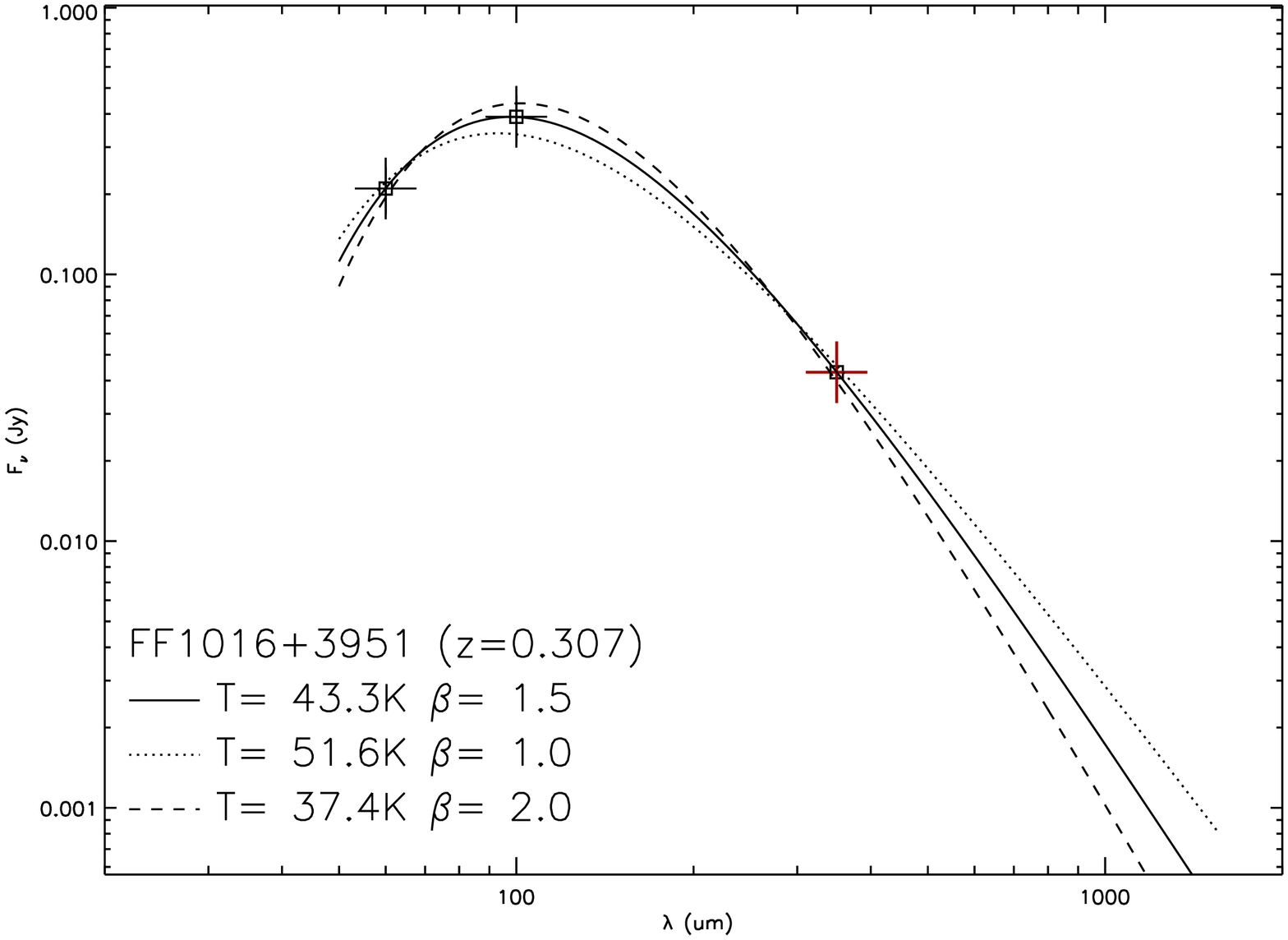}&
\includegraphics[width=1.5in, angle=90]{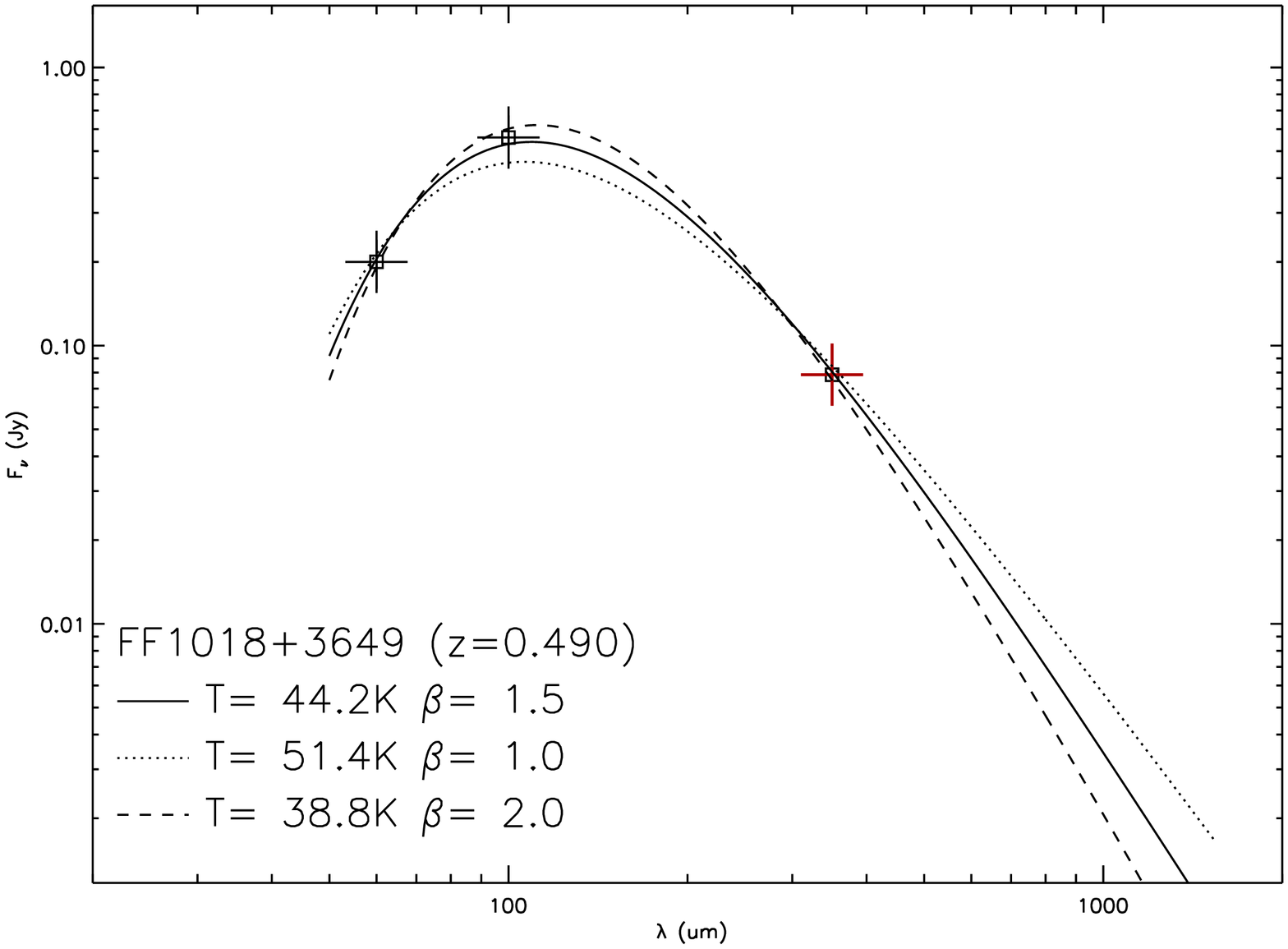}&
\includegraphics[width=1.5in, angle=90]{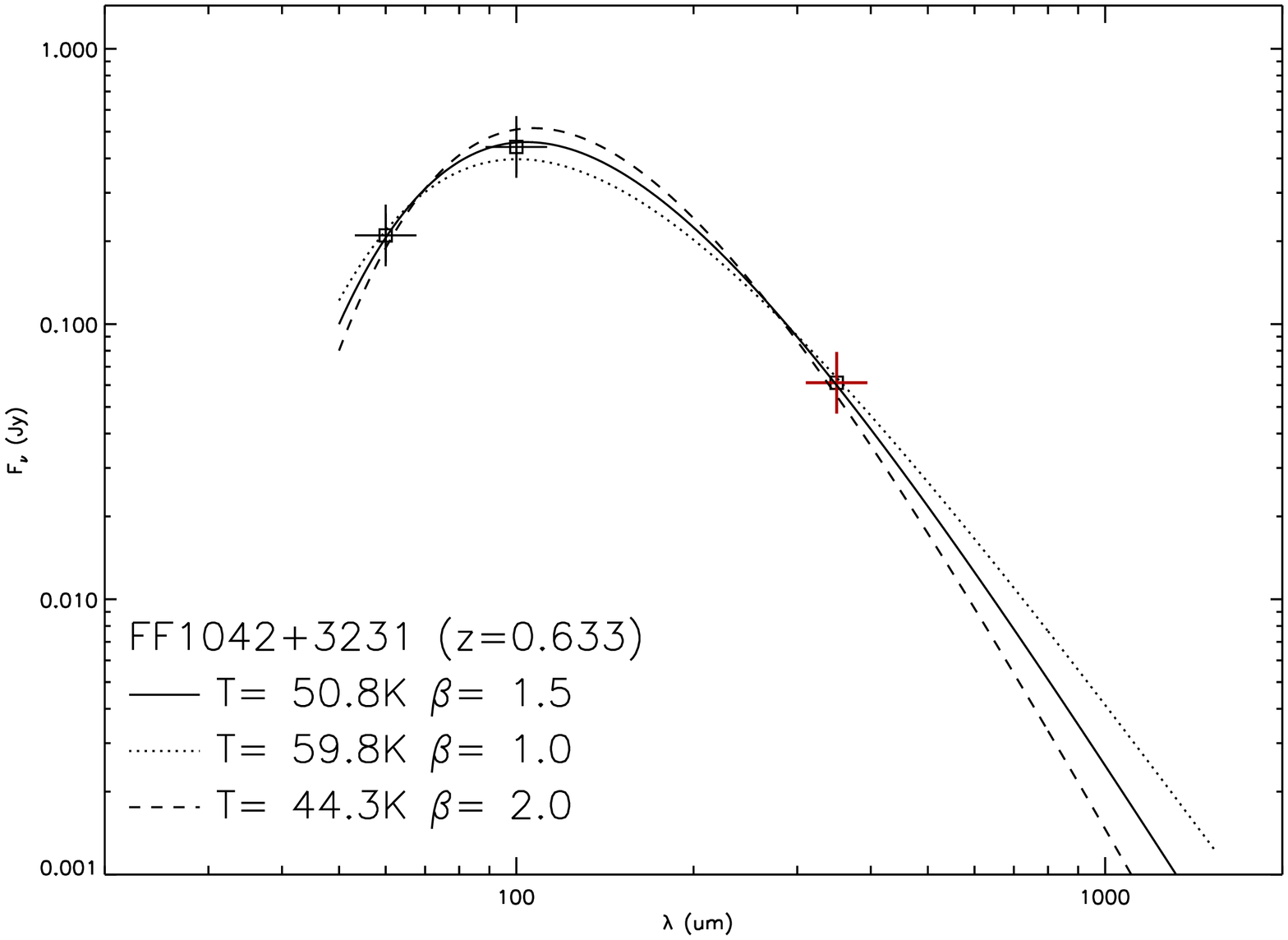}\\
\end{tabular}
\caption{Far-IR/submm SED fits of the SHARC-II detected intermediate-redshift ULIRGs.}\label{sedplot_stanford}
\end{center}
\end{figure*}

\begin{figure*}
\setcounter{figure}{3}
\begin{center}
\begin{tabular}{ccc}
\includegraphics[width=1.5in, angle=90]{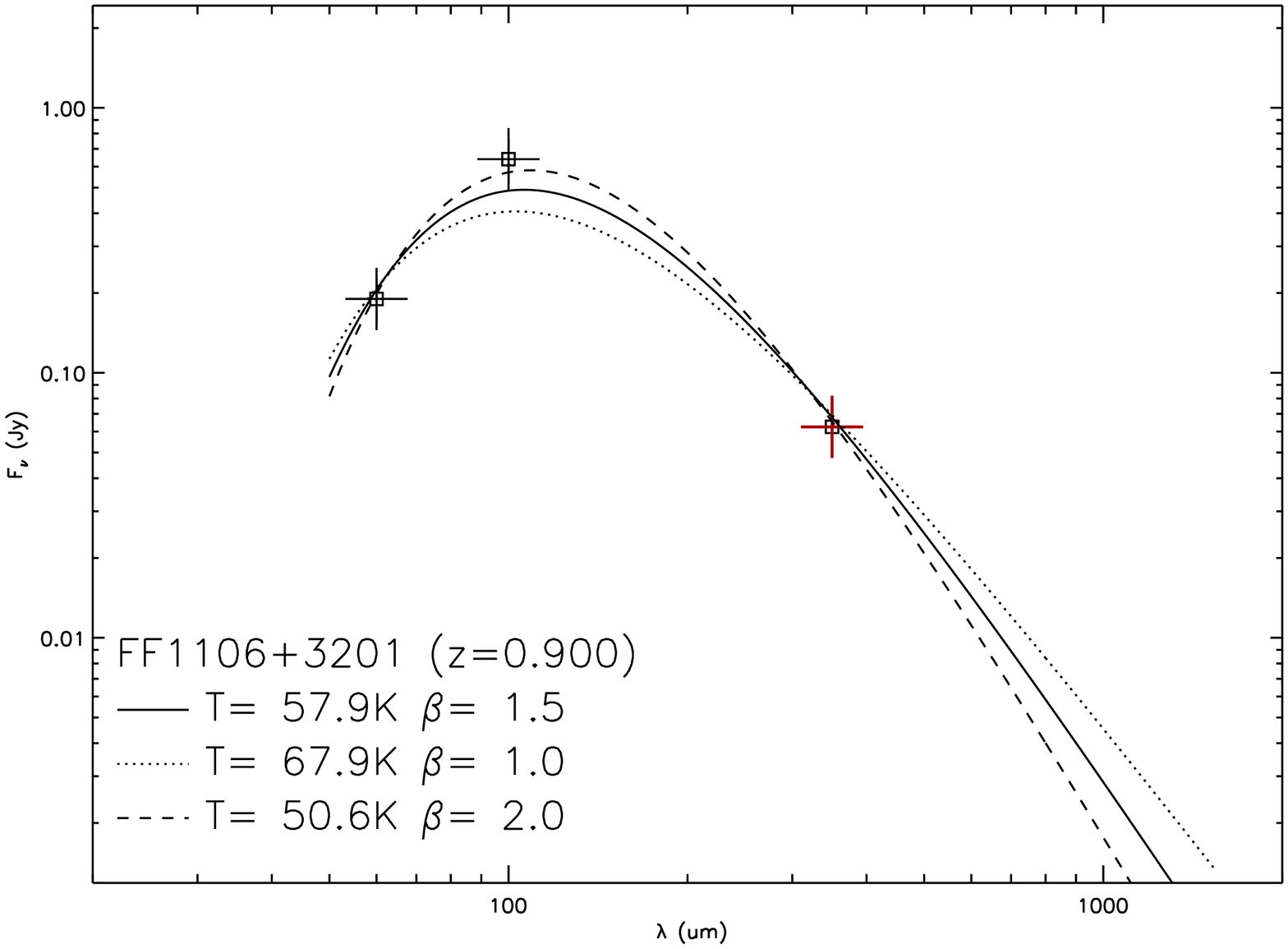}&
\includegraphics[width=1.5in, angle=90]{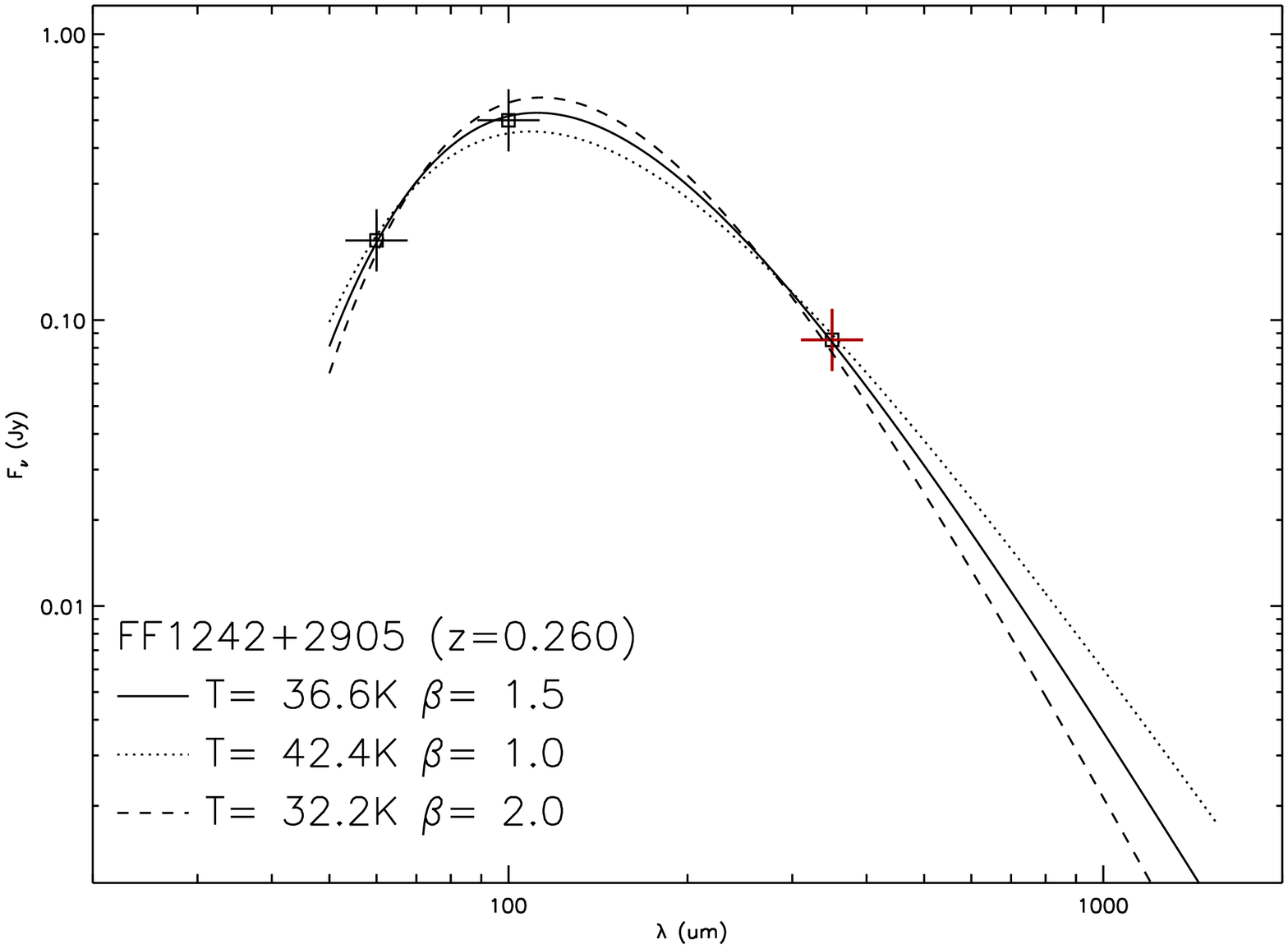}&
\includegraphics[width=1.5in, angle=90]{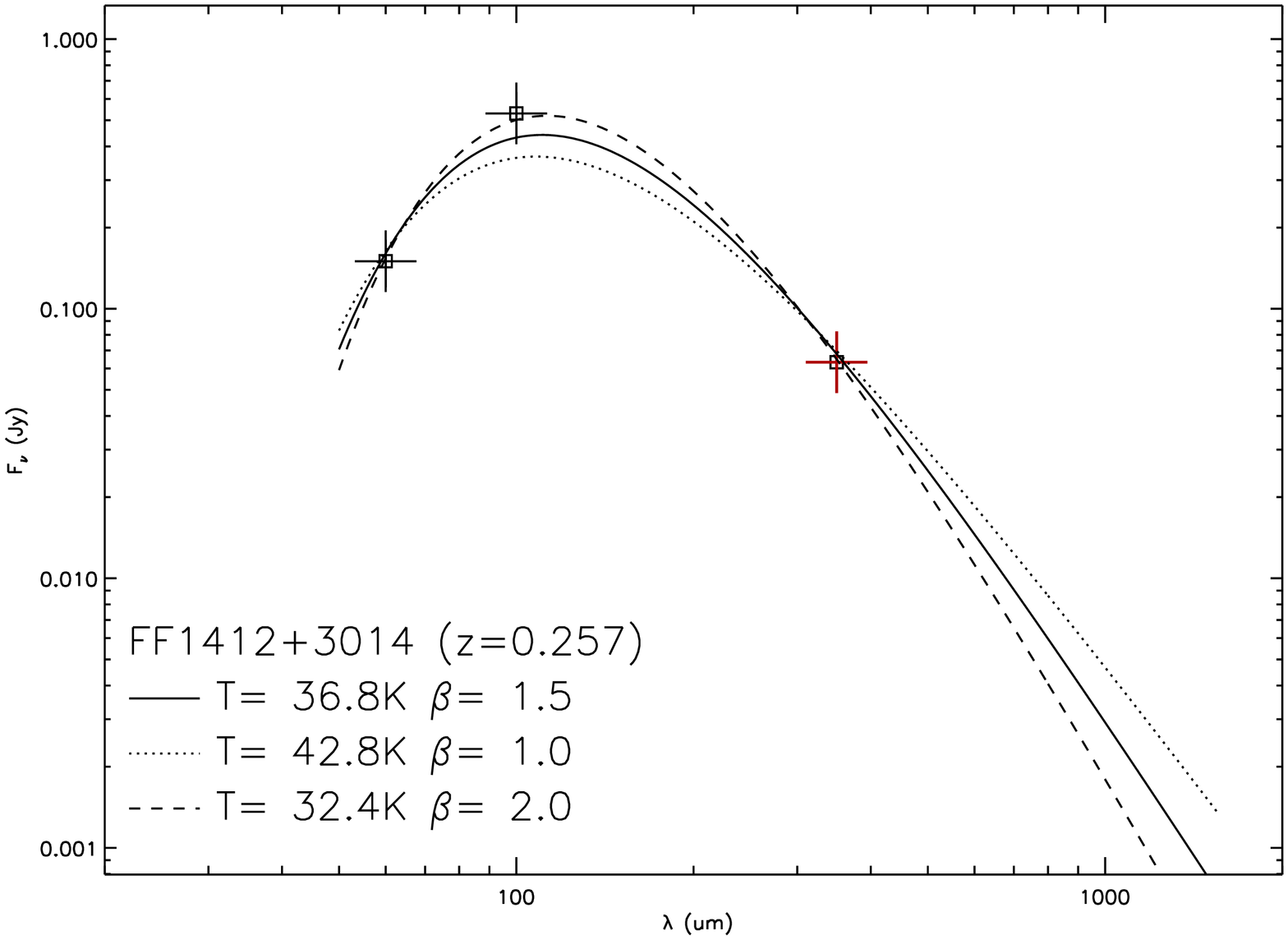}\\
\includegraphics[width=1.5in, angle=90]{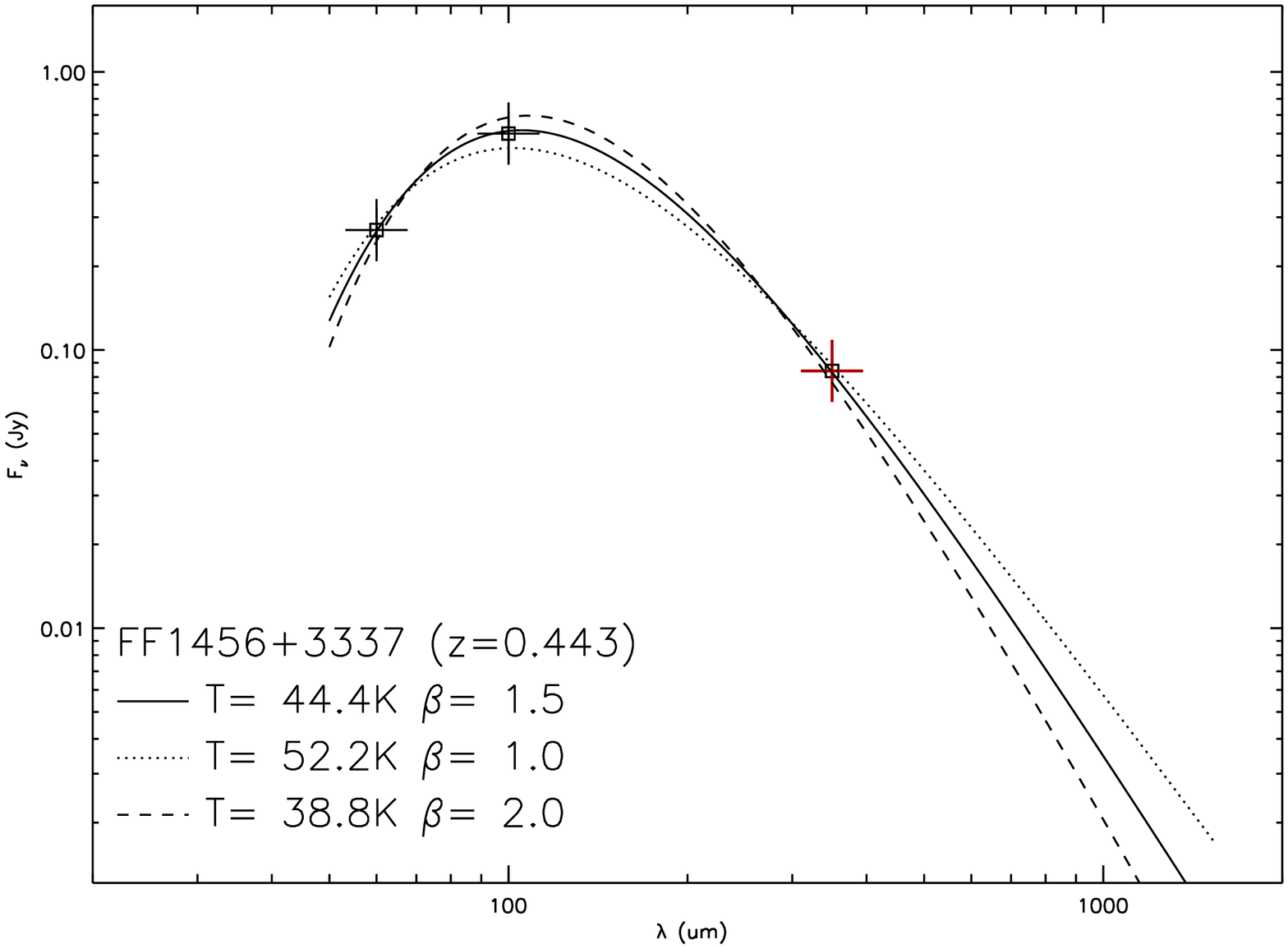}&
\includegraphics[width=1.5in, angle=90]{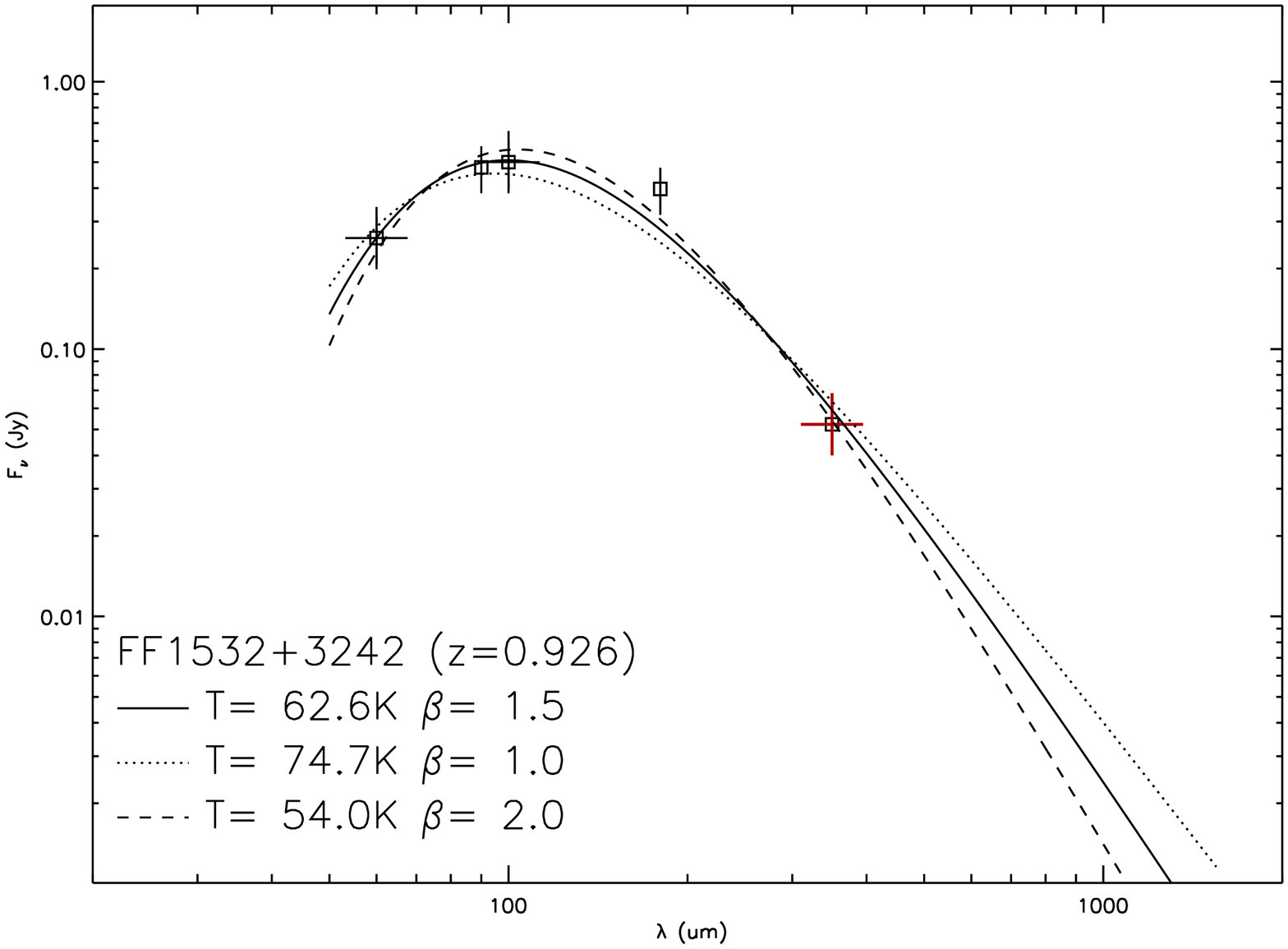}&
\includegraphics[width=1.5in, angle=90]{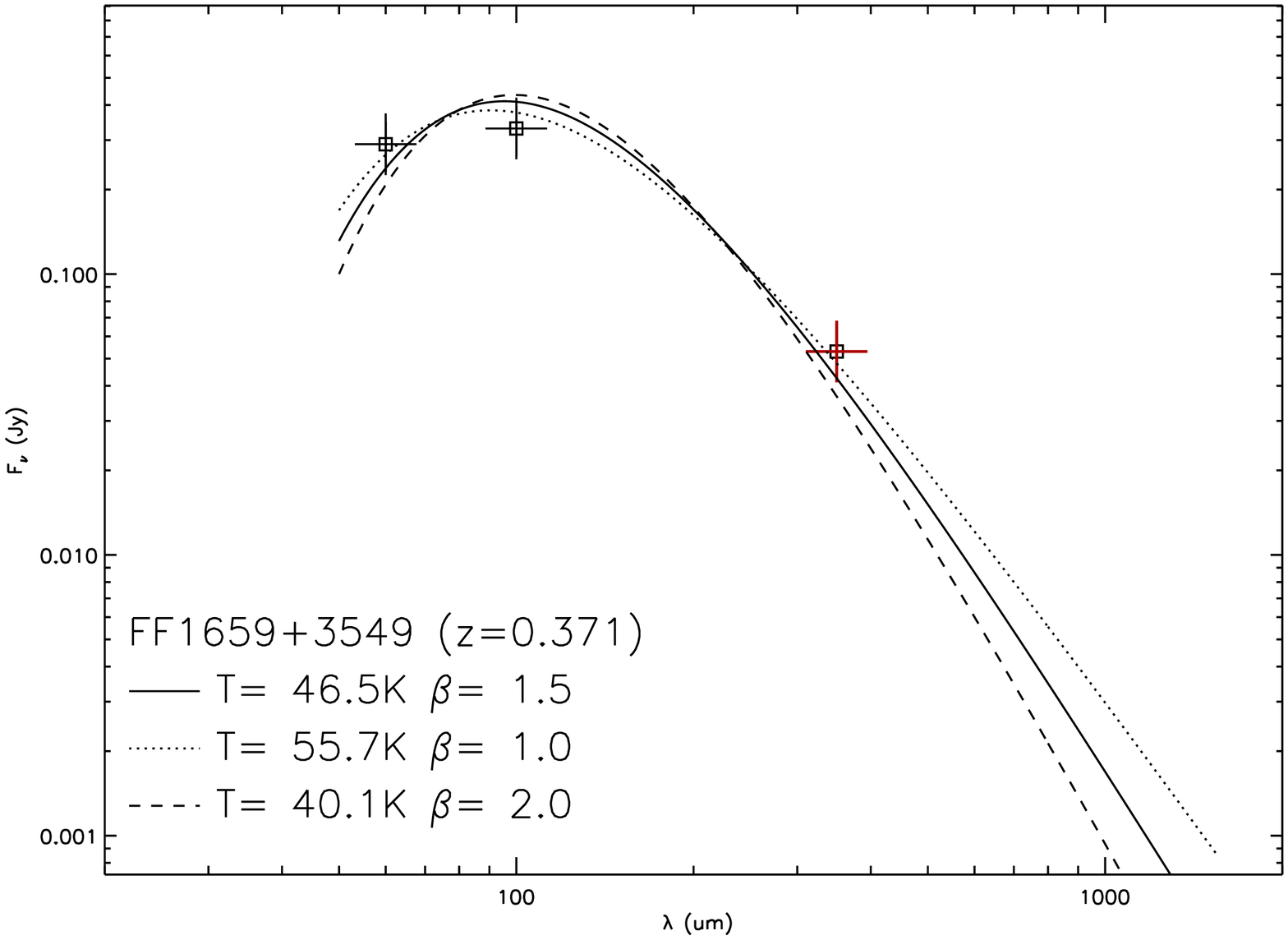}\\
\includegraphics[width=1.5in, angle=90]{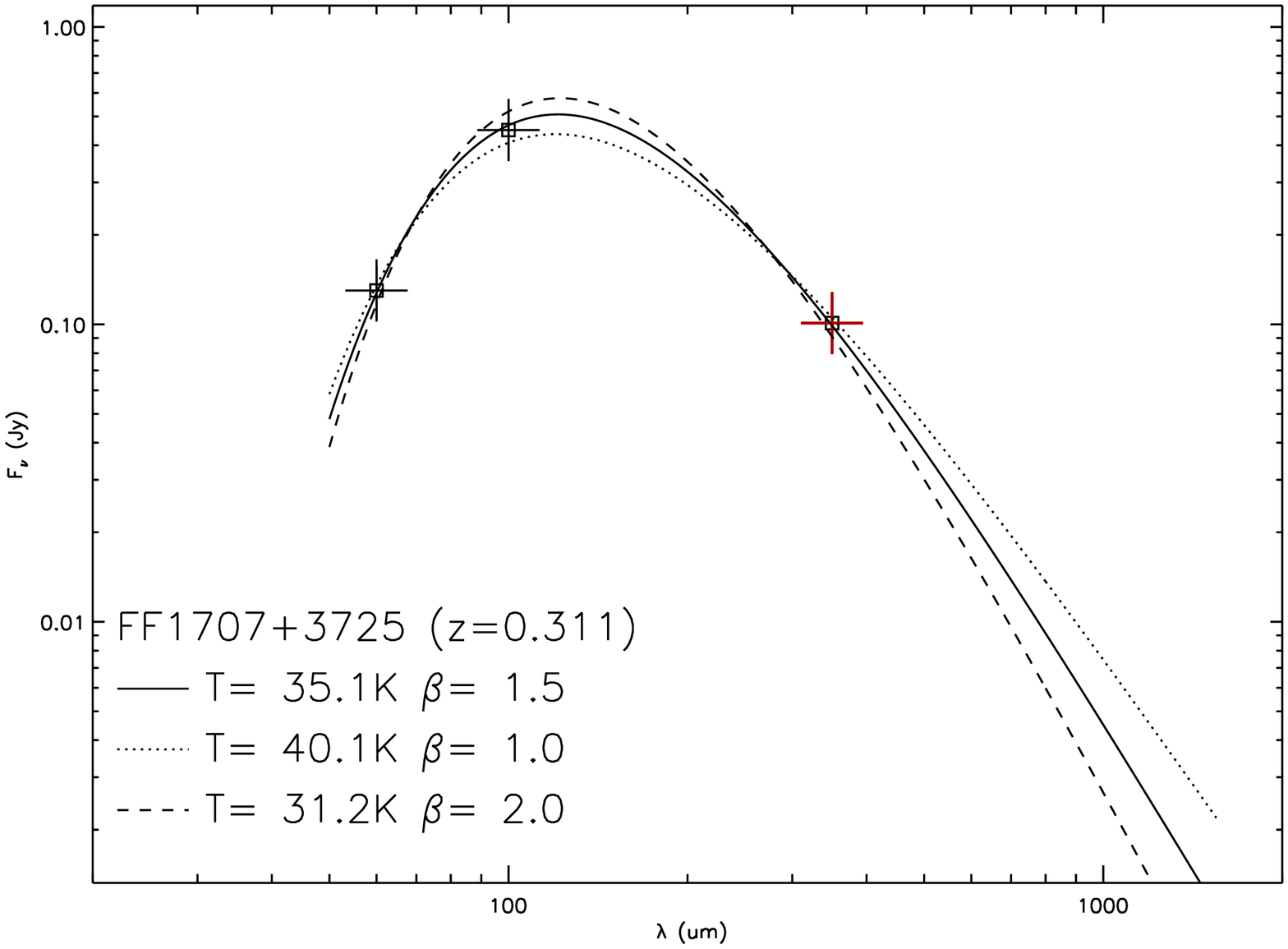}&
\includegraphics[width=1.5in, angle=90]{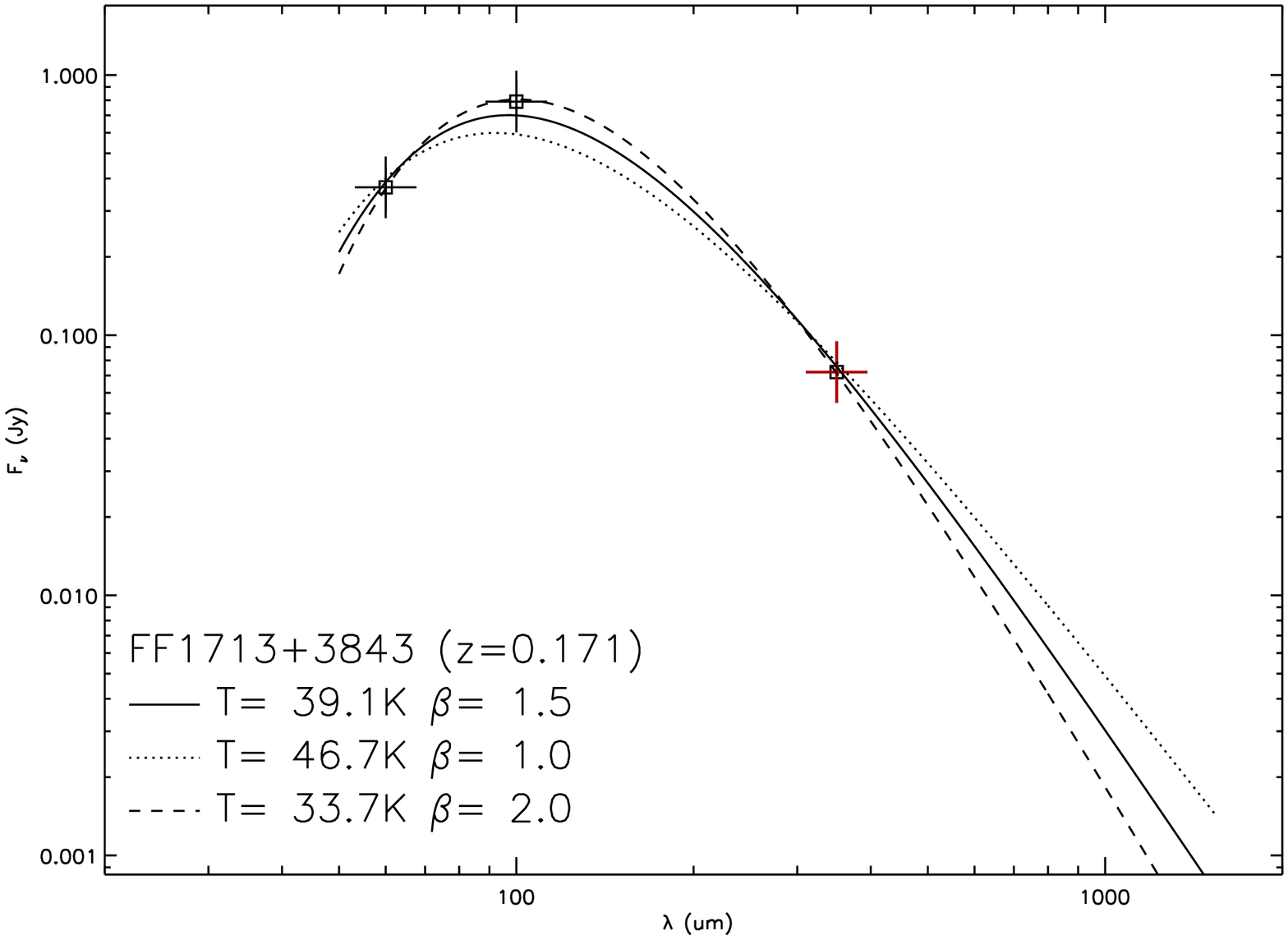}&
\includegraphics[width=1.5in, angle=90]{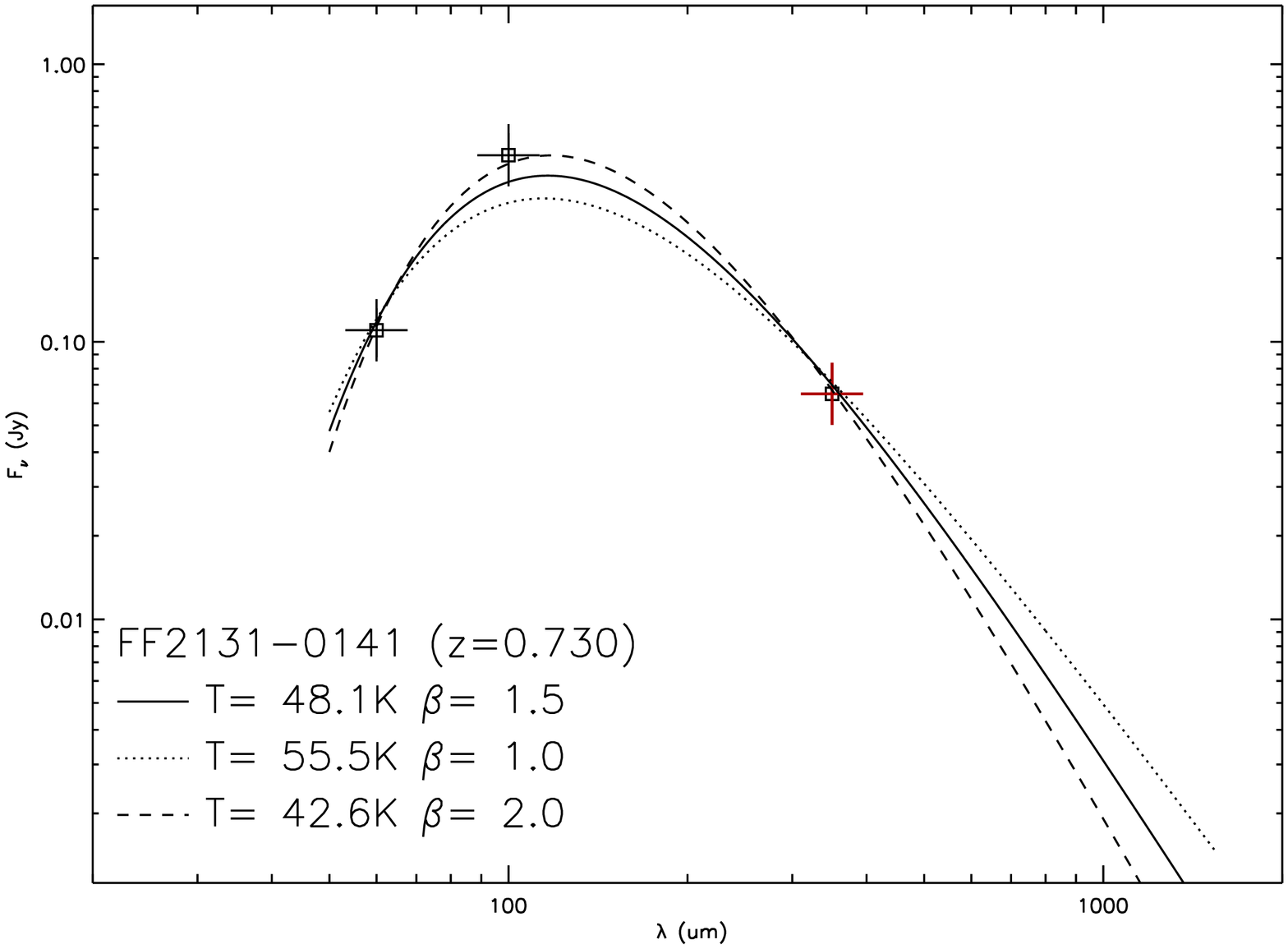}\\
\includegraphics[width=1.5in, angle=90]{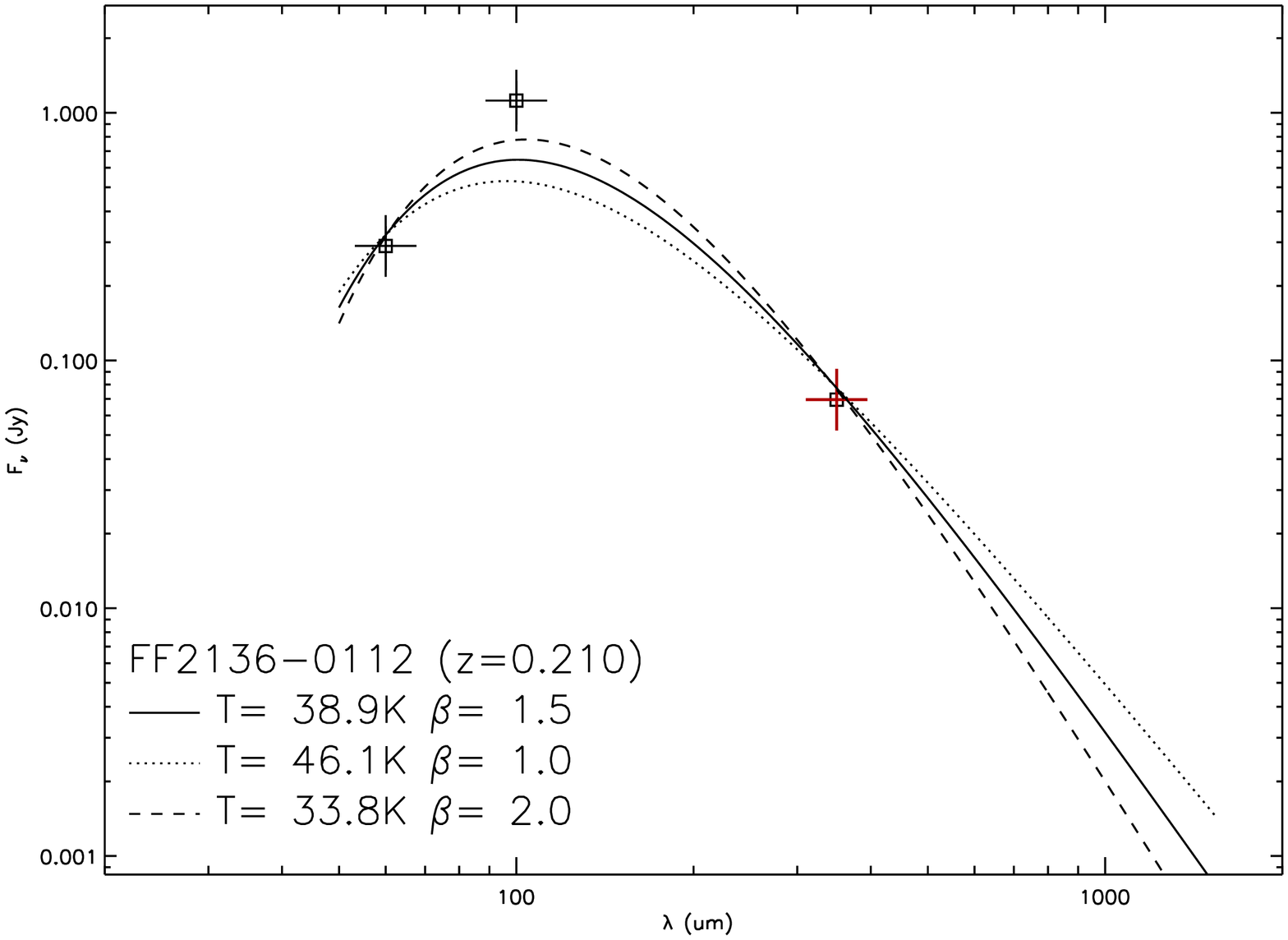}&
\includegraphics[width=1.5in, angle=90]{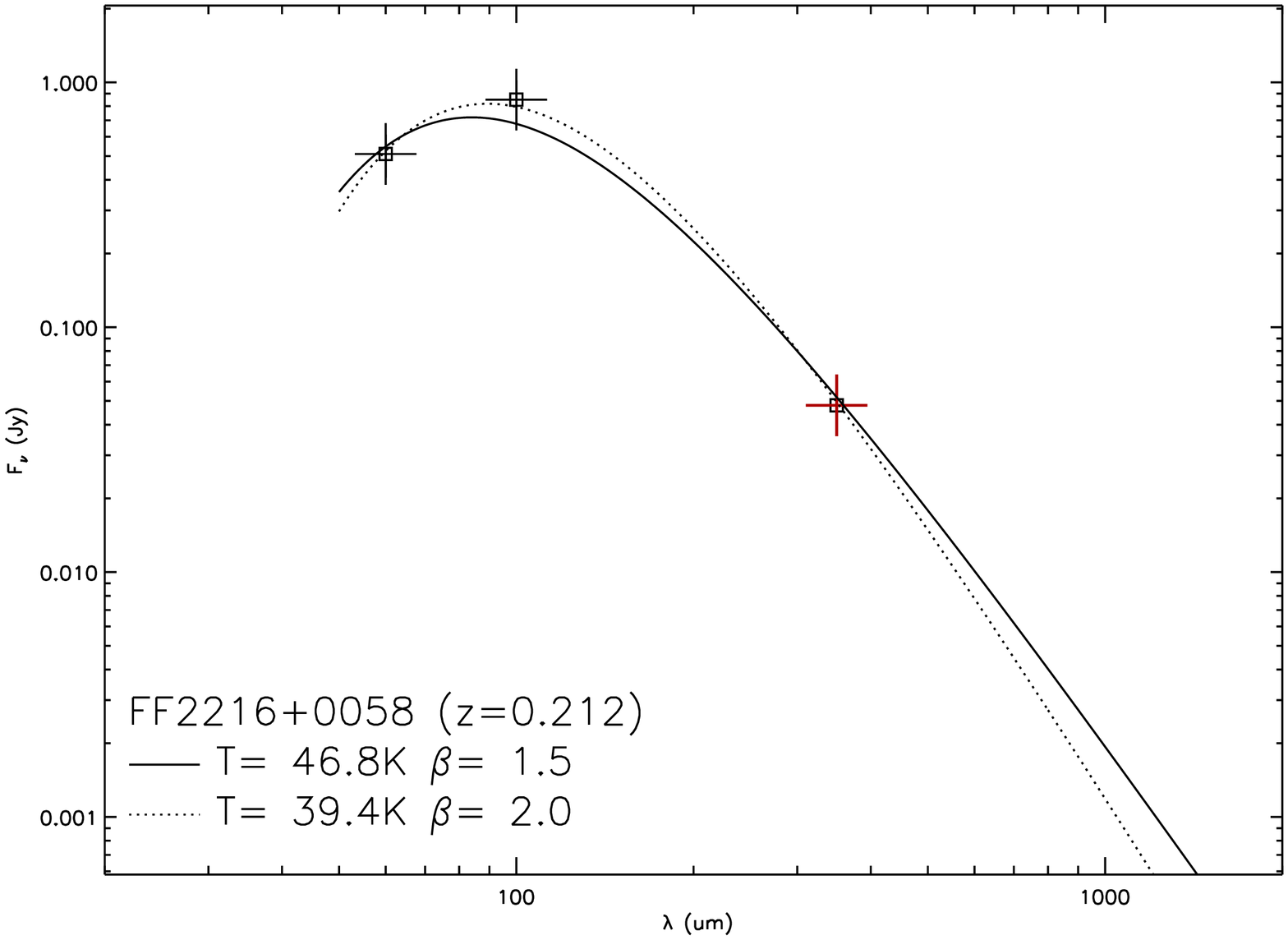}&
\includegraphics[width=1.5in, angle=90]{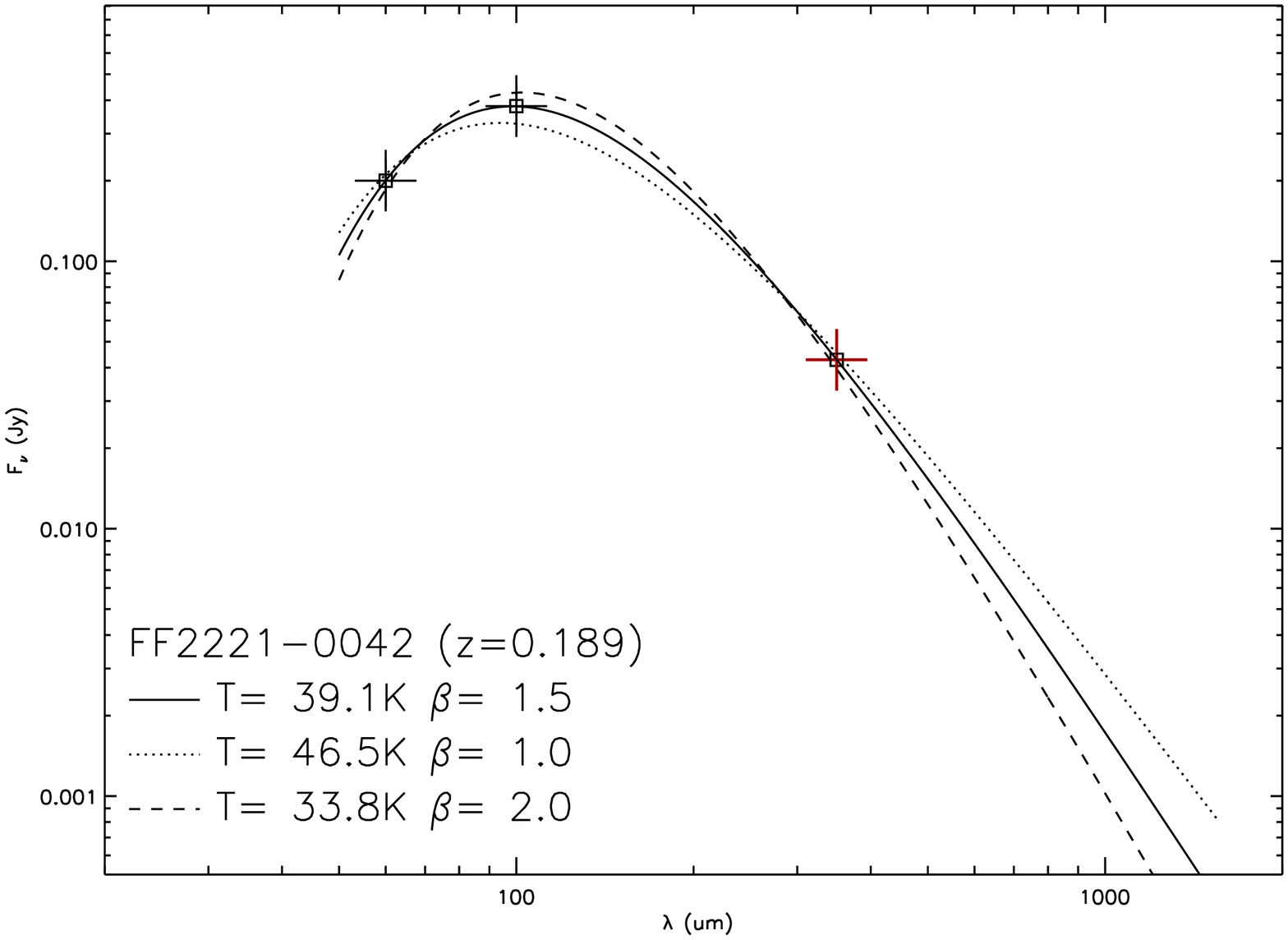}\\
\includegraphics[width=1.5in, angle=90]{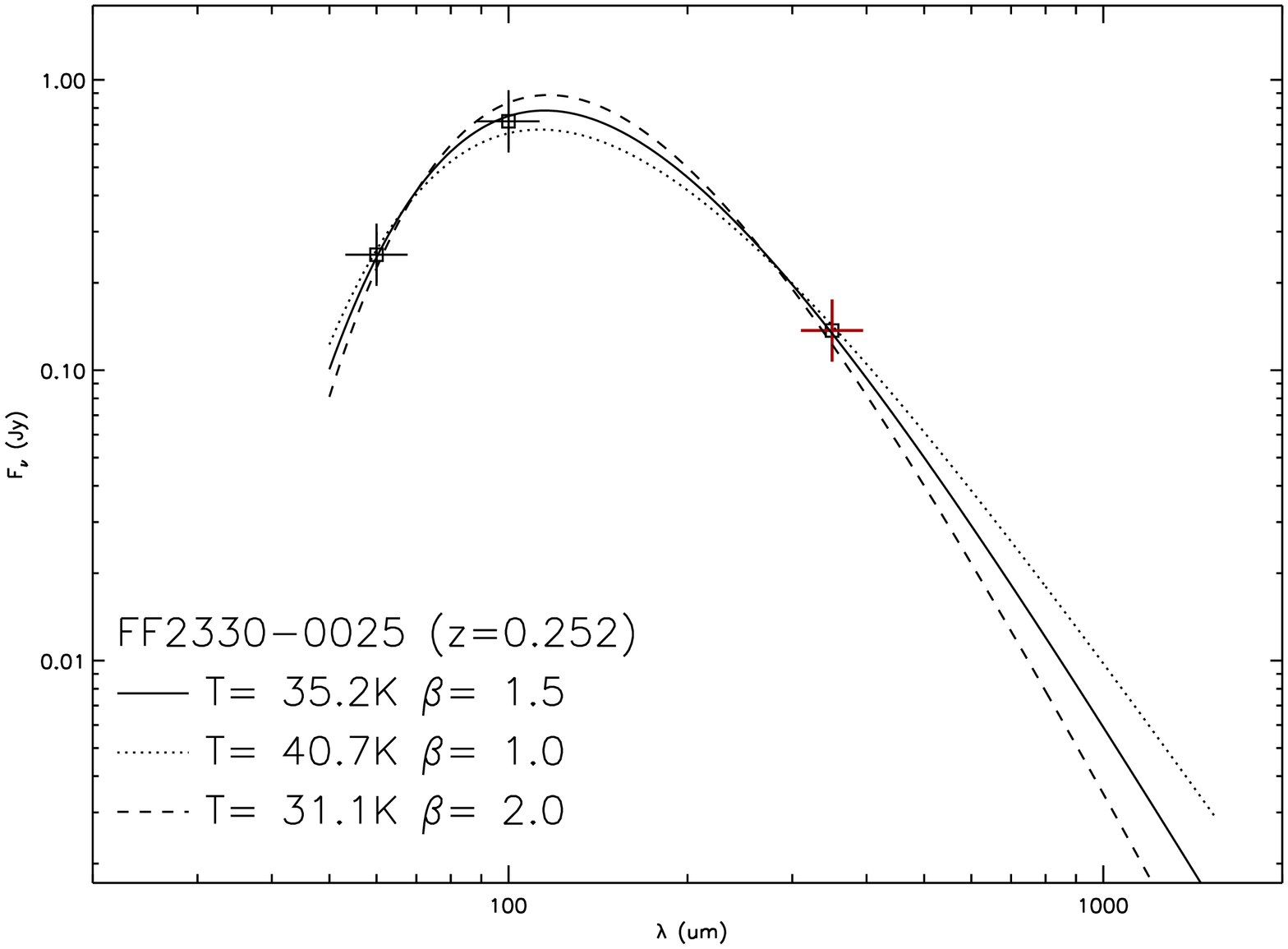}\\
\end{tabular}
\caption{Continued}
\end{center}
\end{figure*}

\end{document}